\documentclass[aps,prd,longbibliography,twocolumn,floatfix,noeprint]{revtex4-2}

\usepackage{amsmath}
\usepackage{amssymb}
\usepackage{amsfonts}
\usepackage{graphicx}
\usepackage{dcolumn}
\usepackage{bm}
\usepackage{latexsym}
\usepackage{slashed}
\usepackage{xcolor}
\usepackage[colorlinks=true,linkcolor=blue,citecolor=blue]{hyperref}
\usepackage{array}

\newcommand{\Rco}{$\mathcal R_{\slashed E}$}

\newcommand{\orcid}[1]{\href{https://orcid.org/#1}{\includegraphics[width=10pt]{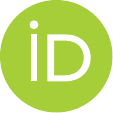}}}

\begin{document}

\title{Particle acceleration up to the synchrotron burn-off limit in relativistic magnetized turbulence}

\author{Martin Lemoine \orcid{0000-0002-2395-7812}} 
\email{mlemoine@apc.in2p3.fr}
\affiliation{Astroparticule \& Cosmologie, CNRS -- Universit\'e Paris-Cité, 75013 Paris, France}
\author{Virginia Bresci \orcid{0000-0001-7237-3373}}
\affiliation{Focused Energy GmbH, 64293 Darmstadt, Germany}
\author{Laurent Gremillet \orcid{0000-0003-0116-5248}}
\email{laurent.gremillet@cea.fr}
\affiliation{CEA, DAM, DIF, F-91297 Arpajon, France}
\affiliation{Universit\'e Paris-Saclay, CEA, LMCE, F-91680 Bruy\`eres-le-Ch\^atel, France}

\date{\today}

\begin{abstract}
In multimessenger high-energy astrophysics, interpreting observed spectra often hinges on understanding the underlying competition between energy gains and radiative losses within the accelerator. To progress along these lines, we report here on numerical particle-in-cell simulations of particle acceleration in relativistic, magnetized turbulent pair plasmas including synchrotron radiative losses. We investigate a regime of weak synchrotron cooling, where the maximal energy predicted by balancing the acceleration and radiation rates resides in the suprathermal tail, i.e., above the thermal bulk and below the confinement energy associated with the outer scale of the turbulence. Our key finding is that the particle energy spectrum, along with the radiated synchrotron spectral power, does not terminate at this maximal energy (or corresponding frequency), but extends significantly beyond with a steepened spectrum, up to the synchrotron burn-off limit where particles cool within a gyrotime. For our adopted parameters (turbulence magnetization parameter $\sigma \approx 1 $ and amplitude $\delta B/B_0\simeq 1$), the particle distribution function follows ${\rm d}n/{\rm d}\gamma\propto \gamma^{-s}$ with $s\simeq 3$ below the predicted maximal energy, then steepens to $s\simeq 4$ above. The particle energy distribution and the radiated synchrotron spectra display strong variability near the cutoff energy down to timescales well below the largest eddy turn-around time. We substantiate our results by demonstrating that the acceleration rate itself displays a broken power-law-like distribution, whose mean value defines the diffusion coefficient and whose maximal value is the gyrofrequency. We perform a detailed analysis of the acceleration mechanism for the highest-energy particles, and demonstrate that they are accelerated to this extreme limit by a generalized Fermi process in ideal electric fields. This process is driven by a gradient of the four-velocity field $u_E$ of the magnetic field lines of relativistic amplitude, $\delta u_E \gtrsim c$, ordered on a scale comparable to, or larger than, the particle gyroradius at the synchrotron burn-off limit. We contend that this is a generic feature of relativistic, large-amplitude turbulence. Lastly, we apply our results to the case of the Crab nebula, which exhibits a hierarchy of characteristic Lorentz factors similar to that studied here. We conclude that stochastic acceleration in this environment is a promising mechanism for explaining the highest-energy part of the synchrotron spectral energy distribution, and its variability. 
\end{abstract}

\maketitle

\section{Introduction}
\label{sec:intr}

Stochastic particle acceleration in magnetized turbulent plasmas is commonly invoked to explain the production of high-energy particles in astrophysical phenomena of energy release, from the solar atmosphere to the remote Universe, e.g.~\cite{2012SSRv..173..535P}. Since turbulence naturally occurs in astrophysical plasmas, and its energy dissipation is controlled by a few parameters, astrophysical stochastic acceleration is, in some sense, universal. In the realm of relativistic and multimessenger astrophysics, this acceleration process is also noted for a few rare virtues: it operates efficiently in the highly magnetized regime, and it tends to concentrate most of the energy in the highest-energy particles, e.g.~\cite{1962SvA.....6..317K,1984A&A...136..227S}.

One key input for theoretical models of multimessenger emissions is the highest energy that a particle can reach in a given accelerator. To determine this maximum energy, standard practice is to equate the characteristic acceleration rate with that of energy losses (or the escape and dynamical rates of the system). In the context of stochastic acceleration, the acceleration rate $\nu_{\rm acc}$ is commonly defined through the momentum diffusion coefficient, $D_{\gamma\gamma}$, as a function of particle Lorentz factor $\gamma$ (for relativistic particles), e.g., $\nu_{\rm acc} \propto D_{\gamma\gamma}/\gamma^2$. Although numerical simulations of particle acceleration in turbulent plasmas have long been used to study the acceleration processes~\cite{2004ApJ...617..667D,06Arzner,2012PhRvL.108x1102K,2013ApJ...777..128L,2014ApJ...783..143D,2017PhRvL.119d5101I,2018JPlPh..84f7201P,2019MNRAS.485..163K,2020ApJ...894..136T,2021MNRAS.506.1128S,2022ApJ...928...25P,2022ApJ...924L..19V,2023ApJ...959...28P,2023ApJ...944..122M},  detailed estimates of the diffusion coefficient have been obtained only recently through kinetic simulations in trans- to fully relativistic turbulence~\cite{2018ApJ...867L..18Z,18Comisso,2019ApJ...886..122C,2019PhRvL.122e5101Z,2020ApJ...893L...7W,2025MNRAS.543.1842W}. However, neither the spectral shape of the particle energy distribution around the maximal energy, as set by energy losses, nor the precise value of this maximal energy has yet been examined in detail. This observation forms the basis of the present paper, which seeks to study the spectral shape at and above the expected cutoff energy, by incorporating synchrotron energy losses in particle-in-cell (PIC) kinetic simulations of relativistic turbulence.

The influence of radiative cooling on relativistic magnetized turbulence has been addressed in a number of studies, yet mostly in the strong cooling regime~\cite{2020MNRAS.493..603Z,2020ApJ...895L..40C,2021PhRvL.127y5102C}, in which case the thermal bulk is itself subject to cooling on timescales shorter than the light crossing time of the simulation box, at magnetization parameters well above unity, or for losses corresponding to inverse Compton cooling on a soft photon background~\cite{2021ApJ...921...87N,2024PhRvL.132h5202G}. Other studies have considered the influence of escape losses on the spectral shape~\cite{2024ApJ...977L..18C,2025arXiv250303820G}  or used cooling as a tool to inhibit particle acceleration in the early stage of turbulence generation~\cite{2024PhRvL.133d5202B}. Here, we will rather be interested in a weak cooling case, in which radiative losses affect only part of the suprathermal population. One prototypical source for the scenario that we will examine is a pulsar wind nebula, where the magnetization of the turbulent (pair) plasma downstream of the termination shock is expected to be of order unity, and the magnetic field weak enough to leave the bulk uncooled for many turn-around times~\cite{2019hepr.confE..33A,2021Univ....7..448A}.

The aforementioned kinetic numerical simulations of particle acceleration in relativistic turbulence have revealed quite a few surprises and deviations from the generic phenomenological framework on which the theoretical estimates of the maximal energy rely. This includes, for instance, the observation of sudden bursts of particle acceleration in isolated regions~\cite{2017PhRvL.119d5101I, 2018JPlPh..84f7201P, 2019MNRAS.485..163K, 2020ApJ...894..136T, 2021ApJ...921...87N} or the generation of power-law tails that harden with increasing turbulence amplitude and magnetization~\cite{2018ApJ...867L..18Z, 18Comisso,2019ApJ...886..122C, 2022PhRvL.129u5101L}. These observations suggest that
particles can be accelerated at widely different rates~\cite{2020MNRAS.499.4972L}, and thus provide further motivation for the present study. The inhomogeneity of the magnetic field strength in a turbulent bath may also modify the shape of the cutoff, since particles may be accelerated in weak-field regions and cool in strong-field regions~\cite{2012MNRAS.427L..40K, 2021ApJ...914...76K}. In relativistic turbulence, Lorentz boosts associated with the fast motion of the acceleration sites can further distort this spectrum. Finally, for particles that are preferentially accelerated along the magnetic field direction, their small pitch angles tend to reduce synchrotron losses and help them reach higher energies. As a clear realization of these possibilities, acceleration in sharp, dynamic curvature bends of the magnetic field increases the longitudinal momentum of the 
particle~\cite{2021PhRvD.104f3020L}, takes place in weaker-than-average magnetic-field regions~\cite{2004ApJ...612..276S, 2019PhPl...26g2306Y, 2020ApJ...898...66Y, 2020ApJ...893L..25B, 2020ApJ...898L..18H, 2023JPlPh..89e1701L,2023MNRAS.525.4985K}, and has been argued to dominate particle acceleration in large-amplitude relativistic turbulence~\cite{2022PhRvL.129u5101L, 2022PhRvD.106b3028B}.

Unsurprisingly, therefore, the numerical simulations reported here demonstrate that particle acceleration can proceed to energies well in excess of the (naive) theoretical maximum energy, approaching or even reaching the synchrotron burn-off limit, albeit with a steepened energy distribution. These deviations from the generic, and commonly used, predictions open interesting possibilities for phenomenological applications in high-energy, multimessenger astrophysics.

This paper is organized as follows. In Sec.~\ref{sec:sims}, we detail the setup of our numerical simulations, while in Sec.~\ref{sec:numres}, we discuss their results regarding particle and radiation energy spectra as well as the dominant acceleration mechanisms. We interpret these results in Sec.~\ref{sec:disc}, computing the statistical distribution of individual acceleration rates. We also apply our findings to the case of pulsar wind nebulae and conclude that stochastic acceleration in the postshock turbulence can account for the spectral shape of the spectrum in the hard x-ray to $\gamma$-ray domain. A summary and conclusions are provided in Sec.~\ref{sec:conc}.

\section{Radiative kinetic simulations}
\label{sec:sims}

\subsection{Radiation reaction and characteristic Lorentz factors}
\label{sec:setup}

Classical radiative losses~\cite{1960ecm..book.....L} are implemented in our PIC simulations following, e.g., ~\cite{2012ApJ...746..148C}. Namely, we compute the synchrotron power $P_{\rm syn}$ radiated by each particle, at each time step, according to
\begin{equation}
    P_{\rm syn}(\gamma) = \frac{1}{4\pi} \sigma c \gamma^2 \left[\left(\mathbf{E}+\boldsymbol{\beta}\times\mathbf{B}\right)^2 - \left(\boldsymbol{\beta}\cdot\mathbf{E}\right)^2\right] \,,
    \label{eq:psyn}
\end{equation}
where $\mathbf{E}$ and $\mathbf{B}$, respectively, denote the electric and magnetic fields in the simulation (or laboratory) frame, $\boldsymbol{\beta}$ the electron/positron velocity in units of $c$, and $\gamma$ the corresponding Lorentz factor. In standard synchrotron radiation, $\sigma = \sigma_{\rm T}$ the Thomson cross section, but we will rescale it further below to adapt it to the units of the numerical simulation. Particles of momentum $\boldsymbol{p}$ are then advanced according to their equation of motion
\begin{equation}
    \frac{{\rm d}\boldsymbol{p}}{{\rm d}t} = q\left(\mathbf{E} + \boldsymbol{\beta}\times\mathbf{B}\right) - \frac{P_{\rm syn}}{c}\boldsymbol{\beta} \,,
    \label{eq:dynrad}
\end{equation}
where $q = \pm e$ is the particle charge. 

The bracketed term in Eq.~\eqref{eq:psyn} can be rewritten using quantities measured in the comoving frame of the magnetic-field lines, written \Rco, in which the ideal component of the electric field (i.e., transverse to the local magnetic field) vanishes. This frame drifts at a normalized velocity $\boldsymbol{\beta_E} \equiv \,\mathbf{E}\times \mathbf{B}/B^2$ relative to the laboratory frame. It exists and is well defined for ideal electric fields, meaning $E^2-B^2<0$ and $\mathbf{E} \cdot \mathbf{B} = 0$. It remains well defined in the presence of a parallel (to $\mathbf B$) component of the electric field, provided $E^2-B^2<0$. Quantities measured in the comoving frame \Rco\ are indicated with prime symbols.

For ideal electric fields (without parallel components), noting that $\gamma^2 [(\mathbf{E} +\boldsymbol{\beta}\times\mathbf{B})^2 - (\boldsymbol{\beta}\cdot\mathbf{E} )^2 ]$ is a Lorentz invariant, one finds
\begin{align}
   \left[\left(\mathbf{E}+\boldsymbol{\beta} \times \mathbf{B}\right)^2 - \left(\boldsymbol{\beta}\cdot\mathbf{E}\right)^2\right] &= \frac{\left(\boldsymbol{u'}\times\mathbf{B'}\right)^2}{\gamma^2} \nonumber \\
   &\simeq \frac{\beta^2}{\gamma_E^4} B^2 \sin^2\theta' \,.
   \label{eq:brckt}
\end{align}
Here, $\gamma_E\equiv (1-\beta_E^2)^{-1/2}$ gives the Lorentz factor of the comoving frame \Rco\ relative to the laboratory frame. 
The particle's normalized four-velocity in this frame is $\boldsymbol{u'}\equiv \boldsymbol{p'}/m_{\rm e} c \equiv \gamma' \boldsymbol{\beta'}$, and $\theta'$ is its pitch angle relative to the magnetic-field direction.

In numerical simulations, electric and magnetic fields are conveniently rescaled by $e/m_{\rm e} \omega_{\rm p}c$, where $\omega_{\rm p}$ is the nonrelativistic plasma frequency of one species (electrons or positrons). We use a tilde symbol to write the corresponding dimensionless quantities: $\tilde{E} \equiv e E/m_{\rm e}\omega_{\rm p}c$ and $\tilde{B} \equiv e B/m_{\rm e}\omega_{\rm p}c$. The implemented radiation cross section is rescaled to express the radiated power in units of $\omega_{\rm p} m_{\rm e} c^2$, according to
\begin{equation}
    P_{\rm syn} = \frac{\gamma^2}{\gamma_{\rm syn}^2}\left[\left(\mathbf{\tilde E} +\boldsymbol{\beta}\times\mathbf{\tilde B}\right)^2 - \left(\boldsymbol{\beta} \cdot \mathbf{\tilde E}\right)^2\right]\,\omega_{\rm p} m_{\rm e} c^2\,,
    \label{eq:psyn2}
\end{equation}
which makes $\gamma_{\rm syn}$ the main parameter governing the strength of radiation in our simulations. 

The synchrotron burn-off Lorentz factor $\gamma_{\rm rad}$ is such that the synchrotron loss rate in a uniform magnetic field, $P_{\rm syn}/m_{\rm e}c^2\gamma$, becomes comparable to the gyrofrequency in the laboratory frame $c/r_{\rm g}$. Throughout, we define gyroradii $r_{\rm g}(\gamma)\equiv  \gamma m_{\rm e} c^2/e\bar{B}$ relative to the average \emph{total} magnetic field in the turbulent bath, $\bar B \equiv \langle (\mathbf{B}_0 + \boldsymbol{\delta}\mathbf{B})^2\rangle^{1/2}$. Here, $\mathbf{B}_0$ denotes the mean guide field, and $\boldsymbol{\delta}\mathbf{B}$ is the random turbulent component. Using these definitions, the radiated power can then be recast as
\begin{align}
   P_{\rm syn}(\gamma) &= m_{\rm e} c^2 \left( \frac{e \bar B}{m_{\rm e}c} \right) \left(\frac{\gamma}{\gamma_{\rm rad}} \right)^2\, \nonumber \\ 
   &=\frac{m_{\rm e} c^2 \gamma_{\rm rad}}{r_{\rm g}(\gamma_{\rm rad})/c}\left(\frac{\gamma}{\gamma_{\rm rad}}\right)^2\,.
   \label{eq:psynrad}
\end{align}
The Lorentz factor $\gamma_{\rm rad}$ is here defined as an average over the turbulent bath. It can then be expressed in terms of $\gamma_{\rm syn}$ through
\begin{align}
   \gamma_{\rm rad} &= \gamma_{\rm syn}\, \frac{\tilde{\bar B}^{1/2}}{\left[\left(\mathbf{\tilde E}+\boldsymbol{\beta}\times\mathbf{\tilde B}\right)^2 - \left(\boldsymbol{\beta}\cdot\mathbf{\tilde E}\right)^2\right]^{1/2}}\nonumber\\
   &\,\simeq\, \gamma_{\rm syn}\,\bar{\gamma}_E^2\,\tilde{\bar B}^{-1/2} \,.
   \label{eq:gradgsyn}
\end{align}
Our simulations employ $\tilde{\bar{B}} \simeq 8$, $\bar \gamma_E^2 \simeq 1.8$, so that $\gamma_{\rm rad}$ is slightly smaller than $\gamma_{\rm syn}$. The exact value of $\gamma_{\rm rad}$ is solely determined by the magnetic-field strength, with $\gamma_{\rm rad}=(6\pi e/\sigma_T B)^{1/2}\simeq 10^8\, B^{-1/2}$, with $B$ in G in the last expression. Depending on the specific astrophysical environment, $\gamma_{\rm rad}$ can thus vary greatly. Our numerical simulations employ values $\gamma_{\rm rad}\sim O(10^3)$ to ensure that this maximum energy is covered by the dynamical range, whose minimum Lorentz factor is of order unity (see below). However, it is important to note that the critical point is not the absolute value of $\gamma_{\rm rad}$, but the hierarchy $\gamma_{\rm th}:\gamma_{\rm rad}$. Simulations could be equally well ran with comparatively larger values of these Lorentz factors, at least in the absence of other physical effects that explicitly depend on the Lorentz factor. In essence, the current values of $\gamma_{\rm rad}$ should not be interpreted as physical values, but rather as a parameter rescaled for numerical convenience.

The finite coherence length scale $\ell_{\rm c}$ of the turbulence determines a Hillas-type limiting Lorentz factor, $\gamma_{\rm c}$~\cite{1984ARA&A..22..425H}, such that the corresponding gyroradius $r_{\rm g}(\gamma_{\rm c}) \equiv 0.3\ell_{\rm c}$. The $0.3$ prefactor differs among studies and is somewhat {\it ad hoc}, but it provides a satisfactory description of the observed spectra shown thereafter. Although turbulence can still accelerate particles to Lorentz factors $\gamma > \gamma_{\rm c}$, it does so at an increasingly slow rate, because the energy diffusion coefficient $D_{\gamma\gamma}$ (written in terms of the Lorentz factor) becomes independent of $\gamma$ when $\gamma > \gamma_{\rm c}$~\cite{2020PhRvD.102b3003D}, implying that the characteristic acceleration rate decreases as $\nu_{\rm acc} \propto D_{\gamma\gamma}/\gamma^2 \propto \gamma^{-2}$. Thus,  $\gamma_{\rm c}$ represents an effective cutoff energy evincing the gradual decoupling of particles from turbulence. Using $\tilde{\bar{B}}\simeq 8$, we obtain
\begin{align}
   \gamma_{\rm c} & \equiv 0.3\,\tilde{\bar{B}}\, \frac{\ell_\mathrm{c}\omega_\mathrm{p}}{c} \nonumber \\
   & \simeq 5\times 10^3 \,\frac{\ell_{\rm c}}{2000\,c/\omega_{\rm p}}\,.
   \label{eq:gammac}
\end{align}

The particle distribution is also characterized by its thermal Lorentz factor $\gamma_{\rm th}$, which in relativistic turbulence at initial magnetization $\sigma$, is of the order of $\gamma_{\rm th}\simeq \sigma$ if the injected plasma is cold~\cite{2018ApJ...867L..18Z,18Comisso,2019ApJ...886..122C, 2019PhRvL.122e5101Z,2020ApJ...893L...7W, 2025MNRAS.543.1842W}. Our simulations assume an initial temperature $k_{\rm B} T/m_{\rm e} c^2 \simeq 2$, therefore $\gamma_{\rm th}$ will be of the order of a few. For reference, we define the magnetization for a plasma of enthalpy density $w$ in a magnetic field of strength ${B}$ as $\sigma \equiv {B}^2/(4\pi w)$. This definition implies that $\sigma = (u_{\rm A}/c)^2$, where $u_{\rm A} \equiv \beta_{\rm A}c/(1-\beta_{\rm A}^2)^{1/2}$ is the Alfvén four-velocity, and $\beta_{\rm A}c$ is the Alfvén three-velocity. Hereafter, $\sigma$ is understood as the total magnetization, defined relative to the average total magnetic field $\bar B$, while $\sigma_0$ is that corresponding to the guide field $B_0$.

As mentioned earlier, we also introduce the \emph{expected} maximum Lorentz factor, $\gamma_{\rm D}$, for which the synchrotron loss rate equals the acceleration rate,
\begin{equation}
    \frac{P_{\rm syn}(\gamma_{\rm D})}{\gamma_{\rm D} m_{\rm e} c^2} \equiv \nu_{\rm acc}(\gamma_{\rm D}) \,.
    \label{eq:gammaD}
\end{equation}
The energy diffusion coefficient measured in PIC simulations scales approximately as $D_{\gamma\gamma}\simeq \kappa^{-1}\gamma^2 c/\ell_{\rm c}$ with $\kappa \simeq 10/ \sigma$~\cite{2018ApJ...867L..18Z, 18Comisso, 2019ApJ...886..122C,2019PhRvL.122e5101Z,2020ApJ...893L...7W,2025MNRAS.543.1842W}. Should stochastic acceleration be governed by a fully diffusive process with diffusion coefficient $D_{\gamma\gamma}$ proportional to the square of the Lorentz factor, the mean particle Lorentz factor would increase exponentially in time as $\langle\gamma\rangle \propto \exp(4D_{\gamma\gamma} t/\gamma^2)$, e.g.~\cite{2024PhRvD.109f3006L}. 
Using, therefore, $\nu_{\rm acc}=4D_{\gamma\gamma}/\gamma^2$, we obtain
\begin{align}
    \gamma_{\rm D} &\simeq 4\,\kappa^{-1} \gamma_{\rm rad} \frac{r_{\rm g}(\gamma_{\rm rad})}{\ell_{\rm c}} \,,
    \label{eq:gammamax}
\end{align}
or, in terms of the fiducial values in our simulations,
\begin{equation}
   \gamma_{\rm D} \simeq 25\, \kappa^{-1}\,\left(\frac{\ell_{\rm c}\omega_{\rm p}/c}{2000}\right)^{-1} \left(\frac{\gamma_{\rm syn}}{500}\right)^2 \left(\frac{{\tilde{\bar {B}}}}{8}\right)^{-2} \left(\frac{{\bar{\gamma_E}}^2}{1.8}\right)^2 \,.
   \label{eq:gammamax-app}
\end{equation}

However, when comparing to actual simulation data, we rely on a more precise estimate of $\gamma_{\rm D}$, obtained from a direct measurement of $\nu_{\rm acc}$ for a large sample of particles, which we use in  Eq.~\eqref{eq:gammaD}. This circumvents the uncertainties tied to the exact value of the coherence scale and, more generally, to the value of the diffusion coefficient itself. The two estimates for $\gamma_{\rm D}$ nevertheless agree with each other to within a factor of the order of unity.

The Bohm scaling, which corresponds to an acceleration rate of the form $\eta_{\rm Bohm}\,c/r_{\rm g}$ (with $\eta_{\rm Bohm}\leq 1$ a numerical prefactor) is often used to estimate the maximal energy in high-energy astrophysical plasmas, in the absence of better knowledge. The corresponding maximum Lorentz factor $\gamma_{\rm Bohm}$ is related to $\gamma_{\rm rad}$ as $\gamma_{\rm Bohm} = \eta_{\rm Bohm}^{1/2} \gamma_{\rm rad}$. In practice, the Bohm scaling provides an upper bound on the acceleration rate, as exceptions are scarce -- e.g., \cite{2006ApJ...642..244M} -- and hence an upper bound on the maximum energy that can be achieved. 

\begin{figure}[t]
   \centering
   \includegraphics[width=0.9\columnwidth]{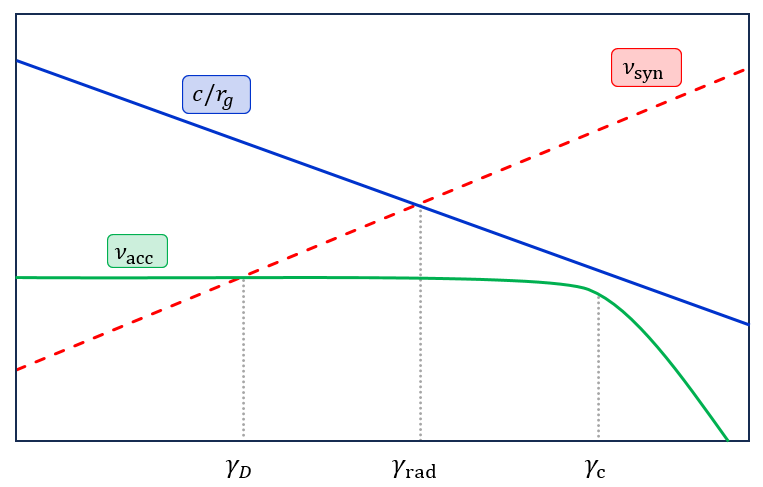}
    \caption{Log-log sketch without specific scales of the acceleration rate $\nu_{\rm acc}$, the synchrotron loss rate $\nu_{\rm syn}$ and the gyrofrequency $c/r_{\rm g}$ versus particle Lorentz factor. This figure defines and illustrates the ordering of maximal Lorentz factors, see text for details.}
    \label{fig:nuscale}
\end{figure}

In a turbulent plasma, the maximum Lorentz factor is, in principle, set by the minimum of $\gamma_{\rm rad}$, $\gamma_{\rm D}$, or $\gamma_{\rm c}$. Obviously, if $\gamma_{\rm D} \gg \gamma_{\rm c}$, synchrotron losses do not play any role. We are here interested in the shape of the spectrum and of the radiated synchrotron flux in the spectral region between $\gamma_{\rm D}$ and $\gamma_{\rm rad}$, as illustrated in Fig.~\ref{fig:nuscale}. Ideally, one seeks a numerical configuration such that 
\begin{equation}
    \gamma_{\rm th}\ll \gamma_{\rm D}\ll\gamma_{\rm rad}\ll \gamma_{\rm c}\,.
    \label{eq:hierarch}
 \end{equation} 
In practice, the limited dynamic range of numerical kinetic simulations compels us to restrict this hierarchy, as we discuss next.

\subsection{Numerical simulation setup}

This hierarchy of characteristic Lorentz factors imposes specific requirements on our simulations. The most important one, obviously, is to reach as large a dynamic range $\ell_{\rm c}\omega_{\rm p}/c$ as possible, since $r_{\rm g}(\gamma_{\rm th}) \sim O(c/\omega_{\rm p})$ while $r_{\rm g}(\gamma_{\rm c})\sim O(\ell_{\rm c})$. For this reason, we conduct most of our study using 2D3V simulations (2D in space and 3D in momentum space).
To probe the range between $\gamma_{\rm D}$ and $\gamma_{\rm rad}$, we maintain $\gamma_{\rm D}$ close to $\gamma_{\rm th}$ and $\gamma_{\rm rad}$ close to $\gamma_{\rm c}$, then vary $\gamma_{\rm syn}$ to study its influence on the spectral shape. 
  
Retaining a large dynamic range between $\gamma_{\rm D}$ and $\gamma_{\rm rad}$ is also necessary to minimize the influence of magnetic reconnection in microscopic current sheets and, more generally, nonideal electric fields on kinetic scales, which can preaccelerate particles up to a few times $\gamma_{\rm th}$~\cite{18Comisso,2019ApJ...886..122C}. Reconnection is also active in long, extended current sheets during the initialization stage of turbulence, when the energy injected on large scales first cascades down to small scales. To reach the stage of well-developed turbulence, we perform simulations with continuous turbulence excitation and study the spectral shapes and particle acceleration on timescales $>3\ell_{\rm c}/c$. However, the integration time of the simulations should not significantly exceed $5$ or $6\,\ell_{\rm c}/c$, as the bulk of the plasma starts to suffer synchrotron cooling on these timescales. We thus halt the simulations at $5$ or $6\,\ell_{\rm c}/c$.

We carry out our simulations with the finite-difference time-domain, relativistic PIC \textsc{calder} code~\cite{Lefebvre_2003}, which includes the turbulence-generation modules introduced in~\cite{2022PhRvD.106b3028B, 2023PhRvR...5b3194B}. While the simulations reported here are run in 2D3V, we have performed an additional large-scale, shorter-duration, fully 3D simulation for a convergence study, described in the Appendix. Turbulence is driven throughout the simulation duration with a set of 24 random ``Langevin antennas'', following~\cite{TenBarge_2014, 2018ApJ...867L..18Z}, which excite random plane wave current perturbations $\boldsymbol{\delta j}\parallel\mathbf{B_0}$ (the guide field $\mathbf{B_0}$ is along the $z$ axis). These currents are characterized by a damping constant $\omega_{\rm I} = 0.7 k c$ and a wave number $\boldsymbol{k}\perp\mathbf{B_0}$ with $k L/(2\pi) = \sqrt{3}$ or $\sqrt{5}$, where $L$ is the box size. This setup yields a coherence scale of $\ell_{\rm c} \simeq 0.5\, L$. 

The electron-positron pair plasma initially obeys a relativistic Maxwell-J\"uttner distribution with a temperature $T=1\,\rm MeV$, and is modeled by 20 or 30 macroparticles per cell (10 or 15 per species). A shorter-duration 2D3V simulation using 100 macroparticles per cell yielded no significant difference in the nonthermal particle energy distributions. The time step is $\delta t =0.99\,\delta x/c = 0.99\,\delta y/c$. Our simulations employ fourth-order shape functions and eight passes of binomial filtering at each time step to remove noise on the grid scale. Particles are advanced using a Boris pusher, taking into account radiation reaction as discussed above. Periodic boundary conditions are used for both particles and fields in all directions.

The 2D spatial domain has a length of $L=8000 \,\delta x$ in both directions, with a mesh size $\delta x = \delta y = 0.5\,c/\omega_{\rm p}$. Here, $\omega_{\rm p}$ represents $1/\sqrt{2}$ of the total (nonrelativistic) plasma frequency, implying $\ell_{\rm c} \simeq 2000\,c/\omega_{\rm p}$. Through trial and error, we have found that this setting provides a fitting configuration at a given computational cost for the problem at hand. A fine sampling of the skin depth notably reduces numerical heating, which would otherwise reduce the scale separation between the thermal bulk and highest-energy particles, as we have verified with additional simulations. In the absence of energy losses, $\ell_{\rm c}\omega_{\rm p}/c$ must exceed a few hundreds to ensure that particle acceleration extends beyond the preacceleration stage in microscopic reconnection layers~\cite{2018ApJ...867L..18Z}. Our chosen values for $L$ and $\tilde{B}$ allow us to reach $\gamma_{\rm c} \sim 5000$. Moreover, we set $\gamma_{\rm syn}$ to $500 - 2000$ so that $\gamma_{\rm D} \sim 10-40$ (see below). The magnetization associated with the guide field is $\sigma_0 \simeq 1.5$.

Our simulations are conducted in several steps. First, we establish the turbulence cascade by exciting large-scale perturbations using the Langevin antennas. We set $\tilde{B}_0=5$ and tune the antenna amplitude to obtain $\delta B/B_0 \simeq 1$. The total magnetization is thus $\sigma \approx 2$ in the initial state, and subsequently decreases as the plasma draws energy from the turbulence. A detailed discussion of the evolution of the total magnetization parameters in the various simulations is provided in the Appendix. The cascade takes about $3\,\ell_{\rm c}/c$ to fully develop. After this time, we proceed to track a substantial number ($\sim 2-3 \times 10^4$) of particles to monitor their energization histories and mark down those reaching energies close to the synchrotron burn-off limit. We then rerun this second stage of the simulation to closely follow the preselected particles, in order to better understand their energization mechanisms. This analysis is presented in Sec.~\ref{sec:acc}.

To ensure that the initialization stage of turbulence has a negligible effect on the spectra studied, we conducted simulations that included two additional subdominant species of electron-positron pairs. In these simulations, the dominant species is introduced at the initial time, forming the plasma bulk that sustains the development of turbulence. The second, subdominant species is injected at $3\,\ell_{\rm c}/c$, by which time the turbulence has fully developed on all scales. In the Appendix, we show that the spectra for the two species exhibit similar behavior, confirming that beyond $3\,\ell_{\rm c}/c$, the spectra are insensitive to how the turbulent cascade is initiated.

As observed in previous simulations, particles are preaccelerated in intense, extended current sheets during the first $2-3\,\ell_{\rm c}/c$. The cascade that builds up during this period generates structures spanning all scales from $\ell_{\rm c}$ down to thermal gyroradii, destroying the coherence of these current sheets, which nevertheless survive on kinetic scales. Their role then narrows down to accelerating particles up to Lorentz factors of the order of a few times $\sigma$, after which particles become susceptible to stochastic acceleration from large-scale plasma motions. Here, we focus on tracking the evolution of particles between $3$ and $5\,\ell_{\rm c}/c$, across an energy range that extends sufficiently far from the thermal pool, in practice, 2 orders of magnitude above it.

\section{Results and discussion}
\label{sec:numres}

This section presents and discusses the results of our numerical exploration. We first present the particle energy distribution at various times, for different configurations, then the radiated spectra, to finally examine in greater detail the processes that bring particles close to the synchrotron burn-off limit.

\subsection{Numerical particle energy spectra}
\label{sec:partspec}

\begin{figure}[ht!]
   \centering
   \includegraphics[width=0.9\columnwidth]{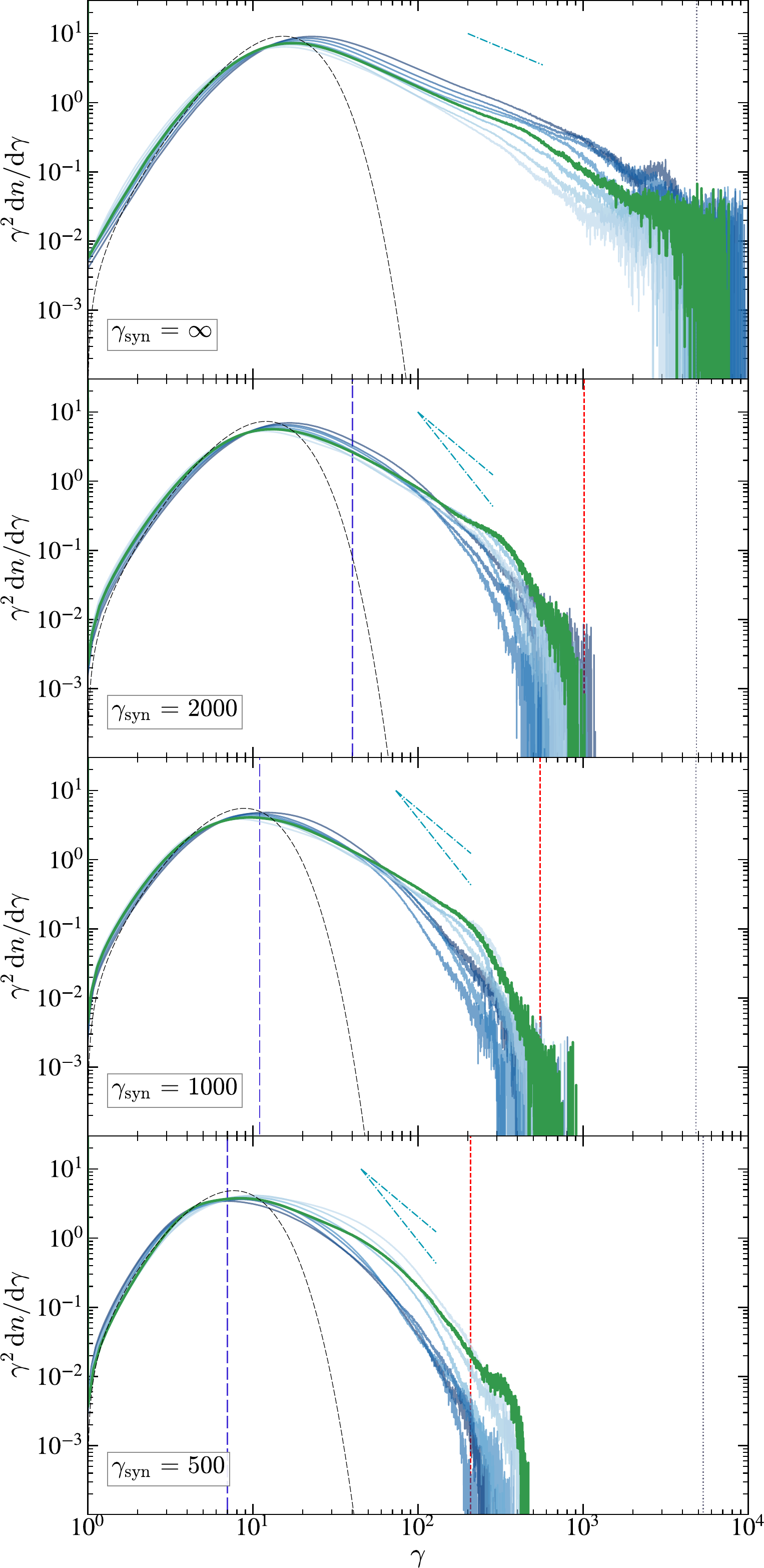}
    \caption{Spectral energy distributions $\gamma^2 {\rm d}n/{\rm d}\gamma$ versus particle Lorentz factor $\gamma$. All four simulations shown here share the same characteristics up to the cooling Lorentz factor $\gamma_{\rm syn}$, the value of which is indicated in the lower left corner. Spectra are plotted in blue colors at different times, with hue from light to dark as time progresses from $t=3\,\ell_{\rm c}/c$ to $5\,\ell_{\rm c}/c$. In each panel, the thick green curve corresponds to the spectrum at $t=4\,\ell_{\rm c}/c$. The dotted gray line indicates the Lorentz factor $\gamma_{\rm c}$ [Eq.~\eqref{eq:gammac}], the dashed blue line the Lorentz factor $\gamma_{\rm D}$ [Eq.~\eqref{eq:gammamax}] and the dashed red line the synchrotron burn-off Lorentz factor $\gamma_{\rm rad}$ [Eq.~\eqref{eq:gradgsyn}]. The short blue dash-dotted lines indicate power-law slopes ${\rm d}n/{\rm d}\gamma \propto \gamma^{-4}$ and $\propto\gamma^{-5}$ when $\gamma_{\rm syn}<\infty$, and $\propto \gamma^{-3}$ for $\gamma_{\rm syn}\rightarrow \infty$. The thin dashed black curves represent Maxwell-Jüttner distributions with $k_B T/m_ec^2 = 3.8$, 3., 2.2 and 1.9 from top to bottom.
    \label{fig:meanspec}}
\end{figure}

Figure~\ref{fig:meanspec} presents a compilation of particle energy distribution $\gamma^2 {\rm d}n/{\rm d}\gamma$ at different times and for different values of $\gamma_{\rm syn}$, which controls the level of synchrotron cooling. In each panel, the thin blue lines show the spectra at time intervals of $\simeq 0.25\,\ell_{\rm c}/c$ between $3$ and $5\,\ell_{\rm c}/c$. The green spectrum highlights the spectrum in the middle of this interval at $4\,\ell_{\rm c}/c$. The gray dotted line indicates $\gamma_{\rm c}$ [Eq.~\eqref{eq:gammac}], which sets the maximum Lorentz factor in the absence of cooling (upper panel). The dashed blue line marks the value of $\gamma_{\rm D}$. Finally, the dashed red line shows $\gamma_{\rm rad}$, which marks the location of the synchrotron burn-off limiting Lorentz factor, see Eq.~\eqref{eq:gradgsyn}. These various Lorentz factors have been computed according to their respective definitions, using the time- and space-averaged magnetic field $\bar{B}$ and Lorentz factor $\bar\gamma_E$. These values vary little over time during the simulations, as discussed in the Appendix, so the average value shown here offers a satisfactory approximation. To obtain an accurate estimate of $\gamma_{\rm D}$, we have combined Eq.~\eqref{eq:gammaD} with direct measurements of the mean acceleration rate, $\bar\nu_{\rm acc}(\gamma)$, and the mean of  $P_{\rm syn}(\gamma)/\gamma^2$ over a large sample of particles with various Lorentz factors. The specific procedure for measuring this mean acceleration rate is discussed further on, in Sec.~\ref{sec:disc}.

This figure highlights several key effects of synchrotron losses: (i) the spectrum extends significantly beyond $\gamma_{\rm D}$; (ii) it reaches or approaches the synchrotron burn-off limit for all examined values of $\gamma_{\rm syn}$; (iii) the spectral shape becomes highly variable in time at the highest energies, within about a decade of the synchrotron burn-off limit.   

The influence of radiative cooling on the spectral shape at low Lorentz factors $\gamma\lesssim \gamma_{\rm th}$ is minimal, as desired. This can be seen from the comparison with a Maxwell-Jüttner distribution, shown as a dotted black line in Fig.~\ref{fig:meanspec}. In the absence of synchrotron losses, the temperature of the plasma bulk is close to $\simeq 4\,$MeV (upper panel). This temperature slightly decreases in the presence of radiative losses, all the more so as $\gamma_{\rm syn}$ decreases (leading to stronger losses), down to $T\simeq 2\,$MeV for $\gamma_{\rm syn}=500$. While the electric fields of reconnection layers contribute to particle energization at, or slightly beyond, the thermal peak, they exert little influence on the high-energy tail, as discussed below. Furthermore, our estimate of $\gamma_{\rm D}$ incorporates part of this effect, as our measurement of the acceleration rate accounts for all ideal electric fields, but not nonideal ones (see Sec.~\ref{sec:disc}), regardless of their scale and nature. This can be also be seen by comparing the trend in the values of $\gamma_{\rm D}$ across the simulations. Assuming $D_{\gamma\gamma} \propto \gamma^2$, as measured at suprathermal energies, we expect $\gamma_{\rm D} \propto \gamma_{\rm syn}^2$ from  Eq.~\eqref{eq:gammamax-app}, which fits well the ratio of $\gamma_{\rm D}$ measured in the two simulations with $\gamma_{\rm syn} = 1000$ and $2000$. At $\gamma_{\rm syn} = 500$, however, $\gamma_{\rm D}$ takes a larger value than anticipated, $\gamma_{\rm D}\simeq 7$ instead of $\simeq 2$, because the mean acceleration rate is measured to be three times larger than that in the other two simulations. This excess likely comes from the contribution of reconnection at energies below or close to the thermal peak.

\begin{figure}[t]
   \centering
   \includegraphics[width=0.9\columnwidth]{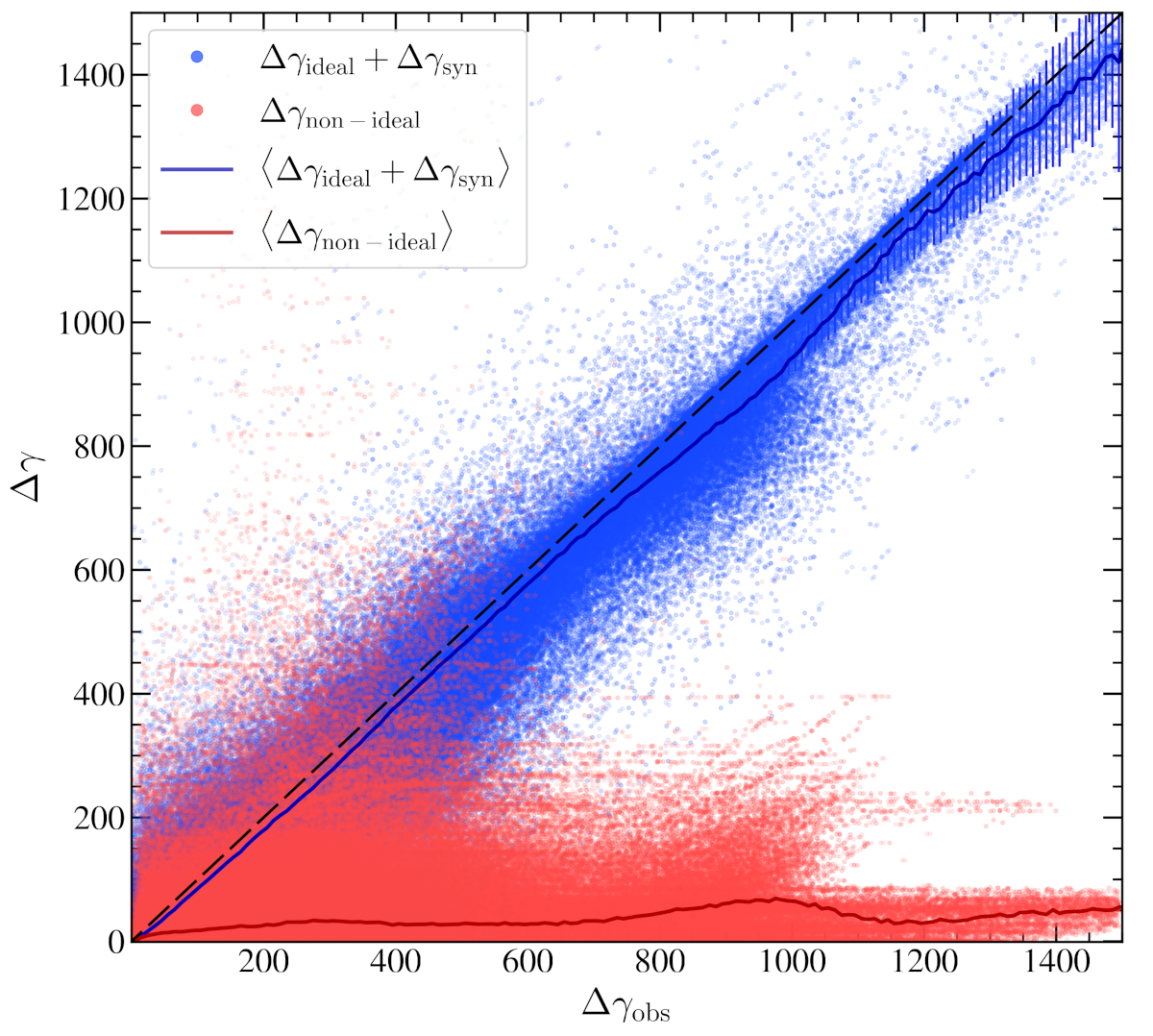}
    \caption{Estimates of the ideal and nonideal energy gains for a large sample of particles in the simulation with $\gamma_{\rm syn}=2000$. For each point, the abscissa indicates the observed jump in Lorentz factor $\Delta\gamma_{\rm obs}$ between two points selected randomly along particle trajectories, while the ordinates show the respective contributions from the nonideal electric field (red symbols) and ideal electric field corrected by the energy lost through radiation (blue symbols). The thick red (respectively, blue) line plots the median of the distribution of red (respectively, blue) symbols at different values of $\Delta \gamma_{\rm obs}$. The thin black line indicates the one-to-one correlation $\Delta\gamma = \Delta\gamma_{\rm obs}$. At large values of $\Delta \gamma_{\rm obs}$, the thin dashed lines indicate the errors resulting from the limited sample size.
    \label{fig:workdist}}
\end{figure}

To further verify that particles are indeed accelerated through their interactions with the turbulence rather than by microscopic current sheets, we tracked a large number $\sim O(10^4)$ of macroparticles in the simulations, recording the time histories of their momenta along with the electromagnetic-field components they experienced. This allows us to reconstruct the energization process and decompose it into the contributions from ideal and nonideal electric fields. We define the ideal electric field as the component of $\mathbf{E}$ perpendicular to $\mathbf{B}$ when $E^2 < B^2$. The remaining component, called nonideal, comprises all components $\mathbf{E_\parallel}$ parallel to $\mathbf{B}$ as well as the total electric field in regions where $E^2 \geq B^2$.

We compare these contributions in Fig.~\ref{fig:workdist}, using the statistics measured in the simulation with $\gamma_{\rm syn}=2000$ (second panel from the top in Fig.~\ref{fig:meanspec}). To do so, we randomly select pairs of points along each trajectory and measure the energy gain (if any) $\Delta \gamma_{\rm obs}$ between these points. We then reconstruct the contributions from the ideal and nonideal electric fields in this interval to calculate the ideal ($\Delta \gamma_{\rm ideal}$) and nonideal ($\Delta \gamma_{\rm nonideal}$) energy gains. We also evaluate the synchrotron energy loss $\Delta\gamma_{\rm syn}<0$ via Eq.~\eqref{eq:psyn2} using the recorded values of the electric and magnetic fields, then include it into the ideal contribution. Next, we plot the cloud of points $(\Delta \gamma_{\rm obs},\,\Delta \gamma_{\rm ideal}+\Delta\gamma_{\rm syn})$ in blue and $(\Delta \gamma_{\rm obs},\,\Delta \gamma_{\rm nonideal})$ in red, see Fig.~\ref{fig:workdist}. The dashed lines show the evolution of the median of each contribution in different bins of $\Delta\gamma_{\rm obs}$. In agreement with earlier studies in the absence of synchrotron cooling~\cite{2019ApJ...886..122C}, the contribution from nonideal electric fields remains limited to the range $\gamma \lesssim 50$, and is dominated by the ideal contribution as soon as $\gamma \gtrsim 20$, i.e., several times above thermal energies.

Particles with Lorentz factors $\gamma \gtrsim 50-100$ have gyroradii that exceed that of thermal particles by a factor $\gtrsim 10$. Therefore, they explore larger spatial scales in the turbulence cascade and become insensitive to the physics of microscopic current sheets. The extension of the spectrum well above these energies, up to $\gamma\sim 10^3$, thus strongly suggests that stochastic particle acceleration in the large-scale turbulence is the process that pushes particles well beyond the expected cutoff at $\gamma_{\rm D}$. We will confirm this further by analyzing the trajectory of some of the highest-energy particles in Sec.~\ref{sec:acc}.

To better understand the spectral shapes, we have unfolded from the spectra shown in Fig.~\ref{fig:meanspec} a functional form that models a power-law extension above a minimum Lorentz factor, with an exponential cutoff at $\gamma_{\rm rad}$. This exercise, discussed in the Appendix, reveals that at Lorentz factors $\gamma \gtrsim 40$, the spectral shape in the simulations with radiative cooling follows an approximate power law ${\rm d}n/{\rm d}\gamma\propto \gamma^{-s}$ with $s \simeq 4$. However, the time variability at the highest energies implies substantial uncertainty ($\simeq \pm 0.5$) on this spectral index. In Fig.~\ref{fig:meanspec}, the thin dash-dotted lines illustrate power-law tails with $s=4$ and $5$ for reference. These indeed seem to bracket relatively well the observed spectra.

In the absence of radiative losses, the energy distribution follows a power law, albeit with spectral index $s\simeq 3$ (see also the Appendix), in agreement with earlier studies for similar relativistic turbulent plasmas \cite{18Comisso}. These spectra become harder and the distribution shifts to larger Lorentz factors with time, in line with expectations. This contrasts with the spectra observed in the other panels, where no systematic trend is observed in time.

We also note that for the two simulations with $\gamma_{\rm syn}=1000$ and $\gamma_{\rm syn}=2000$, the spectra do not cut off abruptly close to $\gamma_{\rm rad}$. Rather, they tend to fall off smoothly down to the range where they become limited by the finite number of particles in the simulation. This is expected insofar as increasing $\gamma_{\rm syn}$ extends the dynamical range of the spectrum, while the total number of particles in the simulation remains constant. By contrast, the simulation with stronger synchrotron cooling, and therefore shorter dynamical range, namely $\gamma_{\rm syn}=500$, extends to $\gamma_{\rm rad}$ where it presents a more pronounced cutoff. This suggests that, in our simulations with $\gamma_{\rm syn}\geq 1000$, the actual spectra likely extend beyond those shown in Fig.~\ref{fig:meanspec}.

\subsection{Radiated spectra}\label{sec:radspec}

We record the radiated synchrotron spectra on the fly, by summing the contribution of each macroparticle $i$ of weight $W_i$ as
\begin{equation}
    \nu L_{\nu} \equiv \sum_i W_i P_{\rm syn}(i) \left[\nu/\nu_{\rm p}(i)\right]^{4/3}\exp\left[-\nu/\nu_{\rm p}(i)\right]\,,
    \label{eq:psynPIC}
\end{equation}
where $\nu\,L_\nu$ is a shorthand notation for the power radiated in a logarithmic interval ${\rm d}\ln\nu$ of frequency $\nu$, and
\begin{equation}
    \nu_{\rm p}(i) \equiv \frac{3}{4\pi}\frac{e {B}'}{m_e c}\sin{\theta'}{\gamma'}^2\gamma_E
    \label{eq:nupk}
\end{equation}
denotes the synchrotron peak frequency of particles with (comoving frame) Lorentz factor $\gamma'$ and pitch angle $\theta'$. $P_{\rm syn}(i)$ is the power emitted by macroparticle $i$, defined in Eq.~\eqref{eq:psyn}. The above implementation is an approximation of the actual synchrotron spectral shape and normalization, which nevertheless suffices for the present purposes. 

\begin{figure}[ht!]
   \centering
   \includegraphics[width=0.89\columnwidth]{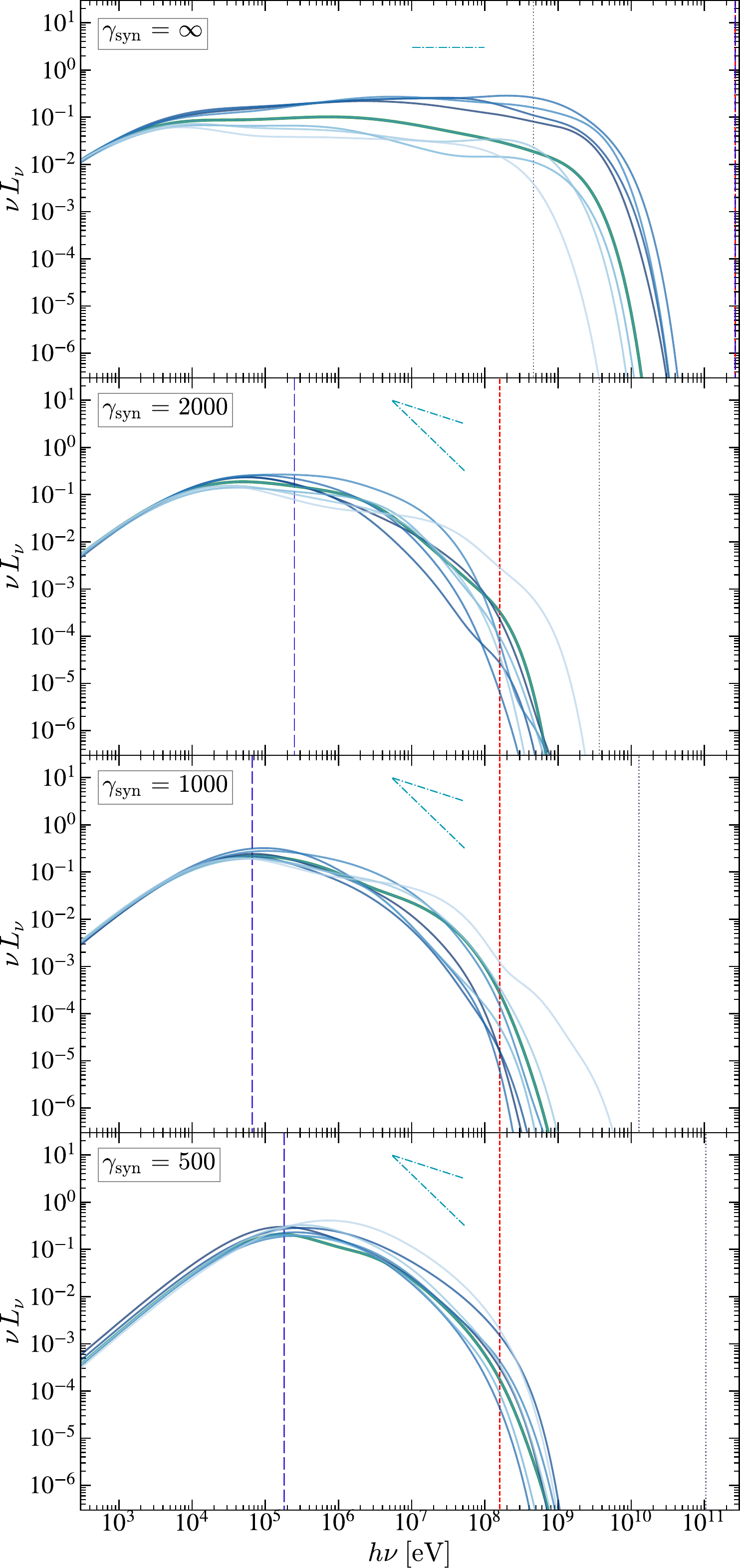}
    \caption{Radiated synchrotron power per logarithmic interval of frequency $\nu L_\nu$ versus energy $h\nu$, for the simulations whose particle energy distributions have been plotted in Fig.~\ref{fig:meanspec}. The vertical lines retain their meaning from Fig.~\ref{fig:meanspec}, although translated into frequency space. Here, the dotted gray line indicates the peak synchrotron frequency $\nu_{\rm c}$ for particles with Lorentz factor $\gamma_{\rm c}$; the dashed blue line indicates the peak synchrotron frequency $\nu_D$ corresponding to the Lorentz factor $\gamma_{\rm D}$, and the dashed red line the peak frequency $\nu_{\rm rad}$ corresponding to the Lorentz factor $\gamma_{\rm rad}$. The thin blue curves show spectra taken at different times, ordered as before from light to dark as time proceeds, while the thick green curve shows the spectrum at $t = 4\,\ell_{\rm c}/c$. The thin dash-dotted lines indicate scalings $\nu L_\nu \propto \nu^{-0.5}$ and $\propto \nu^{-1.5}$ when $\gamma_{\rm syn}<\infty$, and $\propto \nu^0$ for $\gamma_{\rm syn}\rightarrow \infty$.
    \label{fig:meanLnu}}
\end{figure}

Figure~\ref{fig:meanLnu} presents the radiated spectra $\nu L_\nu$ versus photon energy $h\nu$ for the various simulations. This figure offers a direct counterpart to Fig.~\ref{fig:meanspec} for the particle energy spectra. We can define the characteristic frequencies $\nu_D \equiv \nu_{\rm p}(\gamma_{\rm D})$, $\nu_{\rm rad} \equiv \nu_{\rm p}(\gamma_{\rm rad})$, and $\nu_{\rm c} \equiv \nu_{\rm p}(\gamma_{\rm c})$ using Eq.~\eqref{eq:nupk} and the simulation parameters. Although the numerical frequency units are arbitrary, the energy of synchrotron photons at the synchrotron burn-off limit is fixed~\cite{Guilbert_1983,1996ApJ...457..253D,2011ApJ...737L..40U}, because equating the synchrotron cooling time with the gyrotime imposes $\gamma_{\rm rad}=\left(6\pi e/\sigma_{\rm T}B\right)^{1/2}$, and therefore $h\nu_{\rm rad} \simeq \,\gamma_{\rm rad}^2\,\hbar eB/m_e c  = (9/4) m_e c^2/\alpha\simeq 160\,\rm MeV$, where $\alpha=e^2/\hbar c$ denotes the fine structure constant. In Fig.~\ref{fig:meanLnu}, we therefore rescale all frequency units in order to fix $h\nu_{\rm rad}=160\,\rm MeV$. When synchrotron losses are disabled, the units of frequency remain arbitrary. The synchrotron spectra plotted here represent what would be observed if the particles were subjected to a magnetic field in the postacceleration phase, over a time span short enough to prevent synchrotron cooling from modulating the particle distribution function.

These spectra follow the general trends expected on the basis of the particle spectra seen in Fig.~\ref{fig:meanspec}. Namely, the spectral shape follows $\nu L_\nu \propto \nu^{-s_\gamma}$ with $s_\gamma \simeq 0$ for $\gamma_{\rm syn}\rightarrow \infty$, as expected if $s\simeq 3$ since $s_\gamma = (3-s)/2$ for isotropic particle distributions~\cite{1986rpa..book.....R}. The thin dash-dotted lines shown here indicate scalings $s_\gamma = 0.5$ and $s_\gamma=1.5$, which appear to bracket relatively well the observed spectral slopes when synchrotron cooling is present. The radiation-reaction limit imposes a cutoff on (about) all observed spectra, as for the particle energy distributions.

The spectrum corresponding to $\gamma_{\rm syn}=\infty$ hardens with time, as does the particle spectrum, while the spectra obtained with synchrotron cooling display strong variability at the highest energies. To better quantify the variability of spectra in the presence of synchrotron losses, we construct light curves (synchrotron power versus time) in frequency bins close to $\nu_{\rm rad}$, each of width $0.3$ in $\log_{10}(\nu)$.

\begin{figure}[ht!]
    \centering
    \includegraphics[width=0.9\columnwidth]{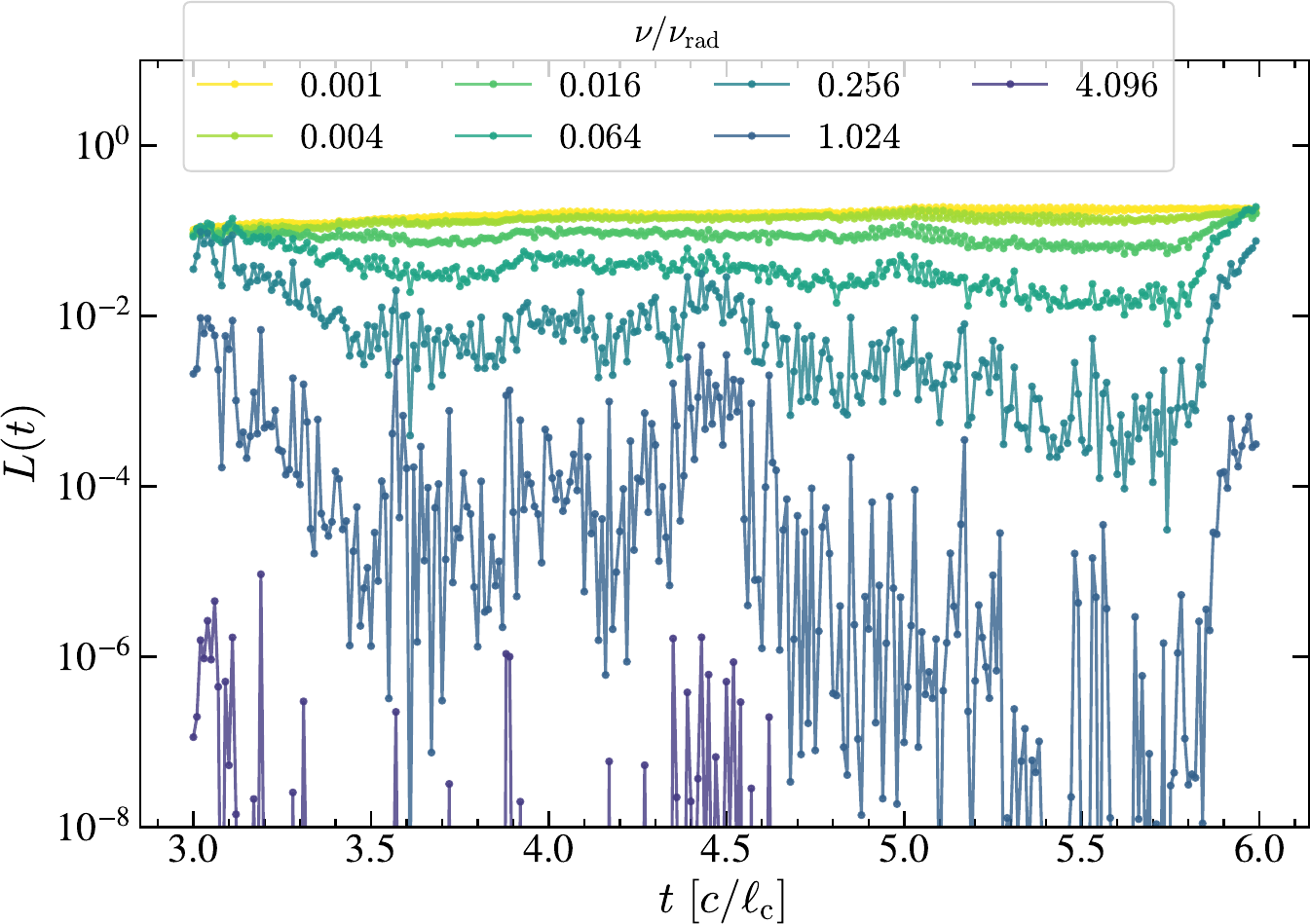}
    \caption{Light curves of the radiated synchrotron power ($\nu L_\nu$) in different frequency bins versus time, between $3\,\ell_{\rm c}/c$ and $6\,\ell_{\rm c}/c$. Each bin is represented by its central value $\nu/\nu_{\rm rad}$, and indicated by the color code. 
    \label{fig:LfLc}}
\end{figure}

To do so, we use an additional simulation that does not include particle tracking but samples the synchrotron spectra at high cadence, every $0.01\ell_{\rm c}/c$, and then rebin each individual spectra on the above frequency bin array. The result, shown in Fig.~\ref{fig:LfLc}, vividly illustrates how time variability evolves with frequency. While the flux appears roughly constant in time at $\nu\ll\nu_{\rm rad}$, the very definition of a mean flux becomes ambiguous close to $\nu_{\rm rad}$, because the spectrum undergoes excursions exceeding an order of magnitude over a timescale well below $\ell_{\rm c}/c$. For instance, at late times, the large-scale flux excursion at $\nu/\nu_{\rm rad} \simeq 0.25$ reaches up to $40\,\%$ of the peak flux at low energies, while that at $\nu/\nu_{\rm rad} \simeq 1$ is only about $1\,\%$ of this mean flux. Note that this simulation extends up to $6\,\ell_{\rm c}/c$. It displays at late times a prominent flare at the highest energies.

As we have checked, the power spectrum of fluctuations in flux evolves from red at $\nu/\nu_{\rm rad}\ll 1$ (fluctuations dominated by long timescales) to white as $\nu\rightarrow \nu_{\rm rad}$, corresponding to equal power per frequency (here understood as Fourier conjugate to time) bin.

We postpone to Sec.~\ref{sec:disc} a discussion of the phenomenological consequences of such spectra and now turn to analyzing in greater detail the acceleration processes at the highest energies.

\subsection{Acceleration mechanisms}
\label{sec:acc}

\begin{figure}[ht!]
   \centering
   \includegraphics[width=0.9\columnwidth]{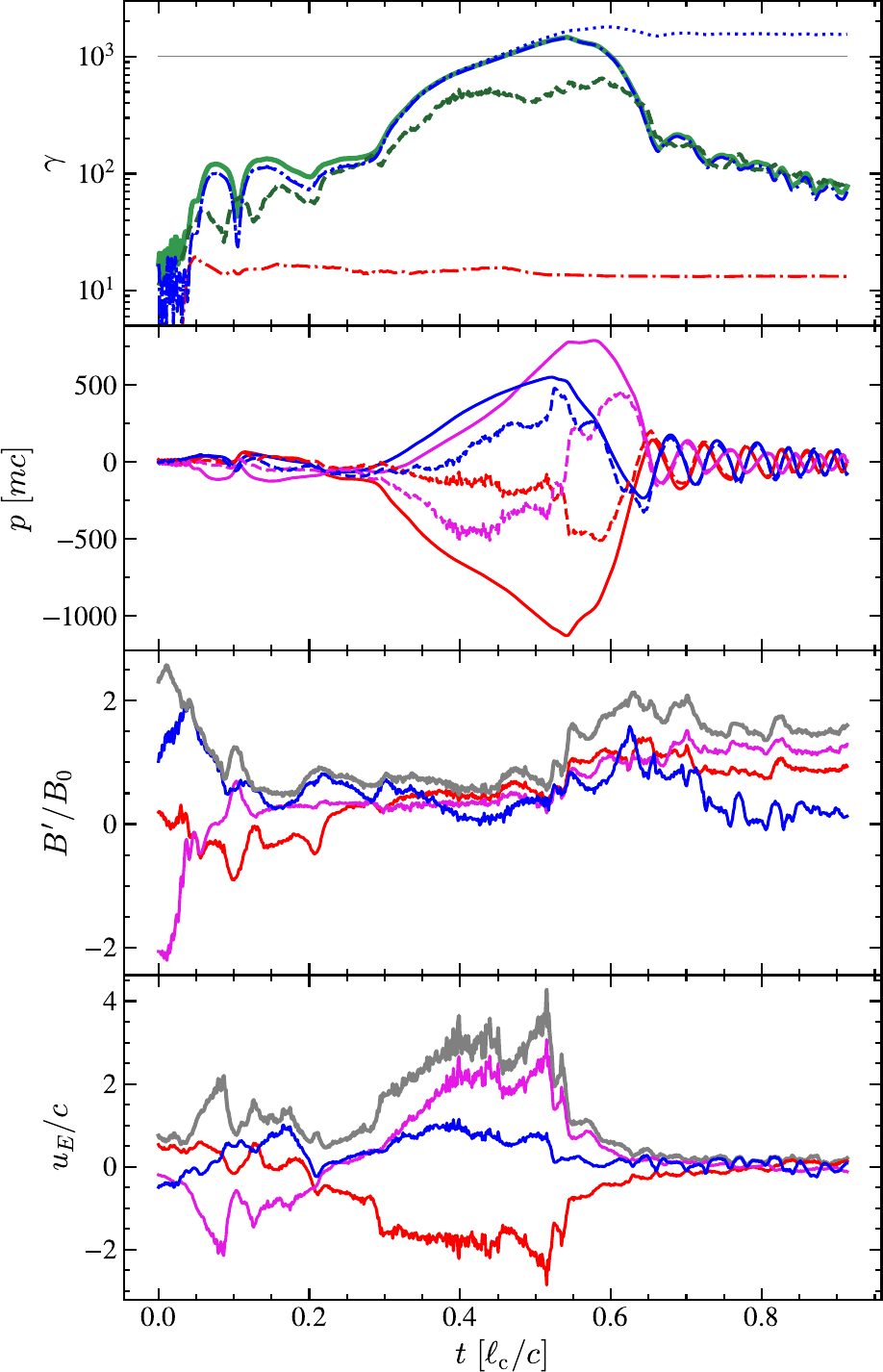}
    \caption{Time history for a particle that approaches the synchrotron burn-off limit in the simulation with $\gamma_{\rm syn} = 2\,000$. Upper panel: evolution of the particle's Lorentz factor in the laboratory (thick green line) and comoving (dashed dark green) frames.
    The dash-dotted and dotted blue lines indicate the cumulative energy gains reconstructed from the ideal electric fields, respectively with and without radiative losses, while the red dash-dotted line shows the contribution of nonideal fields without radiative losses. The thin solid gray line indicates the value of $\gamma_{\rm rad}$ for this simulation. Second panel from the top: evolution of the momentum components in the laboratory (solid lines) and comoving (dashed lines) frames. In this panel and the following, red corresponds to the $x$ component, magenta to the $y$ component and blue to the $z$ component. Third panel from top: comoving magnetic-field components (and total comoving magnetic field in gray) measured along the particle trajectory. Bottom: four-velocity components of $\boldsymbol{u_E}$, with $u_E$ in gray.
    \label{fig:Ldhist}}
\end{figure}

Each simulation has been run twice between times $2\,\ell_{\rm c}/c$ and $5\,\ell_{\rm c}/c$ to first detect and then analyze the time history of the highest-energy particles. Figure~\ref{fig:Ldhist} shows an example of one such particle in the $\gamma_{\rm syn} = 2000$ simulation. The upper panel depicts the evolution of the particle's Lorentz factor in the simulation frame (thick green line) and its reconstruction from the contributions of ideal (in blue) and nonideal (in dash-dotted red) electric fields. The dotted blue line ignores synchrotron cooling, while the dash-dotted blue line accounts for radiative losses, as calculated from Eq.~\eqref{eq:psyn2} using the recorded values of the electric and magnetic fields. The dashed dark green line plots the evolution of the Lorentz factor in the comoving frame \Rco.

As expected from Fig.~\ref{fig:workdist} and earlier studies, the nonideal electric fields provide an initial boost to the particle, bringing it to $\gamma \simeq 10-20$ within a few $0.01\ell_{\rm c}/c$, but its contribution at later times (larger Lorentz factors) becomes negligible, as indicated by the flattening of the red curve. 

The particle's Lorentz factor then increases to values near the synchrotron limit, $\gamma_{\rm rad} \simeq 1000$, within a time span of about $0.2\ell_{\rm c}/c$. This duration corresponds approximately to a gyrotime for a particle of this energy. This particle is thus energized at a rate close to the local Bohm rate, even though the mean acceleration rate $\bar{\nu}_{\rm acc}$, averaged over time and the particle population, is much lower (see below).

The second panel from the top displays the evolution of the momentum components ($p_x$ in red, $p_y$ in magenta and $p_z$ in blue), both in the simulation frame (solid lines) and in the comoving frame \Rco\ (dashed lines). The third panel from the top shows the magnetic-field components as measured in the comoving frame and as recorded along the particle trajectory, with the same conventions as for the momenta components. The gray line indicates the total value of the comoving magnetic field, with all components normalized to the guide field strength $B_0$. Finally, the bottom panel shows the evolution of the four-velocity of the comoving frame, as measured in the simulation frame along the particle's trajectory. Its total value is plotted in gray.

These histories indicate that most of the acceleration phase (in the comoving frame) occurs in a region of weaker-than-average magnetic field, and that, during this period, the particle sees the comoving frame accelerate. In the later acceleration stage, over $0.5 \lesssim t \lesssim 0.6\,\ell_{\rm c}/c$, the particle gains a slight extra amount of energy as it encounters a sudden increase in the mean magnetic field correlated with a sudden slowdown of the \Rco\ velocity. At times $\gtrsim 0.6\ell_{\rm c}/c$, radiation losses win over energy gains, and the particle's Lorentz factor rapidly drops to $\gamma \lesssim 100$, to further undergo well-resolved gyrations around the magnetic field. 

To analyze the energization process in more detail, we rely on the formalism of ``generalized Fermi acceleration'' that tracks the evolution of the particle momentum in the comoving frame \Rco\ \cite{2019PhRvD..99h3006L, 2021PhRvD.104f3020L, 2025PhRvE.112a5205L} and characterizes the energy gain or loss in terms of the inertial forces that the particle experiences in this frame. The energization process is then decomposed in terms of the gradients of the four-velocity $\boldsymbol{u_E} \equiv \gamma_E \boldsymbol{\beta_E}c$. This includes shear acceleration, corresponding to the spatial exploration of regions with net shear $\sigma_{ij} \equiv \partial_i {u_E}_j$ ($i\neq j$), compression ($\Theta_i \equiv \partial_i {u_E}_i$, no implicit sum on $i$ intended) and (Lagrangian) deceleration/acceleration of $\boldsymbol{u_E}$, written $a_i \equiv c^{-1}\left( \gamma_E \partial_t + \boldsymbol{u_E} \cdot \boldsymbol{\nabla} \right){u_E}_i$.
Introducing $\mu'_i\equiv p'_i/p'$, the rate of energy gain of ultrarelativistic particles can be expressed in terms of comoving time $t'$ as  
\begin{align}
    \frac{{\rm d}\ln\gamma'}{{\rm d}t'} = & 
    - \sum_{i=1}^{3} \left\{\mu'_i\,a_i 
    + {\mu'_i}^2\, \Theta_i 
    + \sum_{j=1,\,j\neq i}^{3} \mu'_i \mu'_j \, \sigma_{ij} \right\} \,.
    \label{eq:dyn_e}
\end{align}

Evaluating the various contributions along the three spatial axes reveals that most of the initial energization, occurring over the interval $0.2\ell_{\rm c}/c \lesssim t \lesssim 0.45\,\ell_{\rm c}/c$, originates from the acceleration of $\boldsymbol{u_E}$ in the $+\boldsymbol{y}$ direction. As the particle gyrates toward $-\boldsymbol{y}$ (see the evolution of $\boldsymbol{p'}$), the accelerated motion of the field line toward $+\boldsymbol{y}$ energizes $p'_y$. Subsequent gyration converts this gain in $p'_y$ into $p'_z$. Overall, the comoving particle energy increases by a factor $\simeq 10-20$ in this short period. Synchrotron losses during this phase are mitigated by the weaker-than-average magnetic field and the near alignment of the particle trajectory with the magnetic field. The pitch-angle cosine indeed takes values between $-0.7$ and $-1$. Interestingly, we observe that the particle Lorentz factor evolves exponentially over time, which is a characteristic feature of the above model whenever particles are subject to a constant force [Eq.~(\ref{eq:dyn_e})].

In the later phase, from $t \simeq 0.45$ to $\simeq 0.55\,\ell_{\rm c}/c$, the particle gets an additional energy boost, albeit more modest than the 10- or 20-fold increase observed in the initial phase. This gain results from the sudden deceleration of $\boldsymbol{u_E}$ and an increase in $B'$, corresponding to a compression of the velocity flow in the perpendicular plane, as in a mirror mode. The pitch-angle cosine accordingly falls to values $\simeq 0$, which maximizes energy gain in such a situation.

In the simulation frame, the energy gain follows from Lorentz boosting of the comoving momentum components with the local four-velocity $\boldsymbol{u_E}$. The momentum components in this simulation frame evolve under the conjunct effect of gyration (in the comoving frame), boost at $\boldsymbol{u_E}$ and drift along a perpendicular electric field. Inspection of the electromagnetic-field components in the vicinity of the trajectory does not reveal any particular structure, such as a current sheet.

\begin{figure}[ht]
    \centering
    \includegraphics[width=0.9\columnwidth]{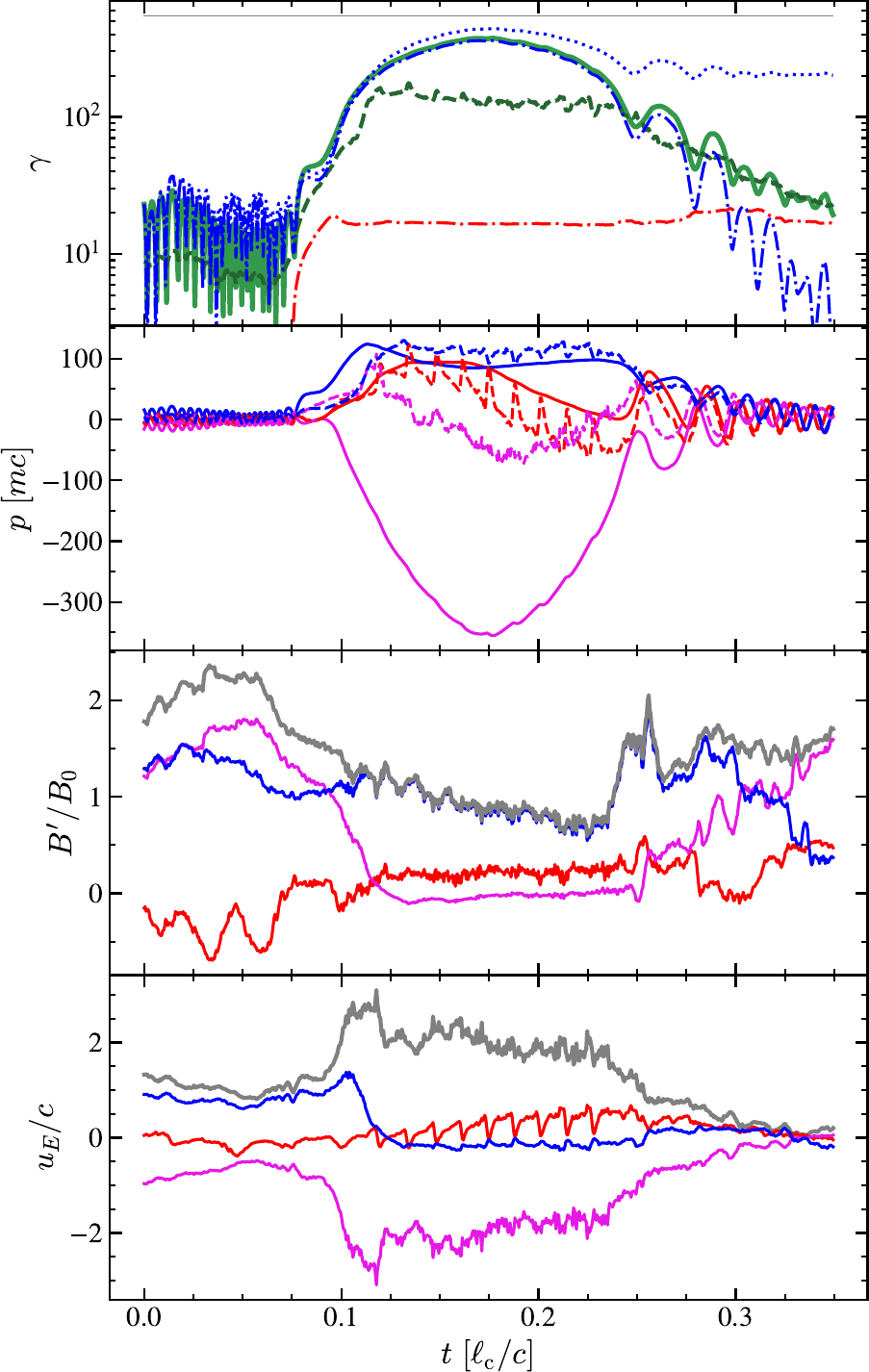}
    \caption{Same as Fig.~\ref{fig:Ldhist} for a particle approaching the synchrotron limit in the $\gamma_{\rm syn} = 500$ simulation. 
    \label{fig:Lhist}}
\end{figure}

Figure~\ref{fig:Lhist} provides a second example, using the same legends and notations as in Fig.~\ref{fig:Ldhist}, for the simulation with $\gamma_{\rm syn}=500$. The particle is here injected near a current sheet and initially boosted by a nonideal electric field up to $\gamma'\simeq 20$. Its energy is then increased tenfold by a perpendicular electric field, bringing the particle close to the synchrotron burn-off limit. Most of this acceleration occurs over $0.07 \lesssim t \lesssim 0.12\,\ell_{\rm c}/c$, after which the particle completes roughly half a gyration before losing its energy through synchrotron radiation. 

The magnetic field is initially polarized in the $y$-$z$ plane, approximately along the diagonal of the first quadrant. It then abruptly rotates toward the $+\boldsymbol{z}$ direction at the onset of the acceleration stage, and remains in this direction for the rest of the plotted time history. At the same time, the velocity field $\boldsymbol{u_E}$ rotates in the $y$-$z$ plane, from the diagonal of the second quadrant to the $-\boldsymbol{y}$ direction. The (comoving frame) pitch-angle cosine of the particle lies between $+0.75$ and $+1$ over most of the plotted time interval.

While these features -- namely, a curved field line accompanied by a rotation of $\boldsymbol{u_E}$ in the plane of curvature and a small pitch angle -- bear the hallmark of curvature-drift acceleration~\cite{2021PhRvD.104f3020L}, the actual energization process turns out to be more involved. The reason is that acceleration proceeds at such a fast rate that the particle gyroradius rapidly exceeds the characteristic curvature radius of the field line, which means that the acceleration is strongly nonadiabatic. A decomposition of the various force terms in the comoving frame confirms that curvature-drift acceleration, which assumes adiabaticity, and which manifests itself as the influence of the shear of $\boldsymbol{u_E}$ along the magnetic-field line, does not contribute substantially to the energy gain. Rather, the particle is accelerated through its exploration of the sheared velocity field $\boldsymbol{u_E}$ along $\boldsymbol{x}$ and, to a lesser degree, by the acceleration of $\boldsymbol{u_E}$ along $\boldsymbol{z}$. The acceleration term $a_z$, parallel to the magnetic-field line, generates an effective gravity force which energizes the $p'_z$ component of the particle as it moves along the magnetic field. The corresponding slight energy gain can be observed in the interval $0.12\,\ell_{\rm c}/c \lesssim t \lesssim 0.13\,\ell_{\rm c}/c$. 

As we record the components of the electromagnetic fields and their gradients along the particle trajectory, but not in its vicinity, it is not possible to draw a map of $\boldsymbol{u_E}$ around this trajectory. However, the shear terms $\partial_x {u_E}_y$ and $\partial_x {u_E}_z$, as well as the acceleration of $\boldsymbol{u_E}$ that the particle experiences along its history can be reconstructed from the gradients of the electromagnetic fields, using the methods developed in Ref.~\cite{2022PhRvD.106b3028B}. 

As in the previous example (Fig.~\ref{fig:Ldhist}), the particle Lorentz factor in the comoving frame evolves exponentially fast with time, at an energy-independent rate. This rate is comparable with, or slightly larger than, that recorded earlier in the simulation with $\gamma_{\rm syn}=2\,000$. 

In the interval $0.12\,\ell_{\rm c}/c\lesssim t\lesssim 0.22\,\ell_{\rm c}/c$, the comoving Lorentz factor $\gamma'$ remains nearly constant while the Lorentz factor in the laboratory frame traces out a half-cycloid, peaking at the midpoint of the interval.
At its maximum, $\gamma$ reaches $\gamma_{\rm rad}$, close to $300$ for this simulation. This evolution corresponds to a half-gyration in the comoving frame before the particle loses energy through radiative losses. These are likely triggered by the increase in comoving magnetic field at $t\sim 0.2\,\ell_{\rm c}/c$, accompanied by a reduction in pitch-angle cosine.
In the laboratory frame, the particle achieves its maximum energy at a phase of the orbit where its perpendicular comoving velocity is aligned with $\boldsymbol{u_E}$ (here along $-\boldsymbol{y}$).

At late times, $t\gtrsim 0.30\,\ell_{\rm c}/c$, the dash-dotted blue line (reconstructed contribution of ideal electric fields and radiative losses) falls below the solid green line (actual particle's Lorentz factor $\gamma$). This is because nonideal electric fields start to contribute again, as evidenced by the slight upward fluctuation of the red dash-dotted line, once $\gamma$ drops to values $\sim O(10-20)$.

\section{Discussion}
\label{sec:disc}

\subsection{Statistics of acceleration rates}
\label{sec:accrate}

\begin{figure}[ht]
   \centering
   \includegraphics[width=0.9\columnwidth]{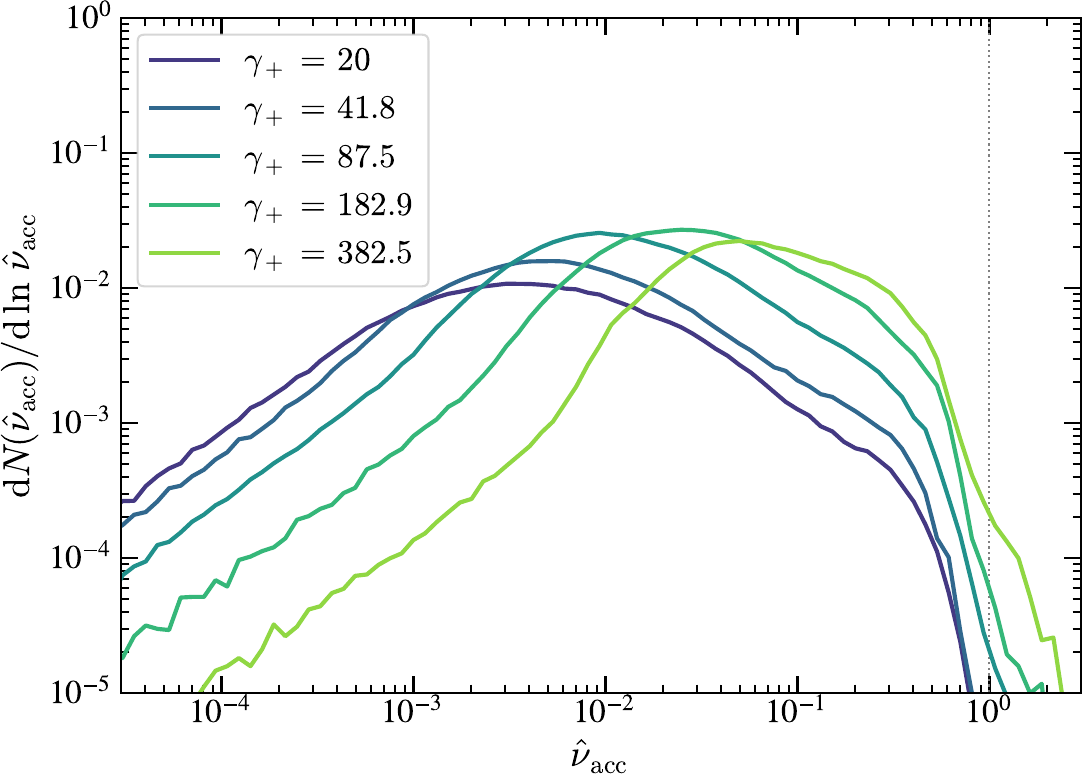}
    \caption{Distributions of normalized acceleration rates $\hat\nu_{\rm acc} = \nu_{\rm acc}r_{\rm g}/c$ with $\nu_{\rm acc} = (\Delta \gamma_{\rm ideal}/\gamma_+)/(t_+-t_-)$ (see text for details) measured in our reference simulation with $\gamma_{\rm syn}=2000$. The different curves are associated with different values of the Lorentz factor $\gamma_+$ at which these rates are measured, as indicated. A value $\hat{\nu}_{\rm acc} = 1$ corresponds to particle acceleration at the Bohm limit. These acceleration rates account for ideal electric fields and synchrotron losses, but discard the effect of nonideal electric fields.
    \label{fig:nudist}}
\end{figure}

Our results thus indicate that the particle spectra extend to Lorentz factors well beyond $\gamma_{\rm D}$, because a fraction of the particles are accelerated at a rate well above the mean characterized by $D_{\gamma \gamma}$. To quantify this process, we have carried out a direct measurement of the acceleration rate statistics in our PIC simulations. Here, we focus on the $\gamma_{\rm syn}=2000$ case. For each tracked particle $i$ (out of $\simeq 17\,000$), we have drawn at random about 100 pairs of points along the particle trajectory, with Lorentz factors $\gamma_-$ and $\gamma_+$ at times $t_-$ and $t_+$, where $t_+>t_-$. For each such pair, we record the reconstructed contribution $\Delta\gamma_{\rm ideal}$ of the perpendicular electric field and the associated radiation loss if $\Delta\gamma\equiv\gamma_+-\gamma_->0$, as in Fig.~\ref{fig:workdist}. The contribution from nonideal electric fields is thus discarded in this measurement. We then compute the relative acceleration rate at which a particle of Lorentz factor $\gamma_+$ has acquired its energy, $\nu = (\Delta\gamma_{\rm ideal}/\gamma_+)/(t_+-t_-)$, which we rescale into the dimensionless quantity $\hat\nu \equiv \nu r_{\rm g}/c$. The hat symbol is used to define quantities in units of the gyrofrequency $c/r_{\rm g}$, here measured at the point $(t_+,\,\gamma_+$) with the local value of the magnetic field.
From this large sample of values of $\hat\nu$, we then construct a histogram of normalized acceleration rates $\hat\nu_{\rm acc}$ for different bins of $\gamma_+$. This histogram, shown in Fig.~\ref{fig:nudist}, quantifies the distribution of acceleration rates for particles reaching a given Lorentz factor $\gamma_+$ within an interval of extent $0.3$ in log space.

This histogram indicates that the mean values of $\hat{\nu}_{\rm acc}$ (written $\bar{\hat\nu}_{\rm acc}$), which approximately correspond to the peak of each individual curve, scale as $\bar{\hat{\nu}}_{\rm acc} \propto \gamma_+$.
This means that the acceleration rate $\bar{\nu}_{\rm acc} \equiv \bar{\hat{\nu}}_{\rm acc}\,c/r_{\rm g}$ is independent of $\gamma_+$, which agrees with the expected scaling $D_{\gamma\gamma}\propto \gamma^2$. We have explicitly verified that $\bar\nu_{\rm acc}$ is independent of $\gamma_+$ to within $\simeq 10\,$\% and measured $\bar\nu_{\rm acc}\simeq 3\times 10^{-4} \,\omega_{\rm p}$ over the probed range of $\gamma_+$. This average value also matches $4D_{\gamma\gamma}/\gamma^2 \simeq 0.4 \sigma c/\ell_{\rm c}\simeq 3\times 10^{-4}\,\omega_{\rm p}$ for $\ell_{\rm c}=2000\,c/\omega_{\rm p}$ and $\sigma = 1.5$ (see Fig.~\ref{fig:parevol}). 

This mean acceleration rate is used in Sec.~\ref{sec:numres} to evaluate $\gamma_{\rm D}$. Specifically, we also measure the average value $\langle P_{\rm syn}/\gamma^2 \rangle\simeq 7.5\times 10^{-6}\,\omega_{\rm p}m_ec^2$ in the same simulation, and then use Eq.~\eqref{eq:gammaD} to derive $\gamma_{\rm D} = \bar\nu_{\rm acc} \,m_e c^2\langle P_{\rm syn}/\gamma^2\rangle^{-1} \simeq 40$. We follow the same procedure to compute $\gamma_{\rm D}$ for other simulations, although we shift the bins of $\gamma_+$ downward to match the trend in $\gamma_{\rm syn}$. We thus find $\bar\nu_{\rm acc} \simeq 3\times 10^{-4}\,\omega_{\rm p}$ for $\gamma_{\rm syn} = 1000$, while $\bar\nu_{\rm acc} \simeq 9\times 10^{-4}\,\omega_{\rm p}$ for $\gamma_{\rm syn}=500$ with some scatter, likely due to the contamination by reconnection at Lorentz factors $\sim \gamma_{\rm D}$ and to the reduced dynamical range. 

Below its mean value, the distribution of acceleration rates shown in Fig.~\ref{fig:nudist} scales approximately as ${\rm d} N/{\rm d} \ln \hat\nu_{\rm acc} \propto \hat{\nu}_{\rm acc}$, corresponding to ${\rm d}N/{\rm d}\hat\nu_{\rm acc} \sim \rm cst$, while above, it extends as a power law ${\rm d}N/{\rm d}\hat{\nu}_{\rm acc} \propto \hat{\nu}_{\rm acc}^{-s_\nu}$ with an index $s_\nu \simeq 2$. This power law eventually cuts off at $\hat\nu_{\rm acc} \simeq 1$, which corresponds to the Bohm scaling $\nu_{\rm acc} = \eta_B c/r_{\rm g}$ with $\eta_B=1$, as defined in Sec.~\ref{sec:setup}. This cutoff illustrates the classical argument that pictures Bohm acceleration as an ideal maximal acceleration rate, because $E < B$ implies $\vert \dot{\gamma} \vert < q B/m_{\rm e}c$, and hence $\nu_{\rm acc} r_{\rm g}/c < E/B <1$. Given that $\langle \mathbf{E}^2 \rangle^{1/2}/\langle\mathbf{B}^2\rangle^{1/2} \simeq 0.5$ in our simulations (see Fig.~\ref{fig:parevol}), it is not surprising that the cutoff in the acceleration rate appears to emerge around $\hat{\nu}_{\rm acc} \approx 0.5$. 

The broad distributions of acceleration rates and their power-law scaling, both clearly apparent in Fig.~\ref{fig:nudist}, are two striking and unexpected features of radiative relativistic turbulence. The fact that these distributions extend up to the Bohm limit for all Lorentz factors explains why some particles can reach energies well above the value $\gamma_D$ predicted by the mean of these distributions, in particular up the synchrotron burn-off limit $\gamma_{\rm rad}$. The power-law scaling with a uniform slope $s_\nu \simeq 2$ hints at a possible scale invariance of the underlying acceleration-rate distribution, which may account for the near-power-law tail of the particle energy spectrum. It also suggests that the observed extended acceleration rate distribution is a generic feature that would hold in situations with hierarchy $\gamma_{\rm rad}:\gamma_{\rm D}:\gamma_{\rm th}$ larger than probed by our simulations.

The upper bound on $\nu_{\rm acc}$ imposed by the Bohm scaling is expressed in the laboratory frame. In the comoving frame of the turbulence, the process can be pictured as follows. First, recall that acceleration is exponential over time as long as the polarity of the force seen by the particle remains constant [Eq.~\eqref{eq:dyn_e}]. However, this polarity changes sign as the particle gyrates around a field line, ultimately bounding the energization rate by the gyrofrequency associated with the Lorentz factor $\gamma_+$ reached at the end of this acceleration stage. In the examples examined in Sec.~\ref{sec:acc}, this change of polarity due to gyration marked the end of the main acceleration stage. 
In compressive modes, the gyration of the particle does not affect the sign of the force, yet the magnetic-field strength grows at least as fast as the particle's Lorentz factor, see Eq.~(A21) of Ref.~\cite{2021PhRvD.104f3020L}. Consequently, synchrotron losses become rapidly prohibitive and limit the maximum Lorentz factor to values smaller than $\gamma_{\rm rad}$. 

The energizing forces that enter Eq.~\eqref{eq:dyn_e} correspond to gradients of the four-velocity field $\boldsymbol{u_E}$. In a turbulent context, these gradients are best defined on a given scale $l$ through coarse graining of the turbulence on this scale~\cite{2025PhRvE.112a5205L}. Their generic value on this scale is hereafter written $\Gamma_l$. The contribution of gradients on small scales, namely, $l\lesssim r_{\rm g}(\gamma)$, to the energization of a particle of Lorentz factor $\gamma$ can generally be neglected, as the gyromotion tends to average out their contributions. Defining $\dot\gamma'\equiv {\rm d}\gamma'/{\rm d}t'$ and noting that the radiated power is a Lorentz invariant, the energization rate in a mode on scale $l \gtrsim r_{\rm g}$ can therefore be approximated by
\begin{equation}
   \frac{\dot\gamma'}{\gamma'}\,\simeq\,\Gamma_l - \frac{q{B}'}{m_{\rm e} c \gamma'_{\rm rad}}\frac{\gamma'}{\gamma'_{\rm rad}} \,.
   \label{eq:comov-syn}
\end{equation}
The synchrotron loss rate has been written in a form analogous to Eq.~\eqref{eq:psynrad}, introducing comoving frame quantities, in particular $\gamma'_{\rm rad}$. Acceleration up to $\gamma'_{\rm rad}$ then becomes possible if the gradient strength $\Gamma_l$ exceeds the gyrofrequency (in the comoving frame) for particles of Lorentz factor $\gamma'_{\rm rad}$,
\begin{equation}
   \Gamma_l \frac{m_{\rm e} c\gamma'_{\rm rad}}{q B'} \gtrsim 1\,.
   \label{eq:cond-grad}
\end{equation}
Now, the characteristic gradient strength is $\Gamma_l \sim \delta u_l /l$, with $\delta u_l$ as the four-velocity fluctuation on scale $l$. Strictly speaking, $\Gamma_l \sim \gamma_E u_E c/l$ for the acceleration term $a_i$ on scale $l$, but we assume $\gamma_E \sim O(1)$ here for simplicity. In the present numerical simulations, $\delta u_l \simeq \bar u_E$ for $l\gtrsim r_{\rm g}(\gamma_{\rm rad})$, since $r_{\rm g}(\gamma_{\rm rad})$ does not lie far below $\gamma_{\rm c}$. Given that, on average, $r'_{\rm g}(\gamma'_{\rm rad}) \sim r_{\rm g}(\gamma_{\rm rad})$, we make no further distinction between these two quantities. Therefore, the above inequality is verified on the scale $l\simeq r'_{\rm g}(\gamma'_{\rm rad})$ because $\bar u_E \gtrsim c$. In other words, the characteristic strength of the gradient on the scale of the gyroradius $r'_{\rm g}(\gamma'_{\rm rad})$ is large enough in relativistic turbulence to sustain particle acceleration up to the synchrotron burn-off limit in the comoving frame.

The analysis of the particle trajectories conducted in Sec.~\ref{sec:acc} directly illustrates the above argument, by exhibiting structures that accelerate particles at a constant rate over approximately half a gyration at maximal energy, close to the synchrotron limit. Importantly, both trajectories experienced a relativistic velocity $u_E/c \sim O(2)$ and a variation in $u_E$ of a comparable amount. A relatively small pitch-angle cosine and weaker-than-average magnetic-field strengths also contributed to reducing the energy losses in these cases.

A generic difficulty in kinetic PIC simulations is extrapolating their results to large spatial scales or, in the present case, to a wider hierarchy between the Lorentz factors $\gamma_{\rm th}$, $\gamma_{\rm D}$, $\gamma_{\rm rad}$ and $\gamma_{\rm c}$. We nonetheless expect our main finding -- that the particle energy distribution extends up to the synchrotron limit -- to hold in such extreme regimes. This is supported by the acceleration rate distributions in  Fig.~\ref{fig:nudist}, which exhibit power laws from $\bar{\hat{\nu}}_{\rm acc}$ to $\sim 1$. As noted above, this suggests the absence of an intrinsic scale that would otherwise preclude a naive extrapolation. 

We can also use the above discussion to predict how our results scale with the turbulence parameters. By borrowing the generic scaling of a Kolmogorov-type cascade $\delta u_l \simeq \bar u_E \,(l/\ell_{\rm c})^{1/3}$, and maximizing the gradient strength by setting the scale $l$ to the particle's gyroradius, one finds that the maximal Lorentz factor becomes
\begin{equation}
   \gamma'\lesssim \gamma'_{\rm rad}\,(\bar u_E/c)^{3/5}\,[r'_{\rm g}(\gamma'_{\rm rad})/\ell_{\rm c}]^{1/5}\,.
   \label{eq:cond-grad-gen}
\end{equation}
This demonstrates that the essential condition for reaching the synchrotron burn-off limit is that of relativistic turbulence, the ratio between $\gamma'_{\rm rad}$ and $\gamma_{\rm c}$ providing a minor correction. This argument further suggests that particle acceleration could become even more extreme in a turbulence without a guide field, as the average $u_E$ would be larger for the same magnetization level.

\subsection{Comparison to theoretical models}
\label{sec:model}

To contextualize our results, we now compare the PIC spectra to those expected from theoretical models of particle acceleration. We first consider a purely diffusive scheme, which is customarily used to model high-energy emission from astrophysical sources. We then integrate the corresponding time-dependent Fokker-Planck transport equation to obtain the energy distribution under conditions similar to those of the simulations. The second model that we consider is the generalized Fermi model presented in Ref.~\cite{2022PhRvL.129u5101L}, which describes particle acceleration in large-amplitude turbulence through random interactions with sharp bends of the magnetic-field lines and compressive modes. This model has been benchmarked against a numerical simulation of incompressible magnetohydrodynamics (MHD) turbulence without a guide field, for a characteristic Alfvén velocity of $\beta_{\rm A} \simeq 0.4$, thus relatively close to the conditions of our simulations.

\begin{figure}[ht]
   \centering
   \includegraphics[width=0.9\columnwidth]{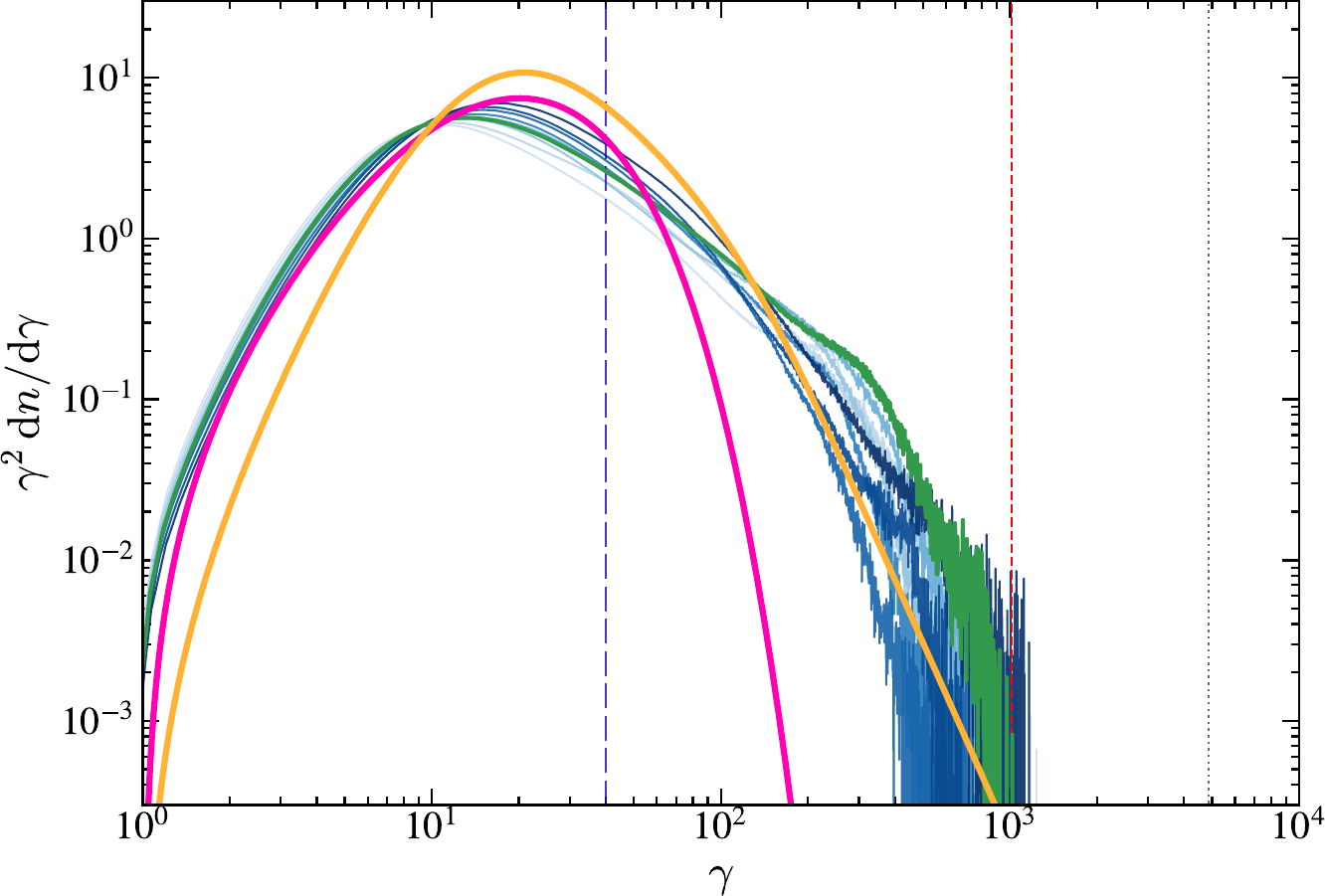}
      \caption{Theoretical energy particle spectra $\gamma^2{\rm d}n/{\rm d}\gamma$ computed with radiative cooling, overlaid on the time-dependent PIC spectra of Fig.~\ref{fig:meanspec} for $\gamma_{\rm syn}=2000$. Conventions for line styles are as in Fig.~\ref{fig:meanspec}. The thick magenta line corresponds to the purely diffusive Fokker-Planck model, while the orange line corresponds to the generalized Fermi model. See text for details. 
      \label{fig:scomp}}
\end{figure}

The standard, purely diffusive solution is obtained by solving the Fokker-Planck equation
\begin{equation}
    \partial_t n_\gamma \,= \,\partial_\gamma\left(\nu_{\rm a}\gamma^2\,\partial_\gamma n_\gamma\right) - 2\partial_\gamma\left(\nu_{\rm a}\gamma\, n_\gamma\right) + \partial_\gamma\left(\nu_{\rm s} \gamma^2\, n_\gamma\right) \,, 
   \label{eq:FP-stoch}
\end{equation} 
where $\nu_{\rm a}$ and $\nu_{\rm s}>0$ are two constants, characterizing acceleration (with diffusion coefficient $D_{\gamma\gamma}=\nu_{\rm a}\gamma^2$ as before) and synchrotron losses (with $\nu_{\rm s} = P_{\rm syn}/\gamma^2)$, respectively. We use the notation $n_\gamma \equiv {\rm d}n/{\rm d}\gamma$. 
This model is called ``purely diffusive'' because the full three-dimensional Fokker Planck operator in (isotropic) momentum space is of the form $\boldsymbol{\nabla}\,\mathbf{D_{pp}}\boldsymbol{\nabla}f$ with $\mathbf{D_{pp}}\propto D_{\gamma\gamma}\mathbf{I}$ ($\mathbf I$ unit matrix), $f(p,\,t)\equiv n_\gamma/4\pi p^2mc$ the distribution function, and $p\simeq \gamma mc$ here; in particular, it does not include a net advection rate. We set $\nu_{\rm a}=0.15$ in units of $c/\ell_{\rm c}$, comparable to the acceleration rate seen in the PIC simulation with $\gamma_{\rm syn}=2000$, then tune $\nu_{\rm s}$ to impose $\gamma_{\rm D}=40$ as in this simulation. Here, we determine this $\gamma_{\rm D}$ using $\nu_{\rm s}\gamma_{\rm D} = 4\nu_{\rm acc}$ as in Eq.~\eqref{eq:gammamax}. The initial distribution function takes on a Maxwell-Jüttner form with temperature $T=1\,\rm MeV$. We then integrate the system in time up to $4\,\ell_{\rm c}/c$, as in PIC simulations. Note that this model does not account for acceleration in reconnecting layers at low energies. Its solution is shown in magenta in Fig.~\ref{fig:scomp}, where it is compared with the time-varying PIC spectra obtained for $\gamma_{\rm sync}= 2000$.

The second theoretical spectrum, plotted as an orange line in Fig.~\ref{fig:scomp}, is obtained by integrating the generalized transport equation from Ref.~\cite{2022PhRvL.129u5101L}, with the addition of synchrotron losses using the same $\nu_{\rm s}$ value as above. The corresponding equation is written
\begin{align}
    \partial_t n_\gamma \,= \,& \int{\rm d}\gamma_1\,
\left[\frac{\varphi\left(\gamma_1\rightarrow\gamma\right)}{t_{\gamma_1}} n_{\gamma_1}(t) - 
\frac{\varphi\left(\gamma\rightarrow\gamma_1\right)}{t_\gamma} n_\gamma(t)\right]\nonumber\\
&\,\,+ \partial_\gamma\left(\nu_{\rm s} \gamma^2\, n_\gamma\right) \,.
    \label{eq:L22}
\end{align}
The function $\varphi\left(\gamma_1\rightarrow\gamma\right)$ encodes the probability of jumping  from $\gamma_1$ to $\gamma$ over a time interval $t_\gamma  = 2\pi r_{\rm g}(\gamma)/c$, and is itself related to the energy-dependent distribution functions of the energizing forces acting on the particle.  We adopt here the same parameters as in Ref.~\cite{2022PhRvL.129u5101L}, but rescale slightly the distribution widths (denoted therein as $\sigma$, not to be confused with the magnetization parameter here) 
of the random forces to match the present value of $\beta_{\rm A}=0.5$. The expected scaling being $\sigma \propto \beta_{\rm A}$, this modification is marginal. 

Because this generalized Fermi model integrates the particle energy distribution in the comoving frame, the spectra must be boosted back to the laboratory frame for comparison. We follow the procedure indicated in Ref.~\cite{2022PhRvL.129u5101L} for that purpose. Specifically, we measure the ratio $\gamma/\gamma'$ for the sample of particles tracked in the PIC simulation to reconstruct the probability density of the boosting factor, which we then apply to the spectra.

All spectra in Fig.~\ref{fig:scomp} are normalized to $\int {\rm d}n/{\rm d}\gamma\,{\rm d}\gamma=1$. Clearly, the Fokker-Planck model predicts a relatively sharp cutoff at $\simeq \gamma_{\rm D}$, which cannot explain the abundance of particles seen in the PIC simulation at Lorentz factors $\gg \gamma_{\rm D}$. This pileup has been commonly understood as a hallmark of stochastic acceleration~\cite{1984A&A...136..227S}. By contrast, the generalized Fermi model leads to a power-law tail, with a slope steeper than in the absence of synchrotron losses, here ${\rm d}n/{\rm d}\gamma \propto \gamma^{-s}$ with $s\simeq 4.5$, close to that observed at high energies in the PIC simulations. In fact, the reasonably good match between this model and the present simulations represents a significant result in itself. 

In this context, we remark that the above theoretical descriptions omit the heating term due to reconnection that would broaden slightly the distributions. Furthermore, the MHD simulations on which the distributions of energizing forces were measured are subrelativistic; it is possible, and generally anticipated on the basis that acceleration scales with the four-velocity $u_E$, that these distributions would become harder in the relativistic case. This would make the spectra harder, and thus in better agreement with the present simulations.

The difference between the Fokker-Planck [Eq.~(\ref{eq:FP-stoch})] and the above generalized Fermi model [Eq.~(\ref{eq:L22})] relates to the statistics of momentum jumps. While the Fokker-Planck equation describes Brownian motion characterized by uncorrelated jumps of small extent at each time step, the generalized transport equation considers an extended distribution of energy jumps, with hard tails as measured in MHD simulations. Consequently, the Fokker-Planck scheme fails to populate the energy region in which losses prevail over average gains, while energy gains remain possible, albeit being less and less probable as energy increases in the case of the generalized Fermi model. Said otherwise, this scheme allows for a distribution of acceleration rates, guaranteeing that a fraction of particles can escape synchrotron losses. 

\subsection{Application to pulsar wind nebulae}

While the present results have broad applications in relativistic astrophysics, they are particularly relevant to the Crab's pulsar wind nebula, famous for its broadband synchrotron emission up to the radiation-reaction limit \cite{1996MNRAS.278..525A,1996ApJ...457..253D, 2009ASSL..357..421K, 2021Univ....7..448A,2023A&A...671A..67D} and its variable behavior at high energies (see~\cite{2014RPPh...77f6901B} and references therein). Its spectral energy distribution is well described by a combination of synchrotron and inverse Compton emissions from a population of ultrarelativistic electron-positron pairs, with energies reaching several PeV. Yet, the acceleration mechanism responsible for this extreme particle distribution remains elusive.

Turbulence has been previously invoked to explain the origin of different spectral components in pulsar wind nebulae \cite{2016JPlPh..82d6301L,2017ApJ...841...78T,2019MNRAS.489.2403L,2020ApJ...896..147L,2023ApJ...953..116L,2024PTEP.2024e3E03T}. Particle acceleration up to the radiation-reaction limit, including during flaring states, is commonly attributed to linear acceleration in the nonideal electric fields of large-scale reconnection layers~\cite{2011ApJ...737L..40U,2012ApJ...746..148C}, although this scenario requires rather extreme magnetization parameters for this type of environment~\cite{2016ApJ...816L...8W}. In this picture, the flaring state is commonly viewed as the emergence of a new, transient hard spectral component superimposed on a steady flux, which is assumed to cut off slightly below the radiation-reaction limit. Alternative models, however, propose that flares arise from stochastic magnetic-field variations near the termination shock,
e.g.~\cite{2012MNRAS.421L..67B}.

Numerical MHD models point to the presence of turbulence throughout the pulsar wind nebula, especially in the vicinity of the termination shock~\cite{2009MNRAS.400.1241C, 2014MNRAS.438..278P, 2016JPlPh..82f6301O}.
Actually, turbulence in this region is required to dissipate the magnetic flux of the wind (the so-called $\sigma$ problem)~\cite{2017ApJ...847...57Z, 2018MNRAS.478.4622T}. From an observational perspective, though, the nebula presents a substantial degree of x-ray synchrotron polarization, suggesting that strongly turbulent regions should be localized~\cite{2023NatAs...7..602B}. The level and degree of pervasiveness of strong turbulence in the nebula thus remain subject to debate.

In this broad (and perhaps too rapidly surveyed) context, our present findings open new avenues for understanding the very high-energy emission of the Crab nebula. In particular, they suggest that synchrotron radiation at the burn-off limit could result from particle acceleration in ideal electric fields driven by turbulent plasma motions. They also suggest that the flaring states could reflect intrinsic variability within the accelerator, rather than the emergence of a new component atop a steady flux.

In fact, observations of flux depressions -- including the occasional disappearance of the Crab nebula from the 100--300~MeV band -- indicate that the mean flux does not set a floor value~\cite{2020MNRAS.496.5227P,2020A&A...638A.147Y}. A recent analysis of the Crab's variability has further revealed that flares occur on average every $0.7\,$yr and can last from several hours to about $0.1\,$yr \cite{2021ApJ...908...65H}.
Our numerical modeling, which predicts strong variability at the highest synchrotron photon energies and a broadband power spectrum of variability timescales, is consistent with these observations.

Interestingly, the hierarchy of Lorentz factors characterizing the present simulations is not dissimilar to that expected in the Crab nebula. To fix ideas, we consider a strongly turbulent region downstream of the termination shock, with reference values of $\sigma \simeq 1\,\hat\sigma$, $\delta B/B_0 \gtrsim 1$, $\ell_{\rm c} \simeq 0.1\,\hat\ell\,\rm pc$ and $B \approx 200\,\hat{B}\,\mu{\rm G}$, which provide reasonable fiducial parameters for this environment~\cite{2019hepr.confE..33A, 2021Univ....7..448A}. Notably, the value of $0.1\,$pc corresponds to the radius of the termination shock in the Crab nebula, and it agrees with the characteristic variability time reported above. Additionally, we posit a characteristic postshock Lorentz factor, prior to turbulent acceleration, of order $10^5$~\cite{2019MNRAS.489.2403L}. With respect to our numerical modeling, this Lorentz factor aligns with $\sim\gamma_{\rm th}$. Direct estimates of the characteristic Lorentz factors then yield (using $\kappa=10$)
\begin{align}
   \gamma_{\rm D} &\simeq 1.9\times 10^8\, \hat \sigma\, \hat{B}^{-2}\,\hat\ell^{-1} \,, \nonumber\\
   \gamma_{\rm rad} &\simeq 8.3\times 10^9\, \hat{B}^{-1/2} \,, \nonumber\\
   \gamma_{\rm c} &\simeq   1.1\times10^{10}\, \hat{B}\,\hat\ell \,.
   \label{eq:gamma-Crab}
\end{align}
Our results can thus be extrapolated to this case as the hierarchy $\gamma_{\rm th}:\gamma_{\rm D}:\gamma_{\rm rad}:\gamma_{\rm c}$ does not substantially differ from the one adopted in our simulations. Our simulations thus indicate that particle acceleration can be effective up to PeV energies in the postshock turbulence, with a spectrum that progressively steepens, or adopts a power-law tail, from $h\nu_D\simeq 100\,{\rm keV}\,\hat\sigma^2\,\hat{B}^{-3}\,\hat\ell^{-2}$ up to the synchrotron burn-off limit at $h\nu_{\rm rad}\sim 100\,\rm MeV$. 

Our simulations assume $\delta B/B_0\simeq 1$ and $\sigma \simeq 1$, which imply $\vert \bar E\vert/\vert\bar B\vert\simeq 0.5$, and hence $\bar u_E\simeq 0.5\,c$. As previously noted, higher values of $\bar u_E$ would allow particles to reach energies closer to the synchrotron burn-off limit, thereby increasing the synchrotron flux at $\simeq 100\,\rm MeV$ and beyond. This may explain why our simulations do not reproduce the extreme features of some giant flares, such as that of April 2011, which extended the synchrotron component up to several times the radiation-reaction limit (Fig.~\ref{fig:LfLc}). Such events may be too rare or may require larger values of $\delta B/B_0$ and/or $\sigma$. 
Furthermore, our simulations consider fixed values of $\ell_{\rm c}$, $\sigma$, and $\delta B/B_0$. Any time variation in these parameters would introduce an extra source of variability, with characteristic timescale $\gtrsim \ell_{\rm c}$. Of course, the total flux variability will eventually be reduced by $N^{1/2}$, where $N$ is the number of regions of size $\ell_{\rm c}$ contributing to the flux. For reference, our simulations typically contain $N\sim 4$ such regions.

To further progress along these lines, one must conduct a campaign of PIC simulations for different values of the parameters $\sigma$ and $\delta B/B_0$. Additionally, we can rely on the transport equation [Eq.~(\ref{eq:L22})] to implement the evolution of the distribution function in a semirealistic model of the nebula, in order to test the above predictions against the observed Crab nebula's spectrum. We stress here that it will be necessary to account for the backreaction of accelerated particles on the turbulence itself, following the methods developed in~\cite{2024PhRvD.109f3006L} and references therein. In fact, for a magnetization parameter $\sigma \sim 1$, particles draw energy from the turbulence about as fast as what is fed into the turbulence bath from external instabilities. This backreaction effectively damps the turbulence on short to intermediate scales. As an additional source of turbulent relaxation, it may therefore have phenomenological virtues in the context of recent IXPE observations. We plan to investigate these issues in a future work.

\section{Conclusions}
\label{sec:conc}

We have analyzed the results of particle-in-cell numerical simulations of particle acceleration in relativistic magnetized turbulence, for a magnetization level $\sigma \simeq 1$ and a fluctuation amplitude $\delta B/B_0\simeq 1$, in the presence of synchrotron cooling. The intensity of the latter is characterized by the Lorentz factor $\gamma_{\rm syn}$, which corresponds (to within a factor of 2) to the synchrotron burn-off limit Lorentz factor $\gamma_{\rm rad}$, at which particles radiate their energy in a gyrotime. In a turbulent relativistic context, this quantity can only be defined in an average sense, as various effects contribute to distort the classical synchrotron radiation picture in a uniform field, notably the Lorentz boosting from the frame of magnetic-field lines to the laboratory frame, the inhomogeneity of the magnetic field, and the possible pitch-angle anisotropy of particles.

Our study aimed to investigate the shape of the suprathermal particle spectrum around the maximal Lorentz factor $\gamma_{\rm D}$ predicted by balancing the average acceleration rate -- defined in terms of the momentum diffusion coefficient -- and the synchrotron cooling rate. Specifically, we have investigated a regime of weak radiative cooling, in which the plasma bulk does not cool on a crossing timescale of the turbulent coherence length $\ell_{\rm c}$. This regime is characterized by the hierarchy $\gamma_{\rm th}<\gamma_{\rm D}\ll \gamma_{\rm c}$, where $\gamma_{\rm th}\sim 5-10$ represents the Lorentz factor of the plasma bulk, and $\gamma_{\rm c}\simeq 5000$ the Lorentz factor at the Hillas-type limit where the particle gyroradius becomes comparable to $\ell_{\rm c}$.

Our main finding is that the particle spectrum does not cut off at $\gamma_{\rm D}$. It extends well beyond, up to the synchrotron burn-off limit at $\sim \gamma_{\rm rad}$, with a steeper slope than in the absence of radiative cooling. Specifically, for our chosen magnetization and turbulence amplitude, the simulation without radiative cooling shows a distribution ${\rm d}n/{\rm d}\gamma \propto \gamma^{-s}$ with $s \simeq 3$, while in the presence of radiative cooling, the spectrum steepens above $\gamma_{\rm D}$, retaining an approximate power law shape with index $s \approx 4$. Furthermore, the spectrum exhibits significant temporal variability at energies approaching the synchrotron limit (i.e., within a decade of it).

Although our 2D3V simulations are already large-scale ($8000^2$ cells), a precise determination of the spectral shape at $\gamma \gg \gamma_{\rm D}$ would necessitate even larger simulations. However, one difficulty that will arise is the statistical noise imposed by the finite number of particles. Indeed, the spectra shown in Fig.~\ref{fig:meanspec} cut off around $\gamma_{\rm rad}$ not because of a sudden steepening, but because the simulation gradually runs out of particles at such high energies. Extending the dynamical range to better capture the shape of this already steep spectrum would thus require many more particles than the several billion present in our simulations.

The simulated spectra can be understood qualitatively by computing the statistical distribution of the rates at which individual particles are accelerated. The corresponding distribution, shown in Fig.~\ref{fig:nudist}, reveals a characteristic broken power-law profile. The mean of this distribution defines a mean acceleration rate $\bar{\nu}_{\rm acc}$, related to the mean diffusion coefficient $D_{\gamma\gamma}$ through $\bar{\nu}_{\rm acc} \simeq 4D_{\gamma\gamma}/\gamma^2$. In agreement with the scaling $D_{\gamma\gamma} \propto \gamma^2$ reported in earlier studies, we find that $\bar{\nu}_{\rm acc}$ is independent of the energy at which we measure it. Interestingly, the acceleration rate is distributed as a power law beyond $\bar{\nu}_{\rm acc}$, up to the gyrofrequency of the particles. This behavior offers a plausible explanation for the extension of the synchrotron spectrum up to the radiation-reaction limit.

By closely tracking the trajectories of some of the particles reaching Lorentz factors close to $\gamma_{\rm rad}$, we have been able to pinpoint the dominant acceleration process. To do so, we have relied on the generalized Fermi formulation of stochastic acceleration~\cite{2025PhRvE.112a5205L}, which describes the evolution of the particle energy in the comoving frame of magnetic-field lines, expressing the energizing forces in terms of spatiotemporal gradients of the velocity field of this frame. This analysis reveals that, as predicted by the model, particles can be accelerated exponentially fast in time whenever exposed to a constant velocity-field gradient. Their orbits take place in weaker-than-average magnetic fields and with small pitch angles, both of which reduce synchrotron losses. Nevertheless, the main factor behind acceleration to the synchrotron burn-off limit is a relativistic velocity fluctuation $\delta u_E/c \sim 1$ ordered on a scale comparable with, or larger than, the gyroradius of particles with Lorentz factor $\gamma_{\rm rad}$. We have demonstrated that such gradients exist on the requisite scale in relativistic turbulence, unless the gyroradius $r_{\rm g}(\gamma_{\rm rad})$ lies orders of magnitude below the coherence scale $\ell_{\rm c}$ of the turbulence; see Eq.~\eqref{eq:cond-grad-gen} for a quantitative assessment.

Our results are naturally relevant to pulsar wind nebulae, which exhibit a similar hierarchy of characteristic Lorentz factors. In the particular case of the Crab nebula, we have found that stochastic acceleration in relativistic turbulence could account for the observed spectral energy distribution at the highest energies, from hard x-rays to $\gamma$-rays close to the radiation-reaction limit at $100\,\rm MeV$. The strong flux variability observed in our simulations could also accommodate the variability recently reported in this $\gamma$-ray domain.

These various observations call for a broad and comprehensive exploration of particle acceleration in relativistic turbulence with moderate synchrotron cooling. Such a study, requiring simulations on unprecedented spatiotemporal scales, would be essential for advancing the modeling of extreme astrophysical objects.

\begin{acknowledgments}
We thank A. Bykov and R. Walter for insightful discussions on the time variability of the Crab nebula. We warmly thank the technical and administrative staff of GENCI at TGCC for their help and support. This project was provided with computer and storage resources by GENCI at TGCC thanks to the Grants No. 2023-A0160411422, No. 2024-A0160411422, and No. 2025-A0160411422 on the Joliot-Curie supercomputer on the ROME partition.  
\end{acknowledgments}

\appendix

\section{Complementary numerical data}
\label{sec:simadd}

\begin{figure}[t]
    \centering
    \includegraphics[width=0.83\columnwidth]{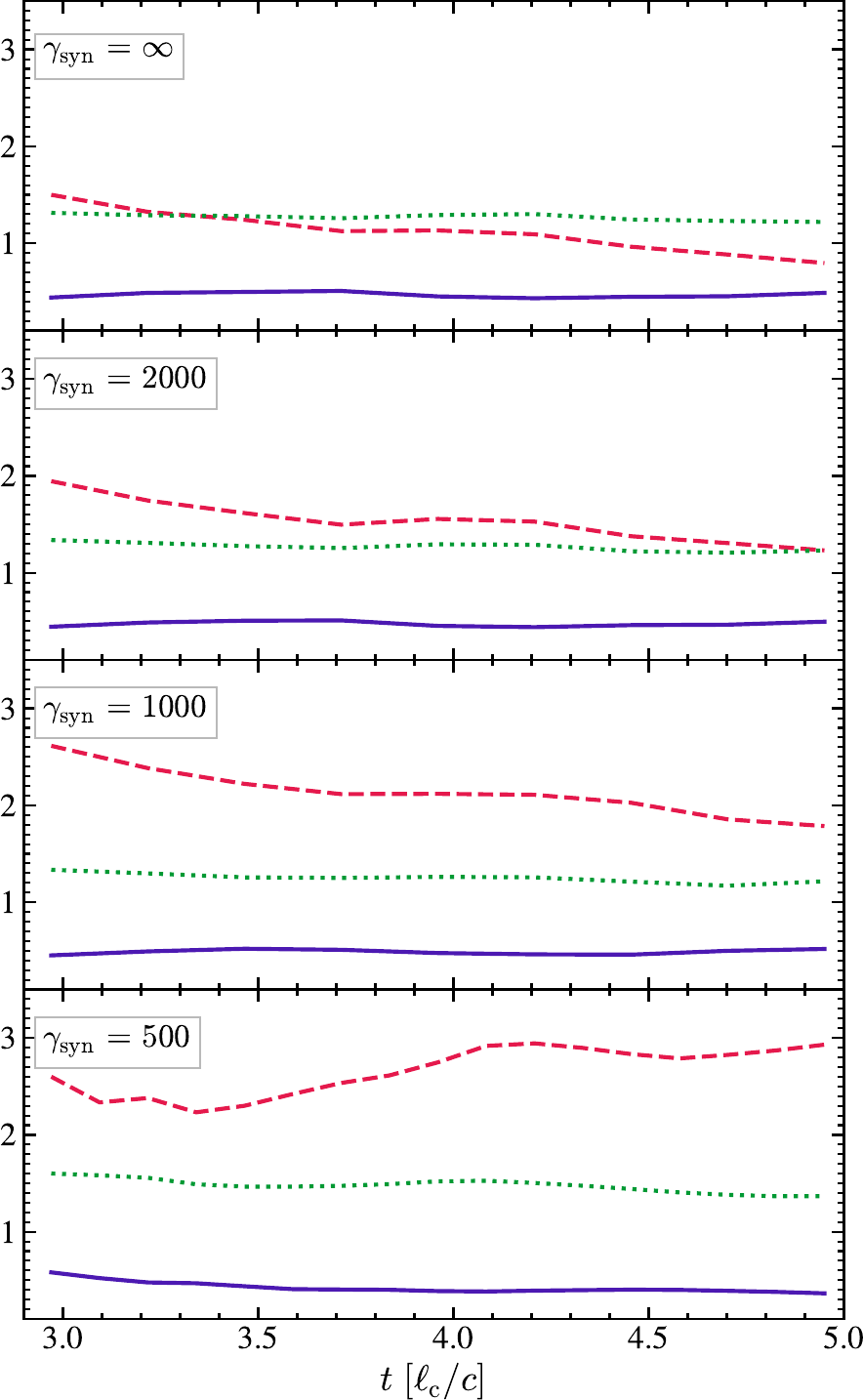}
    \caption{Evolution of the magnetization parameter $\sigma$ (red dashed line), of the magnetic perturbation amplitude $\delta B/B_0$ (green dotted line) and of the mean $\bar u_E/c$ (solid blue line) as a function of time, for the fiducial simulations discussed in the main text.
    \label{fig:parevol}}
\end{figure}

Here, we provide additional details on our numerical simulations. In Fig.~\ref{fig:parevol}, in particular, we show the time evolution of the total magnetization parameter $\sigma$, the magnetic perturbation amplitude $\delta B/B_0$ and the mean four-velocity $\bar u_E$, over the time interval $3\,\ell_{\rm c}/c < t < 5\,\ell_{\rm c}/c$. The four-velocity $\bar u_E$ is computed as the average of $u_E$ over the simulation box at a given time. For reference, the initial plasma magnetization is $\sigma_0 \simeq 1.5$.

\begin{figure}[t]
    \centering
    \includegraphics[width=0.85\columnwidth]{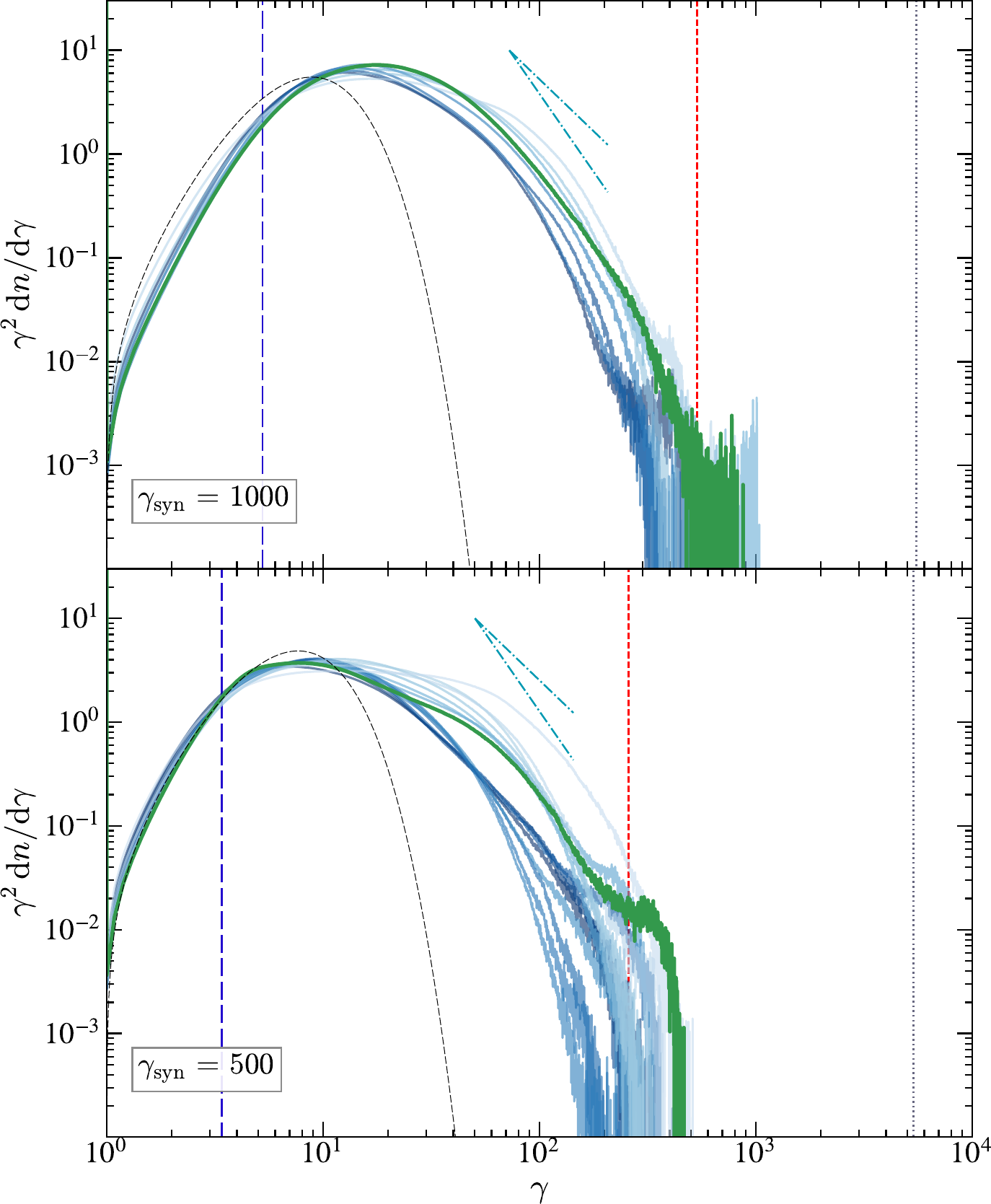}
    \caption{Same as Fig.~\ref{fig:meanspec}, for two subdominant species in the simulations with $\gamma_{\rm syn}= 1000$ (top panel) and $500$ (bottom panel), as indicated. 
    \label{fig:meanspec_34}}
\end{figure}

In Fig.~\ref{fig:meanspec_34}, we present the energy distribution of the subdominant species that were injected at time $t=3\,\ell_{\rm c}/c$ in the two simulations in which they have been introduced, namely, $\gamma_{\rm syn}=1000$ (top panel) and $\gamma_{\rm syn}=500$ (bottom panel). As discussed in Sec.~\ref{sec:sims}, these species are initialized with a small statistical weight, so that they do not contribute to the turbulent plasma dynamics. Their motion is frozen until $t=3\,\ell_{\rm c}/c$, at which point they start to feel the influence of the turbulence. The plotted spectra are comparable to those shown in Fig.~\ref{fig:meanspec} for the dominant species, which have evolved throughout the initialization stage of the turbulence, from $t=0$ to $3\,\ell_{\rm c}/c$. This indicates that the spectra analyzed have not been strongly affected by this initialization stage.

\begin{figure}[h!]
    \centering
    \includegraphics[width=0.85\columnwidth]{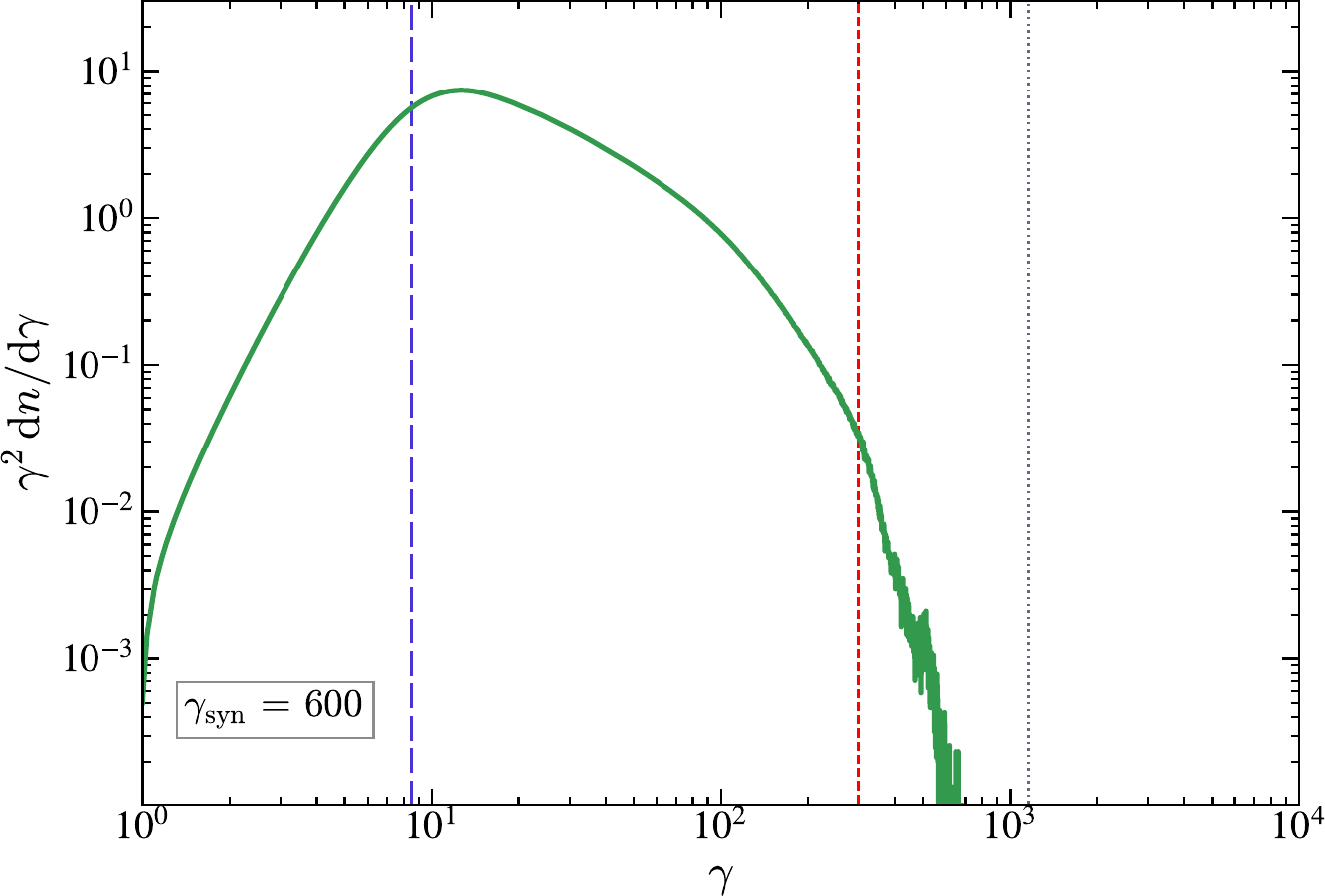}
    \caption{Same as Fig.~\ref{fig:meanspec} for the 3D turbulence simulation with $\gamma_{\rm syn}=600$. The spectrum is recorded at $t=2.8\,\ell_{\rm c}/c$.
    \label{fig:meanspec_3d}}
\end{figure}

Finally, Fig.~\ref{fig:meanspec_3d} depicts the particle energy distribution obtained at time $t=2.8\,\ell_{\rm c}/c$ in a large-scale, fully three-dimensional simulation. The domain contains $1024^3$ cells with a mesh size $\delta x=\delta y=\delta z=1.5\,c/\omega_{\rm p}$, which offers a dynamic range $L\omega_{\rm p}/c = 1728$. The corresponding turbulent coherence length $\ell_{\rm c}\simeq 0.5\,L\simeq 860\,c/\omega_{\rm p}$. This 3D simulation uses a total of 10 particles per cell, and $\gamma_{\rm syn}=600$. The other parameters are similar to those in the 2D3V simulations. The obtained spectrum resembles those observed at comparable times in the 2D3V cases.

\begin{figure}[ht!]
    \centering
    \includegraphics[width=0.8\columnwidth]{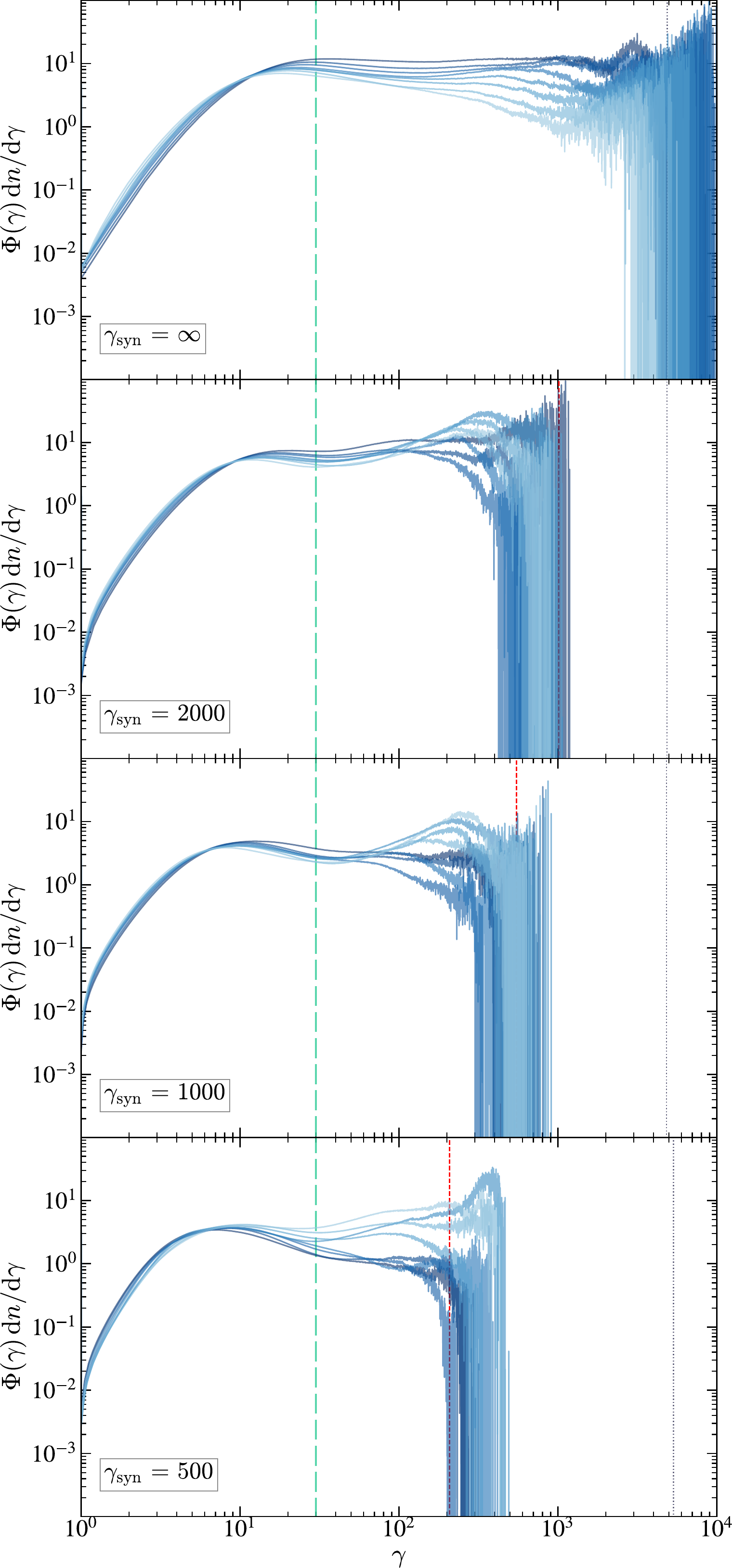}
    \caption{Same as Fig.~\ref{fig:meanspec}, with spectra unfolded by the function $\Phi(\gamma)$ [Eq.~\eqref{eq:defPhi}]. This function takes the approximate value $\simeq 1$ for $\gamma \lesssim \gamma_{\rm lo} = 40$ (indicated by the green dashed line) and above this threshold, evolves into a power-law with an index $s_\Phi$ and an exponential cutoff at $\gamma_{\rm hi}$. In the upper panel, we use $s_\Phi = 3$ and $\gamma_{\rm hi} = \gamma_{\rm c}$, while the other panels correspond to $s_\Phi = 4$ and $\gamma_{\rm hi} = 0.5\gamma_{\rm rad}$ (see text for details). 
    \label{fig:unfold}}
\end{figure}

In all simulations, the magnetization decreases in time as the plasma draws energy from the turbulent cascade. In the absence of radiative losses, the total magnetization drops to values slightly below unity at $5\,\ell_{\rm c}/c$, while it remains slightly above unity in other cases. This behavior directly results from radiative losses, which steepen the high-energy part of the particle distribution and slightly cool the thermal bulk, thereby reducing the total energy content of the plasma. The simulation with $\gamma_{\rm syn}=500$ has been initialized with slightly larger amplitudes of the Langevin antenna and a smaller damping constant, which explains the larger magnetization seen in that simulation. The perturbation amplitude and average $\bar u_E$ velocity remain approximately constant over the time interval of interest in all simulations.

To examine the spectral shapes above $\gamma_{\rm D}$, we have extracted a power-law extension with a high-energy exponential cutoff from the PIC spectra. To do so, we multiply the spectra ${\rm d}n/{\rm d}\gamma$ by the following function
\begin{equation}
    \Phi(\gamma)\,\equiv\, 1 + \exp(-\gamma_{\rm lo}/\gamma)\,(\gamma/\gamma_{\rm lo})^{s_\Phi}\exp(\gamma/\gamma_{\rm hi})\,,
    \label{eq:defPhi}
\end{equation}
where $\gamma_{\rm lo}$, $\gamma_{\rm hi}$ and $s_\Phi$ are free parameters. This ad hoc profile is designed to fulfill the following limits: $\Phi(\gamma)\simeq 1$ at $\gamma\ll \gamma_{\rm lo}$ and $\Phi(\gamma)\simeq (\gamma/\gamma_{\rm lo})^{s_\Phi} \exp(\gamma/\gamma_{\rm hi})$ at $\gamma\gg \gamma_{\rm hi}$. In this description, $\gamma_{\rm lo}$ is the lower bound of our range of interest, $s_\Phi$ is the spectral slope, and $\gamma_{\rm hi}$ is the high-energy cutoff. We choose $\gamma_{\rm lo} = 40$ to focus on the range well beyond the influence of reconnection. 
We show the outcome of this procedure in Fig.~\ref{fig:unfold}, with the following parameters.

For $\gamma_{\rm syn} \rightarrow \infty$ (no synchrotron losses), we take $s_\Phi =3$ and $\gamma_{\rm hi} = \gamma_{\rm c}$. The recovered flat spectrum indicates that the spectral slope indeed follows ${\rm d}n/{\rm d}\gamma \propto \gamma^{-3}$, with a tendency to harden over time.

For the other simulations with radiative cooling, we set $s_\Phi=4$ and place the high-energy cutoff at $\gamma_{\rm hi} = 0.5\gamma_{\rm rad}$. The prefactor of $0.5$ is chosen because $\bar{E}/\bar{B} \simeq 0.5$ in our simulations (see Fig.~\ref{fig:parevol}), although it has a weak influence on the final results. Here as well, the power-law shape appears to fit the observed spectra reasonably well within the temporal variations.


\bibliography{refs}

\begin{thebibliography}{82}%
\makeatletter
\providecommand \@ifxundefined [1]{%
 \@ifx{#1\undefined}
}%
\providecommand \@ifnum [1]{%
 \ifnum #1\expandafter \@firstoftwo
 \else \expandafter \@secondoftwo
 \fi
}%
\providecommand \@ifx [1]{%
 \ifx #1\expandafter \@firstoftwo
 \else \expandafter \@secondoftwo
 \fi
}%
\providecommand \natexlab [1]{#1}%
\providecommand \enquote  [1]{``#1''}%
\providecommand \bibnamefont  [1]{#1}%
\providecommand \bibfnamefont [1]{#1}%
\providecommand \citenamefont [1]{#1}%
\providecommand \href@noop [0]{\@secondoftwo}%
\providecommand \href [0]{\begingroup \@sanitize@url \@href}%
\providecommand \@href[1]{\@@startlink{#1}\@@href}%
\providecommand \@@href[1]{\endgroup#1\@@endlink}%
\providecommand \@sanitize@url [0]{\catcode `\\12\catcode `\$12\catcode
  `\&12\catcode `\#12\catcode `\^12\catcode `\_12\catcode `\%12\relax}%
\providecommand \@@startlink[1]{}%
\providecommand \@@endlink[0]{}%
\providecommand \url  [0]{\begingroup\@sanitize@url \@url }%
\providecommand \@url [1]{\endgroup\@href {#1}{\urlprefix }}%
\providecommand \urlprefix  [0]{URL }%
\providecommand \Eprint [0]{\href }%
\providecommand \doibase [0]{https://doi.org/}%
\providecommand \selectlanguage [0]{\@gobble}%
\providecommand \bibinfo  [0]{\@secondoftwo}%
\providecommand \bibfield  [0]{\@secondoftwo}%
\providecommand \translation [1]{[#1]}%
\providecommand \BibitemOpen [0]{}%
\providecommand \bibitemStop [0]{}%
\providecommand \bibitemNoStop [0]{.\EOS\space}%
\providecommand \EOS [0]{\spacefactor3000\relax}%
\providecommand \BibitemShut  [1]{\csname bibitem#1\endcsname}%
\let\auto@bib@innerbib\@empty
\bibitem [{\citenamefont {{Petrosian}}(2012)}]{2012SSRv..173..535P}%
  \BibitemOpen
  \bibfield  {author} {\bibinfo {author} {\bibfnamefont {V.}~\bibnamefont
  {{Petrosian}}},\ }\bibfield  {title} {\bibinfo {title} {{Stochastic
  acceleration by turbulence}},\ }\href
  {https://doi.org/10.1007/s11214-012-9900-6} {\bibfield  {journal} {\bibinfo
  {journal} {Sp.\ Sc.\ Rev.}\ }\textbf {\bibinfo {volume} {173}},\ \bibinfo
  {pages} {535} (\bibinfo {year} {2012})}\BibitemShut {NoStop}%
\bibitem [{\citenamefont {{Kardashev}}(1962)}]{1962SvA.....6..317K}%
  \BibitemOpen
  \bibfield  {author} {\bibinfo {author} {\bibfnamefont {N.~S.}\ \bibnamefont
  {{Kardashev}}},\ }\bibfield  {title} {\bibinfo {title} {{Nonstationarity of
  spectra of young sources of nonthermal radio emission}},\ }\href@noop {}
  {\bibfield  {journal} {\bibinfo  {journal} {Sov. Astron.}\ }\textbf {\bibinfo
  {volume} {6}},\ \bibinfo {pages} {317} (\bibinfo {year} {1962})}\BibitemShut
  {NoStop}%
\bibitem [{\citenamefont {{Schlickeiser}}(1984)}]{1984A&A...136..227S}%
  \BibitemOpen
  \bibfield  {author} {\bibinfo {author} {\bibfnamefont {R.}~\bibnamefont
  {{Schlickeiser}}},\ }\bibfield  {title} {\bibinfo {title} {{An explanation of
  abrupt cutoffs in the optical-infrared spectra of non-thermal sources - A new
  pile-up mechanism for relativistic electron spectra}},\ }\href
  {http://adsabs.harvard.edu/abs/1984A%26A...136..227S} {\bibfield  {journal}
  {\bibinfo  {journal} {Astron.\ Astrophys.}\ }\textbf {\bibinfo {volume}
  {136}},\ \bibinfo {pages} {227} (\bibinfo {year} {1984})}\BibitemShut
  {NoStop}%
\bibitem [{\citenamefont {{Dmitruk}}\ \emph {et~al.}(2004)\citenamefont
  {{Dmitruk}}, \citenamefont {{Matthaeus}},\ and\ \citenamefont
  {{Seenu}}}]{2004ApJ...617..667D}%
  \BibitemOpen
  \bibfield  {author} {\bibinfo {author} {\bibfnamefont {P.}~\bibnamefont
  {{Dmitruk}}}, \bibinfo {author} {\bibfnamefont {W.~H.}\ \bibnamefont
  {{Matthaeus}}},\ and\ \bibinfo {author} {\bibfnamefont {N.}~\bibnamefont
  {{Seenu}}},\ }\bibfield  {title} {\bibinfo {title} {{Test particle
  energization by current sheets and nonuniform fields in magnetohydrodynamic
  turbulence}},\ }\href {https://doi.org/10.1086/425301} {\bibfield  {journal}
  {\bibinfo  {journal} {Astrophys. J.}\ }\textbf {\bibinfo {volume} {617}},\
  \bibinfo {pages} {667} (\bibinfo {year} {2004})}\BibitemShut {NoStop}%
\bibitem [{\citenamefont {{Arzner}}\ \emph {et~al.}(2006)\citenamefont
  {{Arzner}}, \citenamefont {{Knaepen}}, \citenamefont {{Carati}},
  \citenamefont {{Denewet}},\ and\ \citenamefont {{Vlahos}}}]{06Arzner}%
  \BibitemOpen
  \bibfield  {author} {\bibinfo {author} {\bibfnamefont {K.}~\bibnamefont
  {{Arzner}}}, \bibinfo {author} {\bibfnamefont {B.}~\bibnamefont {{Knaepen}}},
  \bibinfo {author} {\bibfnamefont {D.}~\bibnamefont {{Carati}}}, \bibinfo
  {author} {\bibfnamefont {N.}~\bibnamefont {{Denewet}}},\ and\ \bibinfo
  {author} {\bibfnamefont {L.}~\bibnamefont {{Vlahos}}},\ }\bibfield  {title}
  {\bibinfo {title} {{The effect of coherent structures on stochastic
  acceleration in MHD turbulence}},\ }\href {https://doi.org/10.1086/498341}
  {\bibfield  {journal} {\bibinfo  {journal} {Astrophys. J.}\ }\textbf
  {\bibinfo {volume} {637}},\ \bibinfo {pages} {322} (\bibinfo {year}
  {2006})}\BibitemShut {NoStop}%
\bibitem [{\citenamefont {{Kowal}}\ \emph {et~al.}(2012)\citenamefont
  {{Kowal}}, \citenamefont {{de Gouveia Dal Pino}},\ and\ \citenamefont
  {{Lazarian}}}]{2012PhRvL.108x1102K}%
  \BibitemOpen
  \bibfield  {author} {\bibinfo {author} {\bibfnamefont {G.}~\bibnamefont
  {{Kowal}}}, \bibinfo {author} {\bibfnamefont {E.~M.}\ \bibnamefont {{de
  Gouveia Dal Pino}}},\ and\ \bibinfo {author} {\bibfnamefont {A.}~\bibnamefont
  {{Lazarian}}},\ }\bibfield  {title} {\bibinfo {title} {{Particle Acceleration
  in Turbulence and Weakly Stochastic Reconnection}},\ }\href
  {https://doi.org/10.1103/PhysRevLett.108.241102} {\bibfield  {journal}
  {\bibinfo  {journal} {Phys.\ Rev.\ Lett.}\ }\textbf {\bibinfo {volume}
  {108}},\ \bibinfo {eid} {241102} (\bibinfo {year} {2012})}\BibitemShut
  {NoStop}%
\bibitem [{\citenamefont {{Lynn}}\ \emph {et~al.}(2013)\citenamefont {{Lynn}},
  \citenamefont {{Quataert}}, \citenamefont {{Chandran}},\ and\ \citenamefont
  {{Parrish}}}]{2013ApJ...777..128L}%
  \BibitemOpen
  \bibfield  {author} {\bibinfo {author} {\bibfnamefont {J.~W.}\ \bibnamefont
  {{Lynn}}}, \bibinfo {author} {\bibfnamefont {E.}~\bibnamefont {{Quataert}}},
  \bibinfo {author} {\bibfnamefont {B.~D.~G.}\ \bibnamefont {{Chandran}}},\
  and\ \bibinfo {author} {\bibfnamefont {I.~J.}\ \bibnamefont {{Parrish}}},\
  }\bibfield  {title} {\bibinfo {title} {{The efficiency of second-order Fermi
  acceleration by weakly compressible magnetohydrodynamic turbulence}},\ }\href
  {https://doi.org/10.1088/0004-637X/777/2/128} {\bibfield  {journal} {\bibinfo
   {journal} {Astrophys. J.}\ }\textbf {\bibinfo {volume} {777}},\ \bibinfo
  {eid} {128} (\bibinfo {year} {2013})}\BibitemShut {NoStop}%
\bibitem [{\citenamefont {{Dalena}}\ \emph {et~al.}(2014)\citenamefont
  {{Dalena}}, \citenamefont {{Rappazzo}}, \citenamefont {{Dmitruk}},
  \citenamefont {{Greco}},\ and\ \citenamefont
  {{Matthaeus}}}]{2014ApJ...783..143D}%
  \BibitemOpen
  \bibfield  {author} {\bibinfo {author} {\bibfnamefont {S.}~\bibnamefont
  {{Dalena}}}, \bibinfo {author} {\bibfnamefont {A.~F.}\ \bibnamefont
  {{Rappazzo}}}, \bibinfo {author} {\bibfnamefont {P.}~\bibnamefont
  {{Dmitruk}}}, \bibinfo {author} {\bibfnamefont {A.}~\bibnamefont {{Greco}}},\
  and\ \bibinfo {author} {\bibfnamefont {W.~H.}\ \bibnamefont {{Matthaeus}}},\
  }\bibfield  {title} {\bibinfo {title} {{Test-particle acceleration in a
  hierarchical three-dimensional turbulence model}},\ }\href
  {https://doi.org/10.1088/0004-637X/783/2/143} {\bibfield  {journal} {\bibinfo
   {journal} {Astrophys. J.}\ }\textbf {\bibinfo {volume} {783}},\ \bibinfo
  {eid} {143} (\bibinfo {year} {2014})}\BibitemShut {NoStop}%
\bibitem [{\citenamefont {{Isliker}}\ \emph {et~al.}(2017)\citenamefont
  {{Isliker}}, \citenamefont {{Vlahos}},\ and\ \citenamefont
  {{Constantinescu}}}]{2017PhRvL.119d5101I}%
  \BibitemOpen
  \bibfield  {author} {\bibinfo {author} {\bibfnamefont {H.}~\bibnamefont
  {{Isliker}}}, \bibinfo {author} {\bibfnamefont {L.}~\bibnamefont
  {{Vlahos}}},\ and\ \bibinfo {author} {\bibfnamefont {D.}~\bibnamefont
  {{Constantinescu}}},\ }\bibfield  {title} {\bibinfo {title} {{Fractional
  transport in strongly turbulent plasmas}},\ }\href
  {https://doi.org/10.1103/PhysRevLett.119.045101} {\bibfield  {journal}
  {\bibinfo  {journal} {Phys.\ Rev.\ Lett.}\ }\textbf {\bibinfo {volume}
  {119}},\ \bibinfo {eid} {045101} (\bibinfo {year} {2017})}\BibitemShut
  {NoStop}%
\bibitem [{\citenamefont {{Pecora}}\ \emph {et~al.}(2018)\citenamefont
  {{Pecora}}, \citenamefont {{Servidio}}, \citenamefont {{Greco}},
  \citenamefont {{Matthaeus}}, \citenamefont {{Burgess}}, \citenamefont
  {{Haynes}}, \citenamefont {{Carbone}},\ and\ \citenamefont
  {{Veltri}}}]{2018JPlPh..84f7201P}%
  \BibitemOpen
  \bibfield  {author} {\bibinfo {author} {\bibfnamefont {F.}~\bibnamefont
  {{Pecora}}}, \bibinfo {author} {\bibfnamefont {S.}~\bibnamefont
  {{Servidio}}}, \bibinfo {author} {\bibfnamefont {A.}~\bibnamefont {{Greco}}},
  \bibinfo {author} {\bibfnamefont {W.~H.}\ \bibnamefont {{Matthaeus}}},
  \bibinfo {author} {\bibfnamefont {D.}~\bibnamefont {{Burgess}}}, \bibinfo
  {author} {\bibfnamefont {C.~T.}\ \bibnamefont {{Haynes}}}, \bibinfo {author}
  {\bibfnamefont {V.}~\bibnamefont {{Carbone}}},\ and\ \bibinfo {author}
  {\bibfnamefont {P.}~\bibnamefont {{Veltri}}},\ }\bibfield  {title} {\bibinfo
  {title} {{Ion diffusion and acceleration in plasma turbulence}},\ }\href
  {https://doi.org/10.1017/S0022377818000995} {\bibfield  {journal} {\bibinfo
  {journal} {J.\ Plasma \ Phys.}\ }\textbf {\bibinfo {volume} {84}},\ \bibinfo
  {eid} {725840601} (\bibinfo {year} {2018})}\BibitemShut {NoStop}%
\bibitem [{\citenamefont {{Kimura}}\ \emph {et~al.}(2019)\citenamefont
  {{Kimura}}, \citenamefont {{Tomida}},\ and\ \citenamefont
  {{Murase}}}]{2019MNRAS.485..163K}%
  \BibitemOpen
  \bibfield  {author} {\bibinfo {author} {\bibfnamefont {S.~S.}\ \bibnamefont
  {{Kimura}}}, \bibinfo {author} {\bibfnamefont {K.}~\bibnamefont {{Tomida}}},\
  and\ \bibinfo {author} {\bibfnamefont {K.}~\bibnamefont {{Murase}}},\
  }\bibfield  {title} {\bibinfo {title} {{Acceleration and escape processes of
  high-energy particles in turbulence inside hot accretion flows}},\ }\href
  {https://doi.org/10.1093/mnras/stz329} {\bibfield  {journal} {\bibinfo
  {journal} {Mon.\ Not.\ R.\ Astron.\ Soc.}\ }\textbf {\bibinfo {volume}
  {485}},\ \bibinfo {pages} {163} (\bibinfo {year} {2019})}\BibitemShut
  {NoStop}%
\bibitem [{\citenamefont {{Trotta}}\ \emph {et~al.}(2020)\citenamefont
  {{Trotta}}, \citenamefont {{Franci}}, \citenamefont {{Burgess}},\ and\
  \citenamefont {{Hellinger}}}]{2020ApJ...894..136T}%
  \BibitemOpen
  \bibfield  {author} {\bibinfo {author} {\bibfnamefont {D.}~\bibnamefont
  {{Trotta}}}, \bibinfo {author} {\bibfnamefont {L.}~\bibnamefont {{Franci}}},
  \bibinfo {author} {\bibfnamefont {D.}~\bibnamefont {{Burgess}}},\ and\
  \bibinfo {author} {\bibfnamefont {P.}~\bibnamefont {{Hellinger}}},\
  }\bibfield  {title} {\bibinfo {title} {{Fast acceleration of
  transrelativistic electrons in astrophysical turbulence}},\ }\href
  {https://doi.org/10.3847/1538-4357/ab873c} {\bibfield  {journal} {\bibinfo
  {journal} {Astrophys. J.}\ }\textbf {\bibinfo {volume} {894}},\ \bibinfo
  {eid} {136} (\bibinfo {year} {2020})}\BibitemShut {NoStop}%
\bibitem [{\citenamefont {{Sun}}\ and\ \citenamefont
  {{Bai}}(2021)}]{2021MNRAS.506.1128S}%
  \BibitemOpen
  \bibfield  {author} {\bibinfo {author} {\bibfnamefont {X.}~\bibnamefont
  {{Sun}}}\ and\ \bibinfo {author} {\bibfnamefont {X.-N.}\ \bibnamefont
  {{Bai}}},\ }\bibfield  {title} {\bibinfo {title} {{Particle diffusion and
  acceleration in magnetorotational instability turbulence}},\ }\href
  {https://doi.org/10.1093/mnras/stab1643} {\bibfield  {journal} {\bibinfo
  {journal} {Mon.\ Not.\ R.\ Astron.\ Soc.}\ }\textbf {\bibinfo {volume}
  {506}},\ \bibinfo {pages} {1128} (\bibinfo {year} {2021})}\BibitemShut
  {NoStop}%
\bibitem [{\citenamefont {{Pezzi}}\ \emph {et~al.}(2022)\citenamefont
  {{Pezzi}}, \citenamefont {{Blasi}},\ and\ \citenamefont
  {{Matthaeus}}}]{2022ApJ...928...25P}%
  \BibitemOpen
  \bibfield  {author} {\bibinfo {author} {\bibfnamefont {O.}~\bibnamefont
  {{Pezzi}}}, \bibinfo {author} {\bibfnamefont {P.}~\bibnamefont {{Blasi}}},\
  and\ \bibinfo {author} {\bibfnamefont {W.~H.}\ \bibnamefont {{Matthaeus}}},\
  }\bibfield  {title} {\bibinfo {title} {{Relativistic particle transport and
  acceleration in structured plasma turbulence}},\ }\href
  {https://doi.org/10.3847/1538-4357/ac5332} {\bibfield  {journal} {\bibinfo
  {journal} {Astrophys. J.}\ }\textbf {\bibinfo {volume} {928}},\ \bibinfo
  {eid} {25} (\bibinfo {year} {2022})}\BibitemShut {NoStop}%
\bibitem [{\citenamefont {{Vega}}\ \emph {et~al.}(2022)\citenamefont {{Vega}},
  \citenamefont {{Boldyrev}}, \citenamefont {{Roytershteyn}},\ and\
  \citenamefont {{Medvedev}}}]{2022ApJ...924L..19V}%
  \BibitemOpen
  \bibfield  {author} {\bibinfo {author} {\bibfnamefont {C.}~\bibnamefont
  {{Vega}}}, \bibinfo {author} {\bibfnamefont {S.}~\bibnamefont {{Boldyrev}}},
  \bibinfo {author} {\bibfnamefont {V.}~\bibnamefont {{Roytershteyn}}},\ and\
  \bibinfo {author} {\bibfnamefont {M.}~\bibnamefont {{Medvedev}}},\ }\bibfield
   {title} {\bibinfo {title} {{Turbulence and particle acceleration in a
  relativistic plasma}},\ }\href {https://doi.org/10.3847/2041-8213/ac441e}
  {\bibfield  {journal} {\bibinfo  {journal} {Astrophys. J.}\ }\textbf
  {\bibinfo {volume} {924}},\ \bibinfo {eid} {L19} (\bibinfo {year}
  {2022})}\BibitemShut {NoStop}%
\bibitem [{\citenamefont {{Pugliese}}\ \emph {et~al.}(2023)\citenamefont
  {{Pugliese}}, \citenamefont {{Brodiano}}, \citenamefont {{Andr{\'e}s}},\ and\
  \citenamefont {{Dmitruk}}}]{2023ApJ...959...28P}%
  \BibitemOpen
  \bibfield  {author} {\bibinfo {author} {\bibfnamefont {F.}~\bibnamefont
  {{Pugliese}}}, \bibinfo {author} {\bibfnamefont {M.}~\bibnamefont
  {{Brodiano}}}, \bibinfo {author} {\bibfnamefont {N.}~\bibnamefont
  {{Andr{\'e}s}}},\ and\ \bibinfo {author} {\bibfnamefont {P.}~\bibnamefont
  {{Dmitruk}}},\ }\bibfield  {title} {\bibinfo {title} {{Energization of
  charged test particles in magnetohydrodynamic fields: waves versus turbulence
  picture}},\ }\href {https://doi.org/10.3847/1538-4357/ad055b} {\bibfield
  {journal} {\bibinfo  {journal} {Astrophys. J.}\ }\textbf {\bibinfo {volume}
  {959}},\ \bibinfo {eid} {28} (\bibinfo {year} {2023})}\BibitemShut {NoStop}%
\bibitem [{\citenamefont {{Meringolo}}\ \emph {et~al.}(2023)\citenamefont
  {{Meringolo}}, \citenamefont {{Cruz-Osorio}}, \citenamefont {{Rezzolla}},\
  and\ \citenamefont {{Servidio}}}]{2023ApJ...944..122M}%
  \BibitemOpen
  \bibfield  {author} {\bibinfo {author} {\bibfnamefont {C.}~\bibnamefont
  {{Meringolo}}}, \bibinfo {author} {\bibfnamefont {A.}~\bibnamefont
  {{Cruz-Osorio}}}, \bibinfo {author} {\bibfnamefont {L.}~\bibnamefont
  {{Rezzolla}}},\ and\ \bibinfo {author} {\bibfnamefont {S.}~\bibnamefont
  {{Servidio}}},\ }\bibfield  {title} {\bibinfo {title} {{Microphysical plasma
  relations from special-relativistic turbulence}},\ }\href
  {https://doi.org/10.3847/1538-4357/acaefe} {\bibfield  {journal} {\bibinfo
  {journal} {Astrophys. J.}\ }\textbf {\bibinfo {volume} {944}},\ \bibinfo
  {eid} {122} (\bibinfo {year} {2023})}\BibitemShut {NoStop}%
\bibitem [{\citenamefont {{Zhdankin}}\ \emph {et~al.}(2018)\citenamefont
  {{Zhdankin}}, \citenamefont {{Uzdensky}}, \citenamefont {{Werner}},\ and\
  \citenamefont {{Begelman}}}]{2018ApJ...867L..18Z}%
  \BibitemOpen
  \bibfield  {author} {\bibinfo {author} {\bibfnamefont {V.}~\bibnamefont
  {{Zhdankin}}}, \bibinfo {author} {\bibfnamefont {D.~A.}\ \bibnamefont
  {{Uzdensky}}}, \bibinfo {author} {\bibfnamefont {G.~R.}\ \bibnamefont
  {{Werner}}},\ and\ \bibinfo {author} {\bibfnamefont {M.~C.}\ \bibnamefont
  {{Begelman}}},\ }\bibfield  {title} {\bibinfo {title} {{System-size
  convergence of nonthermal particle acceleration in relativistic plasma
  turbulence}},\ }\href {https://doi.org/10.3847/2041-8213/aae88c} {\bibfield
  {journal} {\bibinfo  {journal} {Astrophys. J.}\ }\textbf {\bibinfo {volume}
  {867}},\ \bibinfo {eid} {L18} (\bibinfo {year} {2018})}\BibitemShut {NoStop}%
\bibitem [{\citenamefont {{Comisso}}\ and\ \citenamefont
  {{Sironi}}(2018)}]{18Comisso}%
  \BibitemOpen
  \bibfield  {author} {\bibinfo {author} {\bibfnamefont {L.}~\bibnamefont
  {{Comisso}}}\ and\ \bibinfo {author} {\bibfnamefont {L.}~\bibnamefont
  {{Sironi}}},\ }\bibfield  {title} {\bibinfo {title} {{Particle acceleration
  in relativistic plasma turbulence}},\ }\href
  {https://doi.org/10.1103/PhysRevLett.121.255101} {\bibfield  {journal}
  {\bibinfo  {journal} {Phys.\ Rev.\ Lett.}\ }\textbf {\bibinfo {volume}
  {121}},\ \bibinfo {eid} {255101} (\bibinfo {year} {2018})}\BibitemShut
  {NoStop}%
\bibitem [{\citenamefont {{Comisso}}\ and\ \citenamefont
  {{Sironi}}(2019)}]{2019ApJ...886..122C}%
  \BibitemOpen
  \bibfield  {author} {\bibinfo {author} {\bibfnamefont {L.}~\bibnamefont
  {{Comisso}}}\ and\ \bibinfo {author} {\bibfnamefont {L.}~\bibnamefont
  {{Sironi}}},\ }\bibfield  {title} {\bibinfo {title} {{The interplay of
  magnetically dominated turbulence and magnetic reconnection in producing
  nonthermal particles}},\ }\href {https://doi.org/10.3847/1538-4357/ab4c33}
  {\bibfield  {journal} {\bibinfo  {journal} {Astrophys. J.}\ }\textbf
  {\bibinfo {volume} {886}},\ \bibinfo {eid} {122} (\bibinfo {year}
  {2019})}\BibitemShut {NoStop}%
\bibitem [{\citenamefont {{Zhdankin}}\ \emph {et~al.}(2019)\citenamefont
  {{Zhdankin}}, \citenamefont {{Uzdensky}}, \citenamefont {{Werner}},\ and\
  \citenamefont {{Begelman}}}]{2019PhRvL.122e5101Z}%
  \BibitemOpen
  \bibfield  {author} {\bibinfo {author} {\bibfnamefont {V.}~\bibnamefont
  {{Zhdankin}}}, \bibinfo {author} {\bibfnamefont {D.~A.}\ \bibnamefont
  {{Uzdensky}}}, \bibinfo {author} {\bibfnamefont {G.~R.}\ \bibnamefont
  {{Werner}}},\ and\ \bibinfo {author} {\bibfnamefont {M.~C.}\ \bibnamefont
  {{Begelman}}},\ }\bibfield  {title} {\bibinfo {title} {{Electron and ion
  energization in relativistic plasma turbulence}},\ }\href
  {https://doi.org/10.1103/PhysRevLett.122.055101} {\bibfield  {journal}
  {\bibinfo  {journal} {Phys.\ Rev.\ Lett.}\ }\textbf {\bibinfo {volume}
  {122}},\ \bibinfo {eid} {055101} (\bibinfo {year} {2019})}\BibitemShut
  {NoStop}%
\bibitem [{\citenamefont {{Wong}}\ \emph {et~al.}(2020)\citenamefont {{Wong}},
  \citenamefont {{Zhdankin}}, \citenamefont {{Uzdensky}}, \citenamefont
  {{Werner}},\ and\ \citenamefont {{Begelman}}}]{2020ApJ...893L...7W}%
  \BibitemOpen
  \bibfield  {author} {\bibinfo {author} {\bibfnamefont {K.}~\bibnamefont
  {{Wong}}}, \bibinfo {author} {\bibfnamefont {V.}~\bibnamefont {{Zhdankin}}},
  \bibinfo {author} {\bibfnamefont {D.~A.}\ \bibnamefont {{Uzdensky}}},
  \bibinfo {author} {\bibfnamefont {G.~R.}\ \bibnamefont {{Werner}}},\ and\
  \bibinfo {author} {\bibfnamefont {M.~C.}\ \bibnamefont {{Begelman}}},\
  }\bibfield  {title} {\bibinfo {title} {{First-principles demonstration of
  diffusive-advective particle acceleration in kinetic simulations of
  relativistic plasma turbulence}},\ }\href
  {https://doi.org/10.3847/2041-8213/ab8122} {\bibfield  {journal} {\bibinfo
  {journal} {Astrophys. J.}\ }\textbf {\bibinfo {volume} {893}},\ \bibinfo
  {eid} {L7} (\bibinfo {year} {2020})}\BibitemShut {NoStop}%
\bibitem [{\citenamefont {{Wong}}\ \emph {et~al.}(2025)\citenamefont {{Wong}},
  \citenamefont {{Zhdankin}}, \citenamefont {{Uzdensky}}, \citenamefont
  {{Werner}},\ and\ \citenamefont {{Begelman}}}]{2025MNRAS.543.1842W}%
  \BibitemOpen
  \bibfield  {author} {\bibinfo {author} {\bibfnamefont {K.~W.}\ \bibnamefont
  {{Wong}}}, \bibinfo {author} {\bibfnamefont {V.}~\bibnamefont {{Zhdankin}}},
  \bibinfo {author} {\bibfnamefont {D.~A.}\ \bibnamefont {{Uzdensky}}},
  \bibinfo {author} {\bibfnamefont {G.~R.}\ \bibnamefont {{Werner}}},\ and\
  \bibinfo {author} {\bibfnamefont {M.~C.}\ \bibnamefont {{Begelman}}},\
  }\bibfield  {title} {\bibinfo {title} {{Energy diffusion and advection
  coefficients in kinetic simulations of relativistic plasma turbulence}},\
  }\href {https://doi.org/10.1093/mnras/staf1589} {\bibfield  {journal}
  {\bibinfo  {journal} {Mon.\ Not.\ R.\ Astron.\ Soc.}\ }\textbf {\bibinfo
  {volume} {543}},\ \bibinfo {pages} {1842} (\bibinfo {year}
  {2025})}\BibitemShut {NoStop}%
\bibitem [{\citenamefont {{Zhdankin}}\ \emph {et~al.}(2020)\citenamefont
  {{Zhdankin}}, \citenamefont {{Uzdensky}}, \citenamefont {{Werner}},\ and\
  \citenamefont {{Begelman}}}]{2020MNRAS.493..603Z}%
  \BibitemOpen
  \bibfield  {author} {\bibinfo {author} {\bibfnamefont {V.}~\bibnamefont
  {{Zhdankin}}}, \bibinfo {author} {\bibfnamefont {D.~A.}\ \bibnamefont
  {{Uzdensky}}}, \bibinfo {author} {\bibfnamefont {G.~R.}\ \bibnamefont
  {{Werner}}},\ and\ \bibinfo {author} {\bibfnamefont {M.~C.}\ \bibnamefont
  {{Begelman}}},\ }\bibfield  {title} {\bibinfo {title} {{Kinetic turbulence in
  shining pair plasma: Intermittent beaming and thermalization by radiative
  cooling}},\ }\href {https://doi.org/10.1093/mnras/staa284} {\bibfield
  {journal} {\bibinfo  {journal} {Mon.\ Not.\ R.\ Astron.\ Soc.}\ }\textbf
  {\bibinfo {volume} {493}},\ \bibinfo {pages} {603} (\bibinfo {year}
  {2020})}\BibitemShut {NoStop}%
\bibitem [{\citenamefont {{Comisso}}\ \emph {et~al.}(2020)\citenamefont
  {{Comisso}}, \citenamefont {{Sobacchi}},\ and\ \citenamefont
  {{Sironi}}}]{2020ApJ...895L..40C}%
  \BibitemOpen
  \bibfield  {author} {\bibinfo {author} {\bibfnamefont {L.}~\bibnamefont
  {{Comisso}}}, \bibinfo {author} {\bibfnamefont {E.}~\bibnamefont
  {{Sobacchi}}},\ and\ \bibinfo {author} {\bibfnamefont {L.}~\bibnamefont
  {{Sironi}}},\ }\bibfield  {title} {\bibinfo {title} {{Hard Synchrotron
  Spectra from Magnetically Dominated Plasma Turbulence}},\ }\href
  {https://doi.org/10.3847/2041-8213/ab93dc} {\bibfield  {journal} {\bibinfo
  {journal} {Astrophys. J.}\ }\textbf {\bibinfo {volume} {895}},\ \bibinfo
  {eid} {L40} (\bibinfo {year} {2020})}\BibitemShut {NoStop}%
\bibitem [{\citenamefont {{Comisso}}\ and\ \citenamefont
  {{Sironi}}(2021)}]{2021PhRvL.127y5102C}%
  \BibitemOpen
  \bibfield  {author} {\bibinfo {author} {\bibfnamefont {L.}~\bibnamefont
  {{Comisso}}}\ and\ \bibinfo {author} {\bibfnamefont {L.}~\bibnamefont
  {{Sironi}}},\ }\bibfield  {title} {\bibinfo {title} {{Pitch-Angle Anisotropy
  Controls Particle Acceleration and Cooling in Radiative Relativistic Plasma
  Turbulence}},\ }\href {https://doi.org/10.1103/PhysRevLett.127.255102}
  {\bibfield  {journal} {\bibinfo  {journal} {Phys.\ Rev.\ Lett.}\ }\textbf
  {\bibinfo {volume} {127}},\ \bibinfo {eid} {255102} (\bibinfo {year}
  {2021})}\BibitemShut {NoStop}%
\bibitem [{\citenamefont {{N{\"a}ttil{\"a}}}\ and\ \citenamefont
  {{Beloborodov}}(2021)}]{2021ApJ...921...87N}%
  \BibitemOpen
  \bibfield  {author} {\bibinfo {author} {\bibfnamefont {J.}~\bibnamefont
  {{N{\"a}ttil{\"a}}}}\ and\ \bibinfo {author} {\bibfnamefont {A.~M.}\
  \bibnamefont {{Beloborodov}}},\ }\bibfield  {title} {\bibinfo {title}
  {{Radiative turbulent flares in magnetically dominated plasmas}},\ }\href
  {https://doi.org/10.3847/1538-4357/ac1c76} {\bibfield  {journal} {\bibinfo
  {journal} {Astrophys. J.}\ }\textbf {\bibinfo {volume} {921}},\ \bibinfo
  {eid} {87} (\bibinfo {year} {2021})}\BibitemShut {NoStop}%
\bibitem [{\citenamefont {{Gro{\v{s}}elj}}\ \emph {et~al.}(2024)\citenamefont
  {{Gro{\v{s}}elj}}, \citenamefont {{Hakobyan}}, \citenamefont {{Beloborodov}},
  \citenamefont {{Sironi}},\ and\ \citenamefont
  {{Philippov}}}]{2024PhRvL.132h5202G}%
  \BibitemOpen
  \bibfield  {author} {\bibinfo {author} {\bibfnamefont {D.}~\bibnamefont
  {{Gro{\v{s}}elj}}}, \bibinfo {author} {\bibfnamefont {H.}~\bibnamefont
  {{Hakobyan}}}, \bibinfo {author} {\bibfnamefont {A.~M.}\ \bibnamefont
  {{Beloborodov}}}, \bibinfo {author} {\bibfnamefont {L.}~\bibnamefont
  {{Sironi}}},\ and\ \bibinfo {author} {\bibfnamefont {A.}~\bibnamefont
  {{Philippov}}},\ }\bibfield  {title} {\bibinfo {title} {{Radiative
  particle-in-cell simulations of turbulent Comptonization in magnetized
  black-hole coronae}},\ }\href
  {https://doi.org/10.1103/PhysRevLett.132.085202} {\bibfield  {journal}
  {\bibinfo  {journal} {Phys.\ Rev.\ Lett.}\ }\textbf {\bibinfo {volume}
  {132}},\ \bibinfo {eid} {085202} (\bibinfo {year} {2024})}\BibitemShut
  {NoStop}%
\bibitem [{\citenamefont {{Comisso}}\ \emph {et~al.}(2024)\citenamefont
  {{Comisso}}, \citenamefont {{Farrar}},\ and\ \citenamefont
  {{Muzio}}}]{2024ApJ...977L..18C}%
  \BibitemOpen
  \bibfield  {author} {\bibinfo {author} {\bibfnamefont {L.}~\bibnamefont
  {{Comisso}}}, \bibinfo {author} {\bibfnamefont {G.~R.}\ \bibnamefont
  {{Farrar}}},\ and\ \bibinfo {author} {\bibfnamefont {M.~S.}\ \bibnamefont
  {{Muzio}}},\ }\bibfield  {title} {\bibinfo {title} {{Ultra-high-energy cosmic
  rays accelerated by magnetically dominated turbulence}},\ }\href
  {https://doi.org/10.3847/2041-8213/ad955f} {\bibfield  {journal} {\bibinfo
  {journal} {Astrophys. J.}\ }\textbf {\bibinfo {volume} {977}},\ \bibinfo
  {eid} {L18} (\bibinfo {year} {2024})}\BibitemShut {NoStop}%
\bibitem [{\citenamefont {{Gorbunov}}\ \emph {et~al.}(2025)\citenamefont
  {{Gorbunov}}, \citenamefont {{Gro{\v{s}}elj}},\ and\ \citenamefont
  {{Bacchini}}}]{2025arXiv250303820G}%
  \BibitemOpen
  \bibfield  {author} {\bibinfo {author} {\bibfnamefont {E.~A.}\ \bibnamefont
  {{Gorbunov}}}, \bibinfo {author} {\bibfnamefont {D.}~\bibnamefont
  {{Gro{\v{s}}elj}}},\ and\ \bibinfo {author} {\bibfnamefont {F.}~\bibnamefont
  {{Bacchini}}},\ }\bibfield  {title} {\bibinfo {title} {{Leaking outside the
  box: kinetic turbulence with cosmic-ray escape}},\ }\href
  {https://doi.org/10.48550/arXiv.2503.03820} {\bibfield  {journal} {\bibinfo
  {journal} {arXiv e-prints}\ ,\ \bibinfo {eid} {arXiv:2503.03820}} (\bibinfo
  {year} {2025})}\BibitemShut {NoStop}%
\bibitem [{\citenamefont {{Bacchini}}\ \emph {et~al.}(2024)\citenamefont
  {{Bacchini}}, \citenamefont {{Zhdankin}}, \citenamefont {{Gorbunov}},
  \citenamefont {{Werner}}, \citenamefont {{Arzamasskiy}}, \citenamefont
  {{Begelman}},\ and\ \citenamefont {{Uzdensky}}}]{2024PhRvL.133d5202B}%
  \BibitemOpen
  \bibfield  {author} {\bibinfo {author} {\bibfnamefont {F.}~\bibnamefont
  {{Bacchini}}}, \bibinfo {author} {\bibfnamefont {V.}~\bibnamefont
  {{Zhdankin}}}, \bibinfo {author} {\bibfnamefont {E.~A.}\ \bibnamefont
  {{Gorbunov}}}, \bibinfo {author} {\bibfnamefont {G.~R.}\ \bibnamefont
  {{Werner}}}, \bibinfo {author} {\bibfnamefont {L.}~\bibnamefont
  {{Arzamasskiy}}}, \bibinfo {author} {\bibfnamefont {M.~C.}\ \bibnamefont
  {{Begelman}}},\ and\ \bibinfo {author} {\bibfnamefont {D.~A.}\ \bibnamefont
  {{Uzdensky}}},\ }\bibfield  {title} {\bibinfo {title} {{Collisionless
  magnetorotational turbulence in pair plasmas: steady-state dynamics, particle
  acceleration, and radiative cooling}},\ }\href
  {https://doi.org/10.1103/PhysRevLett.133.045202} {\bibfield  {journal}
  {\bibinfo  {journal} {Phys.\ Rev.\ Lett.}\ }\textbf {\bibinfo {volume}
  {133}},\ \bibinfo {eid} {045202} (\bibinfo {year} {2024})}\BibitemShut
  {NoStop}%
\bibitem [{\citenamefont {{Amato}}(2019)}]{2019hepr.confE..33A}%
  \BibitemOpen
  \bibfield  {author} {\bibinfo {author} {\bibfnamefont {E.}~\bibnamefont
  {{Amato}}},\ }\bibfield  {title} {\bibinfo {title} {{The Theory Of Pulsar
  Wind Nebulae: Recent Progress}},\ }in\ \href
  {https://doi.org/10.22323/1.354.0033} {\emph {\bibinfo {booktitle} {High
  Energy Phenomena in Relativistic Outflows VII}}}\ (\bibinfo {year} {2019})\
  p.~\bibinfo {pages} {33},\ \Eprint {https://arxiv.org/abs/2001.04442}
  {arXiv:2001.04442 [astro-ph.HE]} \BibitemShut {NoStop}%
\bibitem [{\citenamefont {{Amato}}\ and\ \citenamefont
  {{Olmi}}(2021)}]{2021Univ....7..448A}%
  \BibitemOpen
  \bibfield  {author} {\bibinfo {author} {\bibfnamefont {E.}~\bibnamefont
  {{Amato}}}\ and\ \bibinfo {author} {\bibfnamefont {B.}~\bibnamefont
  {{Olmi}}},\ }\bibfield  {title} {\bibinfo {title} {{The Crab pulsar and
  nebula as seen in gamma-rays}},\ }\href
  {https://doi.org/10.3390/universe7110448} {\bibfield  {journal} {\bibinfo
  {journal} {Universe}\ }\textbf {\bibinfo {volume} {7}},\ \bibinfo {pages}
  {448} (\bibinfo {year} {2021})}\BibitemShut {NoStop}%
\bibitem [{\citenamefont {{Lemoine}}(2022)}]{2022PhRvL.129u5101L}%
  \BibitemOpen
  \bibfield  {author} {\bibinfo {author} {\bibfnamefont {M.}~\bibnamefont
  {{Lemoine}}},\ }\bibfield  {title} {\bibinfo {title} {{First-principles Fermi
  acceleration in magnetized turbulence}},\ }\href
  {https://doi.org/10.1103/PhysRevLett.129.215101} {\bibfield  {journal}
  {\bibinfo  {journal} {Phys.\ Rev.\ Lett.}\ }\textbf {\bibinfo {volume}
  {129}},\ \bibinfo {eid} {215101} (\bibinfo {year} {2022})}\BibitemShut
  {NoStop}%
\bibitem [{\citenamefont {{Lemoine}}\ and\ \citenamefont
  {{Malkov}}(2020)}]{2020MNRAS.499.4972L}%
  \BibitemOpen
  \bibfield  {author} {\bibinfo {author} {\bibfnamefont {M.}~\bibnamefont
  {{Lemoine}}}\ and\ \bibinfo {author} {\bibfnamefont {M.~A.}\ \bibnamefont
  {{Malkov}}},\ }\bibfield  {title} {\bibinfo {title} {{Power-law spectra from
  stochastic acceleration}},\ }\href {https://doi.org/10.1093/mnras/staa3131}
  {\bibfield  {journal} {\bibinfo  {journal} {Mon.\ Not.\ R.\ Astron.\ Soc.}\
  }\textbf {\bibinfo {volume} {499}},\ \bibinfo {pages} {4972} (\bibinfo {year}
  {2020})}\BibitemShut {NoStop}%
\bibitem [{\citenamefont {{Kumar}}\ \emph {et~al.}(2012)\citenamefont
  {{Kumar}}, \citenamefont {{Hern{\'a}ndez}}, \citenamefont {{Bo{\v{s}}njak}},\
  and\ \citenamefont {{Barniol Duran}}}]{2012MNRAS.427L..40K}%
  \BibitemOpen
  \bibfield  {author} {\bibinfo {author} {\bibfnamefont {P.}~\bibnamefont
  {{Kumar}}}, \bibinfo {author} {\bibfnamefont {R.~A.}\ \bibnamefont
  {{Hern{\'a}ndez}}}, \bibinfo {author} {\bibfnamefont {{\v{Z}}.}~\bibnamefont
  {{Bo{\v{s}}njak}}},\ and\ \bibinfo {author} {\bibfnamefont {R.}~\bibnamefont
  {{Barniol Duran}}},\ }\bibfield  {title} {\bibinfo {title} {{Maximum
  synchrotron frequency for shock-accelerated particles}},\ }\href
  {https://doi.org/10.1111/j.1745-3933.2012.01341.x} {\bibfield  {journal}
  {\bibinfo  {journal} {Mon.\ Not.\ R.\ Astron.\ Soc.}\ }\textbf {\bibinfo
  {volume} {427}},\ \bibinfo {pages} {L40} (\bibinfo {year}
  {2012})}\BibitemShut {NoStop}%
\bibitem [{\citenamefont {{Khangulyan}}\ \emph {et~al.}(2021)\citenamefont
  {{Khangulyan}}, \citenamefont {{Aharonian}}, \citenamefont {{Romoli}},\ and\
  \citenamefont {{Taylor}}}]{2021ApJ...914...76K}%
  \BibitemOpen
  \bibfield  {author} {\bibinfo {author} {\bibfnamefont {D.}~\bibnamefont
  {{Khangulyan}}}, \bibinfo {author} {\bibfnamefont {F.}~\bibnamefont
  {{Aharonian}}}, \bibinfo {author} {\bibfnamefont {C.}~\bibnamefont
  {{Romoli}}},\ and\ \bibinfo {author} {\bibfnamefont {A.}~\bibnamefont
  {{Taylor}}},\ }\bibfield  {title} {\bibinfo {title} {{Extension of the
  synchrotron radiation of electrons to very high energies in clumpy
  environments}},\ }\href {https://doi.org/10.3847/1538-4357/abfcbf} {\bibfield
   {journal} {\bibinfo  {journal} {Astrophys. J.}\ }\textbf {\bibinfo {volume}
  {914}},\ \bibinfo {eid} {76} (\bibinfo {year} {2021})}\BibitemShut {NoStop}%
\bibitem [{\citenamefont {{Lemoine}}(2021)}]{2021PhRvD.104f3020L}%
  \BibitemOpen
  \bibfield  {author} {\bibinfo {author} {\bibfnamefont {M.}~\bibnamefont
  {{Lemoine}}},\ }\bibfield  {title} {\bibinfo {title} {{Particle acceleration
  in strong MHD turbulence}},\ }\href
  {https://doi.org/10.1103/PhysRevD.104.063020} {\bibfield  {journal} {\bibinfo
   {journal} {Phys.\ Rev.\ D}\ }\textbf {\bibinfo {volume} {104}},\ \bibinfo
  {eid} {063020} (\bibinfo {year} {2021})}\BibitemShut {NoStop}%
\bibitem [{\citenamefont {{Schekochihin}}\ \emph {et~al.}(2004)\citenamefont
  {{Schekochihin}}, \citenamefont {{Cowley}}, \citenamefont {{Taylor}},
  \citenamefont {{Maron}},\ and\ \citenamefont
  {{McWilliams}}}]{2004ApJ...612..276S}%
  \BibitemOpen
  \bibfield  {author} {\bibinfo {author} {\bibfnamefont {A.~A.}\ \bibnamefont
  {{Schekochihin}}}, \bibinfo {author} {\bibfnamefont {S.~C.}\ \bibnamefont
  {{Cowley}}}, \bibinfo {author} {\bibfnamefont {S.~F.}\ \bibnamefont
  {{Taylor}}}, \bibinfo {author} {\bibfnamefont {J.~L.}\ \bibnamefont
  {{Maron}}},\ and\ \bibinfo {author} {\bibfnamefont {J.~C.}\ \bibnamefont
  {{McWilliams}}},\ }\bibfield  {title} {\bibinfo {title} {{Simulations of the
  small-scale turbulent dynamo}},\ }\href {https://doi.org/10.1086/422547}
  {\bibfield  {journal} {\bibinfo  {journal} {Astrophys. J.}\ }\textbf
  {\bibinfo {volume} {612}},\ \bibinfo {pages} {276} (\bibinfo {year}
  {2004})}\BibitemShut {NoStop}%
\bibitem [{\citenamefont {{Yang}}\ \emph {et~al.}(2019)\citenamefont {{Yang}},
  \citenamefont {{Wan}}, \citenamefont {{Matthaeus}}, \citenamefont {{Shi}},
  \citenamefont {{Parashar}}, \citenamefont {{Lu}},\ and\ \citenamefont
  {{Chen}}}]{2019PhPl...26g2306Y}%
  \BibitemOpen
  \bibfield  {author} {\bibinfo {author} {\bibfnamefont {Y.}~\bibnamefont
  {{Yang}}}, \bibinfo {author} {\bibfnamefont {M.}~\bibnamefont {{Wan}}},
  \bibinfo {author} {\bibfnamefont {W.~H.}\ \bibnamefont {{Matthaeus}}},
  \bibinfo {author} {\bibfnamefont {Y.}~\bibnamefont {{Shi}}}, \bibinfo
  {author} {\bibfnamefont {T.~N.}\ \bibnamefont {{Parashar}}}, \bibinfo
  {author} {\bibfnamefont {Q.}~\bibnamefont {{Lu}}},\ and\ \bibinfo {author}
  {\bibfnamefont {S.}~\bibnamefont {{Chen}}},\ }\bibfield  {title} {\bibinfo
  {title} {{Role of magnetic field curvature in magnetohydrodynamic
  turbulence}},\ }\href {https://doi.org/10.1063/1.5099360} {\bibfield
  {journal} {\bibinfo  {journal} {Phys.\ Plasmas}\ }\textbf {\bibinfo {volume}
  {26}},\ \bibinfo {eid} {072306} (\bibinfo {year} {2019})}\BibitemShut
  {NoStop}%
\bibitem [{\citenamefont {{Yuen}}\ and\ \citenamefont
  {{Lazarian}}(2020)}]{2020ApJ...898...66Y}%
  \BibitemOpen
  \bibfield  {author} {\bibinfo {author} {\bibfnamefont {K.~H.}\ \bibnamefont
  {{Yuen}}}\ and\ \bibinfo {author} {\bibfnamefont {A.}~\bibnamefont
  {{Lazarian}}},\ }\bibfield  {title} {\bibinfo {title} {{Curvature of magnetic
  field lines in compressible magnetized turbulence: Statistics, magnetization
  predictions, gradient curvature, modes, and self-gravitating media}},\ }\href
  {https://doi.org/10.3847/1538-4357/ab9360} {\bibfield  {journal} {\bibinfo
  {journal} {Astrophys. J.}\ }\textbf {\bibinfo {volume} {898}},\ \bibinfo
  {eid} {66} (\bibinfo {year} {2020})}\BibitemShut {NoStop}%
\bibitem [{\citenamefont {{Bandyopadhyay}}\ \emph {et~al.}(2020)\citenamefont
  {{Bandyopadhyay}}, \citenamefont {{Yang}}, \citenamefont {{Matthaeus}},
  \citenamefont {{Chasapis}}, \citenamefont {{Parashar}}, \citenamefont
  {{Russell}}, \citenamefont {{Strangeway}}, \citenamefont {{Torbert}},
  \citenamefont {{Giles}}, \citenamefont {{Gershman}}, \citenamefont
  {{Pollock}}, \citenamefont {{Moore}},\ and\ \citenamefont
  {{Burch}}}]{2020ApJ...893L..25B}%
  \BibitemOpen
  \bibfield  {author} {\bibinfo {author} {\bibfnamefont {R.}~\bibnamefont
  {{Bandyopadhyay}}}, \bibinfo {author} {\bibfnamefont {Y.}~\bibnamefont
  {{Yang}}}, \bibinfo {author} {\bibfnamefont {W.~H.}\ \bibnamefont
  {{Matthaeus}}}, \bibinfo {author} {\bibfnamefont {A.}~\bibnamefont
  {{Chasapis}}}, \bibinfo {author} {\bibfnamefont {T.~N.}\ \bibnamefont
  {{Parashar}}}, \bibinfo {author} {\bibfnamefont {C.~T.}\ \bibnamefont
  {{Russell}}}, \bibinfo {author} {\bibfnamefont {R.~J.}\ \bibnamefont
  {{Strangeway}}}, \bibinfo {author} {\bibfnamefont {R.~B.}\ \bibnamefont
  {{Torbert}}}, \bibinfo {author} {\bibfnamefont {B.~L.}\ \bibnamefont
  {{Giles}}}, \bibinfo {author} {\bibfnamefont {D.~J.}\ \bibnamefont
  {{Gershman}}}, \bibinfo {author} {\bibfnamefont {C.~J.}\ \bibnamefont
  {{Pollock}}}, \bibinfo {author} {\bibfnamefont {T.~E.}\ \bibnamefont
  {{Moore}}},\ and\ \bibinfo {author} {\bibfnamefont {J.~L.}\ \bibnamefont
  {{Burch}}},\ }\bibfield  {title} {\bibinfo {title} {{In situ measurement of
  curvature of magnetic field in turbulent space plasmas: a statistical
  study}},\ }\href {https://doi.org/10.3847/2041-8213/ab846e} {\bibfield
  {journal} {\bibinfo  {journal} {Astrophys. J.}\ }\textbf {\bibinfo {volume}
  {893}},\ \bibinfo {eid} {L25} (\bibinfo {year} {2020})}\BibitemShut {NoStop}%
\bibitem [{\citenamefont {{Huang}}\ \emph {et~al.}(2020)\citenamefont
  {{Huang}}, \citenamefont {{Zhang}}, \citenamefont {{Sahraoui}}, \citenamefont
  {{Yuan}}, \citenamefont {{Deng}}, \citenamefont {{Jiang}}, \citenamefont
  {{Xu}}, \citenamefont {{Wei}}, \citenamefont {{He}},\ and\ \citenamefont
  {{Zhang}}}]{2020ApJ...898L..18H}%
  \BibitemOpen
  \bibfield  {author} {\bibinfo {author} {\bibfnamefont {S.~Y.}\ \bibnamefont
  {{Huang}}}, \bibinfo {author} {\bibfnamefont {J.}~\bibnamefont {{Zhang}}},
  \bibinfo {author} {\bibfnamefont {F.}~\bibnamefont {{Sahraoui}}}, \bibinfo
  {author} {\bibfnamefont {Z.~G.}\ \bibnamefont {{Yuan}}}, \bibinfo {author}
  {\bibfnamefont {X.~H.}\ \bibnamefont {{Deng}}}, \bibinfo {author}
  {\bibfnamefont {K.}~\bibnamefont {{Jiang}}}, \bibinfo {author} {\bibfnamefont
  {S.~B.}\ \bibnamefont {{Xu}}}, \bibinfo {author} {\bibfnamefont {Y.~Y.}\
  \bibnamefont {{Wei}}}, \bibinfo {author} {\bibfnamefont {L.~H.}\ \bibnamefont
  {{He}}},\ and\ \bibinfo {author} {\bibfnamefont {Z.~H.}\ \bibnamefont
  {{Zhang}}},\ }\bibfield  {title} {\bibinfo {title} {{Observations of magnetic
  field line curvature and its role in the space plasma turbulence}},\ }\href
  {https://doi.org/10.3847/2041-8213/aba263} {\bibfield  {journal} {\bibinfo
  {journal} {Astrophys. J.}\ }\textbf {\bibinfo {volume} {898}},\ \bibinfo
  {eid} {L18} (\bibinfo {year} {2020})}\BibitemShut {NoStop}%
\bibitem [{\citenamefont {{Lemoine}}(2023)}]{2023JPlPh..89e1701L}%
  \BibitemOpen
  \bibfield  {author} {\bibinfo {author} {\bibfnamefont {M.}~\bibnamefont
  {{Lemoine}}},\ }\bibfield  {title} {\bibinfo {title} {{Particle transport
  through localized interactions with sharp magnetic field bends in MHD
  turbulence}},\ }\href {https://doi.org/10.1017/S0022377823000946} {\bibfield
  {journal} {\bibinfo  {journal} {J. Plasma Phys.}\ }\textbf {\bibinfo {volume}
  {89}},\ \bibinfo {eid} {175890501} (\bibinfo {year} {2023})}\BibitemShut
  {NoStop}%
\bibitem [{\citenamefont {{Kempski}}\ \emph {et~al.}(2023)\citenamefont
  {{Kempski}}, \citenamefont {{Fielding}}, \citenamefont {{Quataert}},
  \citenamefont {{Galishnikova}}, \citenamefont {{Kunz}}, \citenamefont
  {{Philippov}},\ and\ \citenamefont {{Ripperda}}}]{2023MNRAS.525.4985K}%
  \BibitemOpen
  \bibfield  {author} {\bibinfo {author} {\bibfnamefont {P.}~\bibnamefont
  {{Kempski}}}, \bibinfo {author} {\bibfnamefont {D.~B.}\ \bibnamefont
  {{Fielding}}}, \bibinfo {author} {\bibfnamefont {E.}~\bibnamefont
  {{Quataert}}}, \bibinfo {author} {\bibfnamefont {A.~K.}\ \bibnamefont
  {{Galishnikova}}}, \bibinfo {author} {\bibfnamefont {M.~W.}\ \bibnamefont
  {{Kunz}}}, \bibinfo {author} {\bibfnamefont {A.~A.}\ \bibnamefont
  {{Philippov}}},\ and\ \bibinfo {author} {\bibfnamefont {B.}~\bibnamefont
  {{Ripperda}}},\ }\bibfield  {title} {\bibinfo {title} {{Cosmic ray transport
  in large-amplitude turbulence with small-scale field reversals}},\ }\href
  {https://doi.org/10.1093/mnras/stad2609} {\bibfield  {journal} {\bibinfo
  {journal} {Mon.\ Not.\ R.\ Astron.\ Soc.}\ }\textbf {\bibinfo {volume}
  {525}},\ \bibinfo {pages} {4985} (\bibinfo {year} {2023})}\BibitemShut
  {NoStop}%
\bibitem [{\citenamefont {{Bresci}}\ \emph {et~al.}(2022)\citenamefont
  {{Bresci}}, \citenamefont {{Lemoine}}, \citenamefont {{Gremillet}},
  \citenamefont {{Comisso}}, \citenamefont {{Sironi}},\ and\ \citenamefont
  {{Demidem}}}]{2022PhRvD.106b3028B}%
  \BibitemOpen
  \bibfield  {author} {\bibinfo {author} {\bibfnamefont {V.}~\bibnamefont
  {{Bresci}}}, \bibinfo {author} {\bibfnamefont {M.}~\bibnamefont {{Lemoine}}},
  \bibinfo {author} {\bibfnamefont {L.}~\bibnamefont {{Gremillet}}}, \bibinfo
  {author} {\bibfnamefont {L.}~\bibnamefont {{Comisso}}}, \bibinfo {author}
  {\bibfnamefont {L.}~\bibnamefont {{Sironi}}},\ and\ \bibinfo {author}
  {\bibfnamefont {C.}~\bibnamefont {{Demidem}}},\ }\bibfield  {title} {\bibinfo
  {title} {{Nonresonant particle acceleration in strong turbulence: comparison
  to kinetic and MHD simulations}},\ }\href
  {https://doi.org/10.1103/PhysRevD.106.023028} {\bibfield  {journal} {\bibinfo
   {journal} {Phys.\ Rev.\ D}\ }\textbf {\bibinfo {volume} {106}},\ \bibinfo
  {eid} {023028} (\bibinfo {year} {2022})}\BibitemShut {NoStop}%
\bibitem [{\citenamefont {{Landau}}\ and\ \citenamefont
  {{Lifshitz}}(1960)}]{1960ecm..book.....L}%
  \BibitemOpen
  \bibfield  {author} {\bibinfo {author} {\bibfnamefont {L.~D.}\ \bibnamefont
  {{Landau}}}\ and\ \bibinfo {author} {\bibfnamefont {E.~M.}\ \bibnamefont
  {{Lifshitz}}},\ }\href@noop {} {\emph {\bibinfo {title} {{Electrodynamics of
  Continuous Media}}}}\ (\bibinfo  {publisher} {Pergamon Press},\ \bibinfo
  {address} {London},\ \bibinfo {year} {1960})\BibitemShut {NoStop}%
\bibitem [{\citenamefont {{Cerutti}}\ \emph {et~al.}(2012)\citenamefont
  {{Cerutti}}, \citenamefont {{Uzdensky}},\ and\ \citenamefont
  {{Begelman}}}]{2012ApJ...746..148C}%
  \BibitemOpen
  \bibfield  {author} {\bibinfo {author} {\bibfnamefont {B.}~\bibnamefont
  {{Cerutti}}}, \bibinfo {author} {\bibfnamefont {D.~A.}\ \bibnamefont
  {{Uzdensky}}},\ and\ \bibinfo {author} {\bibfnamefont {M.~C.}\ \bibnamefont
  {{Begelman}}},\ }\bibfield  {title} {\bibinfo {title} {{Extreme particle
  acceleration in magnetic reconnection layers: application to the gamma-ray
  flares in the Crab nebula}},\ }\href
  {https://doi.org/10.1088/0004-637X/746/2/148} {\bibfield  {journal} {\bibinfo
   {journal} {Astrophys. J.}\ }\textbf {\bibinfo {volume} {746}},\ \bibinfo
  {eid} {148} (\bibinfo {year} {2012})}\BibitemShut {NoStop}%
\bibitem [{\citenamefont {{Hillas}}(1984)}]{1984ARA&A..22..425H}%
  \BibitemOpen
  \bibfield  {author} {\bibinfo {author} {\bibfnamefont {A.~M.}\ \bibnamefont
  {{Hillas}}},\ }\bibfield  {title} {\bibinfo {title} {{The origin of
  ultra-high-energy cosmic rays}},\ }\href
  {https://doi.org/10.1146/annurev.aa.22.090184.002233} {\bibfield  {journal}
  {\bibinfo  {journal} {Ann.\ Rev.\ Astron.\ Astrophys.}\ }\textbf {\bibinfo
  {volume} {22}},\ \bibinfo {pages} {425} (\bibinfo {year} {1984})}\BibitemShut
  {NoStop}%
\bibitem [{\citenamefont {{Demidem}}\ \emph {et~al.}(2020)\citenamefont
  {{Demidem}}, \citenamefont {{Lemoine}},\ and\ \citenamefont
  {{Casse}}}]{2020PhRvD.102b3003D}%
  \BibitemOpen
  \bibfield  {author} {\bibinfo {author} {\bibfnamefont {C.}~\bibnamefont
  {{Demidem}}}, \bibinfo {author} {\bibfnamefont {M.}~\bibnamefont
  {{Lemoine}}},\ and\ \bibinfo {author} {\bibfnamefont {F.}~\bibnamefont
  {{Casse}}},\ }\bibfield  {title} {\bibinfo {title} {{Particle acceleration in
  relativistic turbulence: A theoretical appraisal}},\ }\href
  {https://doi.org/10.1103/PhysRevD.102.023003} {\bibfield  {journal} {\bibinfo
   {journal} {Phys.\ Rev.\ D}\ }\textbf {\bibinfo {volume} {102}},\ \bibinfo
  {eid} {023003} (\bibinfo {year} {2020})}\BibitemShut {NoStop}%
\bibitem [{\citenamefont {{Lemoine}}\ \emph {et~al.}(2024)\citenamefont
  {{Lemoine}}, \citenamefont {{Murase}},\ and\ \citenamefont
  {{Rieger}}}]{2024PhRvD.109f3006L}%
  \BibitemOpen
  \bibfield  {author} {\bibinfo {author} {\bibfnamefont {M.}~\bibnamefont
  {{Lemoine}}}, \bibinfo {author} {\bibfnamefont {K.}~\bibnamefont
  {{Murase}}},\ and\ \bibinfo {author} {\bibfnamefont {F.}~\bibnamefont
  {{Rieger}}},\ }\bibfield  {title} {\bibinfo {title} {{Nonlinear aspects of
  stochastic particle acceleration}},\ }\href
  {https://doi.org/10.1103/PhysRevD.109.063006} {\bibfield  {journal} {\bibinfo
   {journal} {Phys.\ Rev.\ D}\ }\textbf {\bibinfo {volume} {109}},\ \bibinfo
  {eid} {063006} (\bibinfo {year} {2024})}\BibitemShut {NoStop}%
\bibitem [{\citenamefont {{Malkov}}\ and\ \citenamefont
  {{Diamond}}(2006)}]{2006ApJ...642..244M}%
  \BibitemOpen
  \bibfield  {author} {\bibinfo {author} {\bibfnamefont {M.~A.}\ \bibnamefont
  {{Malkov}}}\ and\ \bibinfo {author} {\bibfnamefont {P.~H.}\ \bibnamefont
  {{Diamond}}},\ }\bibfield  {title} {\bibinfo {title} {{Nonlinear shock
  acceleration beyond the Bohm limit}},\ }\href
  {https://doi.org/10.1086/500445} {\bibfield  {journal} {\bibinfo  {journal}
  {Astrophys. J.}\ }\textbf {\bibinfo {volume} {642}},\ \bibinfo {pages} {244}
  (\bibinfo {year} {2006})}\BibitemShut {NoStop}%
\bibitem [{\citenamefont {{Lefebvre}}\ \emph {et~al.}(2003)\citenamefont
  {{Lefebvre}}, \citenamefont {{Cochet}}, \citenamefont {{Fritzler}},
  \citenamefont {{Malka}}, \citenamefont {{Al{\'e}onard}}, \citenamefont
  {{Chemin}}, \citenamefont {{Darbon}}, \citenamefont {{Disdier}},
  \citenamefont {{Faure}}, \citenamefont {{Fedotoff}}, \citenamefont
  {{Landoas}}, \citenamefont {{Malka}}, \citenamefont {{M{\'e}ot}},
  \citenamefont {{Morel}}, \citenamefont {{Rabec LeGloahec}}, \citenamefont
  {{Rouyer}}, \citenamefont {{Rubbelynck}}, \citenamefont {{Tikhonchuk}},
  \citenamefont {{Wrobel}}, \citenamefont {{Audebert}},\ and\ \citenamefont
  {{Rousseaux}}}]{Lefebvre_2003}%
  \BibitemOpen
  \bibfield  {author} {\bibinfo {author} {\bibfnamefont {E.}~\bibnamefont
  {{Lefebvre}}}, \bibinfo {author} {\bibfnamefont {N.}~\bibnamefont
  {{Cochet}}}, \bibinfo {author} {\bibfnamefont {S.}~\bibnamefont
  {{Fritzler}}}, \bibinfo {author} {\bibfnamefont {V.}~\bibnamefont {{Malka}}},
  \bibinfo {author} {\bibfnamefont {M.-M.}\ \bibnamefont {{Al{\'e}onard}}},
  \bibinfo {author} {\bibfnamefont {J.-F.}\ \bibnamefont {{Chemin}}}, \bibinfo
  {author} {\bibfnamefont {S.}~\bibnamefont {{Darbon}}}, \bibinfo {author}
  {\bibfnamefont {L.}~\bibnamefont {{Disdier}}}, \bibinfo {author}
  {\bibfnamefont {J.}~\bibnamefont {{Faure}}}, \bibinfo {author} {\bibfnamefont
  {A.}~\bibnamefont {{Fedotoff}}}, \bibinfo {author} {\bibfnamefont
  {O.}~\bibnamefont {{Landoas}}}, \bibinfo {author} {\bibfnamefont
  {G.}~\bibnamefont {{Malka}}}, \bibinfo {author} {\bibfnamefont
  {V.}~\bibnamefont {{M{\'e}ot}}}, \bibinfo {author} {\bibfnamefont
  {P.}~\bibnamefont {{Morel}}}, \bibinfo {author} {\bibfnamefont
  {M.}~\bibnamefont {{Rabec LeGloahec}}}, \bibinfo {author} {\bibfnamefont
  {A.}~\bibnamefont {{Rouyer}}}, \bibinfo {author} {\bibfnamefont
  {C.}~\bibnamefont {{Rubbelynck}}}, \bibinfo {author} {\bibfnamefont
  {V.}~\bibnamefont {{Tikhonchuk}}}, \bibinfo {author} {\bibfnamefont
  {R.}~\bibnamefont {{Wrobel}}}, \bibinfo {author} {\bibfnamefont
  {P.}~\bibnamefont {{Audebert}}},\ and\ \bibinfo {author} {\bibfnamefont
  {C.}~\bibnamefont {{Rousseaux}}},\ }\bibfield  {title} {\bibinfo {title}
  {{Electron and photon production from relativistic laser plasma
  interactions}},\ }\href {https://doi.org/10.1088/0029-5515/43/7/317}
  {\bibfield  {journal} {\bibinfo  {journal} {Nucl. Fusion}\ }\textbf {\bibinfo
  {volume} {43}},\ \bibinfo {pages} {629} (\bibinfo {year} {2003})}\BibitemShut
  {NoStop}%
\bibitem [{\citenamefont {{Bresci}}\ \emph {et~al.}(2023)\citenamefont
  {{Bresci}}, \citenamefont {{Lemoine}},\ and\ \citenamefont
  {{Gremillet}}}]{2023PhRvR...5b3194B}%
  \BibitemOpen
  \bibfield  {author} {\bibinfo {author} {\bibfnamefont {V.}~\bibnamefont
  {{Bresci}}}, \bibinfo {author} {\bibfnamefont {M.}~\bibnamefont
  {{Lemoine}}},\ and\ \bibinfo {author} {\bibfnamefont {L.}~\bibnamefont
  {{Gremillet}}},\ }\bibfield  {title} {\bibinfo {title} {{Particle
  acceleration at magnetized, relativistic, turbulent shock fronts}},\ }\href
  {https://doi.org/10.1103/PhysRevResearch.5.023194} {\bibfield  {journal}
  {\bibinfo  {journal} {Phys. Rev. Res.}\ }\textbf {\bibinfo {volume} {5}},\
  \bibinfo {eid} {023194} (\bibinfo {year} {2023})}\BibitemShut {NoStop}%
\bibitem [{\citenamefont {TenBarge}\ \emph {et~al.}(2014)\citenamefont
  {TenBarge}, \citenamefont {Howes}, \citenamefont {Dorland},\ and\
  \citenamefont {Hammett}}]{TenBarge_2014}%
  \BibitemOpen
  \bibfield  {author} {\bibinfo {author} {\bibfnamefont {J.}~\bibnamefont
  {TenBarge}}, \bibinfo {author} {\bibfnamefont {G.}~\bibnamefont {Howes}},
  \bibinfo {author} {\bibfnamefont {W.}~\bibnamefont {Dorland}},\ and\ \bibinfo
  {author} {\bibfnamefont {G.}~\bibnamefont {Hammett}},\ }\bibfield  {title}
  {\bibinfo {title} {{An oscillating Langevin antenna for driving plasma
  turbulence simulations}},\ }\href {https://doi.org/10.1016/j.cpc.2013.10.022}
  {\bibfield  {journal} {\bibinfo  {journal} {Comp. Phys. Commun.}\ }\textbf
  {\bibinfo {volume} {185}},\ \bibinfo {pages} {578} (\bibinfo {year}
  {2014})}\BibitemShut {NoStop}%
\bibitem [{\citenamefont {{Guilbert}}\ \emph {et~al.}(1983)\citenamefont
  {{Guilbert}}, \citenamefont {{Fabian}},\ and\ \citenamefont
  {{Rees}}}]{Guilbert_1983}%
  \BibitemOpen
  \bibfield  {author} {\bibinfo {author} {\bibfnamefont {P.~W.}\ \bibnamefont
  {{Guilbert}}}, \bibinfo {author} {\bibfnamefont {A.~C.}\ \bibnamefont
  {{Fabian}}},\ and\ \bibinfo {author} {\bibfnamefont {M.~J.}\ \bibnamefont
  {{Rees}}},\ }\bibfield  {title} {\bibinfo {title} {{Spectral and variability
  constraints on compact sources}},\ }\href
  {https://doi.org/10.1093/mnras/205.3.593} {\bibfield  {journal} {\bibinfo
  {journal} {Mon.\ Not.\ Roy.\ Astron.\ Soc.}\ }\textbf {\bibinfo {volume}
  {205}},\ \bibinfo {pages} {593} (\bibinfo {year} {1983})}\BibitemShut
  {NoStop}%
\bibitem [{\citenamefont {{de Jager}}\ \emph {et~al.}(1996)\citenamefont {{de
  Jager}}, \citenamefont {{Harding}}, \citenamefont {{Michelson}},
  \citenamefont {{Nel}}, \citenamefont {{Nolan}}, \citenamefont {{Sreekumar}},\
  and\ \citenamefont {{Thompson}}}]{1996ApJ...457..253D}%
  \BibitemOpen
  \bibfield  {author} {\bibinfo {author} {\bibfnamefont {O.~C.}\ \bibnamefont
  {{de Jager}}}, \bibinfo {author} {\bibfnamefont {A.~K.}\ \bibnamefont
  {{Harding}}}, \bibinfo {author} {\bibfnamefont {P.~F.}\ \bibnamefont
  {{Michelson}}}, \bibinfo {author} {\bibfnamefont {H.~I.}\ \bibnamefont
  {{Nel}}}, \bibinfo {author} {\bibfnamefont {P.~L.}\ \bibnamefont {{Nolan}}},
  \bibinfo {author} {\bibfnamefont {P.}~\bibnamefont {{Sreekumar}}},\ and\
  \bibinfo {author} {\bibfnamefont {D.~J.}\ \bibnamefont {{Thompson}}},\
  }\bibfield  {title} {\bibinfo {title} {{Gamma-Ray observations of the Crab
  nebula: a study of the synchro-Compton spectrum}},\ }\href
  {https://doi.org/10.1086/176726} {\bibfield  {journal} {\bibinfo  {journal}
  {Astrophys. J.}\ }\textbf {\bibinfo {volume} {457}},\ \bibinfo {pages} {253}
  (\bibinfo {year} {1996})}\BibitemShut {NoStop}%
\bibitem [{\citenamefont {{Uzdensky}}\ \emph {et~al.}(2011)\citenamefont
  {{Uzdensky}}, \citenamefont {{Cerutti}},\ and\ \citenamefont
  {{Begelman}}}]{2011ApJ...737L..40U}%
  \BibitemOpen
  \bibfield  {author} {\bibinfo {author} {\bibfnamefont {D.~A.}\ \bibnamefont
  {{Uzdensky}}}, \bibinfo {author} {\bibfnamefont {B.}~\bibnamefont
  {{Cerutti}}},\ and\ \bibinfo {author} {\bibfnamefont {M.~C.}\ \bibnamefont
  {{Begelman}}},\ }\bibfield  {title} {\bibinfo {title} {{Reconnection-powered
  linear accelerator and gamma-ray flares in the Crab nebula}},\ }\href
  {https://doi.org/10.1088/2041-8205/737/2/L40} {\bibfield  {journal} {\bibinfo
   {journal} {Astrophys. J.}\ }\textbf {\bibinfo {volume} {737}},\ \bibinfo
  {eid} {L40} (\bibinfo {year} {2011})}\BibitemShut {NoStop}%
\bibitem [{\citenamefont {{Rybicki}}\ and\ \citenamefont
  {{Lightman}}(1985)}]{1986rpa..book.....R}%
  \BibitemOpen
  \bibfield  {author} {\bibinfo {author} {\bibfnamefont {G.~B.}\ \bibnamefont
  {{Rybicki}}}\ and\ \bibinfo {author} {\bibfnamefont {A.~P.}\ \bibnamefont
  {{Lightman}}},\ }\href
  {https://doi.org/https://doi.org/10.1002/9783527618170.fmatter} {\emph
  {\bibinfo {title} {{Radiative Processes in Astrophysics}}}}\ (\bibinfo
  {publisher} {John Wiley and Sons, Ltd},\ \bibinfo {year} {1985})\BibitemShut
  {NoStop}%
\bibitem [{\citenamefont {{Lemoine}}(2019)}]{2019PhRvD..99h3006L}%
  \BibitemOpen
  \bibfield  {author} {\bibinfo {author} {\bibfnamefont {M.}~\bibnamefont
  {{Lemoine}}},\ }\bibfield  {title} {\bibinfo {title} {{Generalized Fermi
  acceleration}},\ }\href {https://doi.org/10.1103/PhysRevD.99.083006}
  {\bibfield  {journal} {\bibinfo  {journal} {Phys.\ Rev.\ D}\ }\textbf
  {\bibinfo {volume} {99}},\ \bibinfo {eid} {083006} (\bibinfo {year}
  {2019})}\BibitemShut {NoStop}%
\bibitem [{\citenamefont {{Lemoine}}(2025)}]{2025PhRvE.112a5205L}%
  \BibitemOpen
  \bibfield  {author} {\bibinfo {author} {\bibfnamefont {M.}~\bibnamefont
  {{Lemoine}}},\ }\bibfield  {title} {\bibinfo {title} {{Effective theory for
  stochastic particle acceleration, with application to magnetized
  turbulence}},\ }\href {https://doi.org/10.1103/3xxg-x5dg} {\bibfield
  {journal} {\bibinfo  {journal} {\pre}\ }\textbf {\bibinfo {volume} {112}},\
  \bibinfo {eid} {015205} (\bibinfo {year} {2025})}\BibitemShut {NoStop}%
\bibitem [{\citenamefont {{Atoyan}}\ and\ \citenamefont
  {{Aharonian}}(1996)}]{1996MNRAS.278..525A}%
  \BibitemOpen
  \bibfield  {author} {\bibinfo {author} {\bibfnamefont {A.~M.}\ \bibnamefont
  {{Atoyan}}}\ and\ \bibinfo {author} {\bibfnamefont {F.~A.}\ \bibnamefont
  {{Aharonian}}},\ }\bibfield  {title} {\bibinfo {title} {{On the mechanisms of
  gamma radiation in the Crab Nebula}},\ }\href
  {https://doi.org/10.1093/mnras/278.2.525} {\bibfield  {journal} {\bibinfo
  {journal} {Mon.\ Not.\ R.\ Astron.\ Soc.}\ }\textbf {\bibinfo {volume}
  {278}},\ \bibinfo {pages} {525} (\bibinfo {year} {1996})}\BibitemShut
  {NoStop}%
\bibitem [{\citenamefont {{Kirk}}\ \emph {et~al.}(2009)\citenamefont {{Kirk}},
  \citenamefont {{Lyubarsky}},\ and\ \citenamefont
  {{Petri}}}]{2009ASSL..357..421K}%
  \BibitemOpen
  \bibfield  {author} {\bibinfo {author} {\bibfnamefont {J.~G.}\ \bibnamefont
  {{Kirk}}}, \bibinfo {author} {\bibfnamefont {Y.}~\bibnamefont
  {{Lyubarsky}}},\ and\ \bibinfo {author} {\bibfnamefont {J.}~\bibnamefont
  {{Petri}}},\ }\bibfield  {title} {\bibinfo {title} {{The theory of pulsar
  winds and nebulae}},\ }in\ \href
  {https://doi.org/10.1007/978-3-540-76965-1_16} {\emph {\bibinfo {booktitle}
  {Astr. Sp. Sc. Lib.}}},\ \bibinfo {series} {Astrophysics and Space Science
  Library}, Vol.\ \bibinfo {volume} {357},\ \bibinfo {editor} {edited by\
  \bibinfo {editor} {\bibfnamefont {W.}~\bibnamefont {{Becker}}}}\ (\bibinfo
  {year} {2009})\ p.\ \bibinfo {pages} {421},\ \Eprint
  {https://arxiv.org/abs/astro-ph/0703116} {arXiv:astro-ph/0703116 [astro-ph]}
  \BibitemShut {NoStop}%
\bibitem [{\citenamefont {{Dirson}}\ and\ \citenamefont
  {{Horns}}(2023)}]{2023A&A...671A..67D}%
  \BibitemOpen
  \bibfield  {author} {\bibinfo {author} {\bibfnamefont {L.}~\bibnamefont
  {{Dirson}}}\ and\ \bibinfo {author} {\bibfnamefont {D.}~\bibnamefont
  {{Horns}}},\ }\bibfield  {title} {\bibinfo {title} {{Phenomenological
  modelling of the Crab Nebula's broadband energy spectrum and its apparent
  extension}},\ }\href {https://doi.org/10.1051/0004-6361/202243578} {\bibfield
   {journal} {\bibinfo  {journal} {Astron.\ Astrophys.}\ }\textbf {\bibinfo
  {volume} {671}},\ \bibinfo {eid} {A67} (\bibinfo {year} {2023})}\BibitemShut
  {NoStop}%
\bibitem [{\citenamefont {{B{\"u}hler}}\ and\ \citenamefont
  {{Blandford}}(2014)}]{2014RPPh...77f6901B}%
  \BibitemOpen
  \bibfield  {author} {\bibinfo {author} {\bibfnamefont {R.}~\bibnamefont
  {{B{\"u}hler}}}\ and\ \bibinfo {author} {\bibfnamefont {R.}~\bibnamefont
  {{Blandford}}},\ }\bibfield  {title} {\bibinfo {title} {{The surprising Crab
  pulsar and its nebula: a review}},\ }\href
  {https://doi.org/10.1088/0034-4885/77/6/066901} {\bibfield  {journal}
  {\bibinfo  {journal} {Rep. Prog. Phys.}\ }\textbf {\bibinfo {volume} {77}},\
  \bibinfo {eid} {066901} (\bibinfo {year} {2014})}\BibitemShut {NoStop}%
\bibitem [{\citenamefont {{Lemoine}}(2016)}]{2016JPlPh..82d6301L}%
  \BibitemOpen
  \bibfield  {author} {\bibinfo {author} {\bibfnamefont {M.}~\bibnamefont
  {{Lemoine}}},\ }\bibfield  {title} {\bibinfo {title} {{A corrugated
  termination shock in pulsar wind nebulae?}},\ }\href
  {https://doi.org/10.1017/S0022377816000659} {\bibfield  {journal} {\bibinfo
  {journal} {J.\ Plasma \ Phys.}\ }\textbf {\bibinfo {volume} {82}},\ \bibinfo
  {eid} {635820401} (\bibinfo {year} {2016})}\BibitemShut {NoStop}%
\bibitem [{\citenamefont {{Tanaka}}\ and\ \citenamefont
  {{Asano}}(2017)}]{2017ApJ...841...78T}%
  \BibitemOpen
  \bibfield  {author} {\bibinfo {author} {\bibfnamefont {S.~J.}\ \bibnamefont
  {{Tanaka}}}\ and\ \bibinfo {author} {\bibfnamefont {K.}~\bibnamefont
  {{Asano}}},\ }\bibfield  {title} {\bibinfo {title} {{On the radio-emitting
  particles of the Crab nebula: stochastic acceleration model}},\ }\href
  {https://doi.org/10.3847/1538-4357/aa6f13} {\bibfield  {journal} {\bibinfo
  {journal} {Astrophys. J.}\ }\textbf {\bibinfo {volume} {841}},\ \bibinfo
  {eid} {78} (\bibinfo {year} {2017})}\BibitemShut {NoStop}%
\bibitem [{\citenamefont {{Lyutikov}}\ \emph {et~al.}(2019)\citenamefont
  {{Lyutikov}}, \citenamefont {{Temim}}, \citenamefont {{Komissarov}},
  \citenamefont {{Slane}}, \citenamefont {{Sironi}},\ and\ \citenamefont
  {{Comisso}}}]{2019MNRAS.489.2403L}%
  \BibitemOpen
  \bibfield  {author} {\bibinfo {author} {\bibfnamefont {M.}~\bibnamefont
  {{Lyutikov}}}, \bibinfo {author} {\bibfnamefont {T.}~\bibnamefont {{Temim}}},
  \bibinfo {author} {\bibfnamefont {S.}~\bibnamefont {{Komissarov}}}, \bibinfo
  {author} {\bibfnamefont {P.}~\bibnamefont {{Slane}}}, \bibinfo {author}
  {\bibfnamefont {L.}~\bibnamefont {{Sironi}}},\ and\ \bibinfo {author}
  {\bibfnamefont {L.}~\bibnamefont {{Comisso}}},\ }\bibfield  {title} {\bibinfo
  {title} {{Interpreting Crab nebula's synchrotron spectrum: two acceleration
  mechanisms}},\ }\href {https://doi.org/10.1093/mnras/stz2023} {\bibfield
  {journal} {\bibinfo  {journal} {Mon.\ Not.\ R.\ Astron.\ Soc.}\ }\textbf
  {\bibinfo {volume} {489}},\ \bibinfo {pages} {2403} (\bibinfo {year}
  {2019})}\BibitemShut {NoStop}%
\bibitem [{\citenamefont {{Luo}}\ \emph {et~al.}(2020)\citenamefont {{Luo}},
  \citenamefont {{Lyutikov}}, \citenamefont {{Temim}},\ and\ \citenamefont
  {{Comisso}}}]{2020ApJ...896..147L}%
  \BibitemOpen
  \bibfield  {author} {\bibinfo {author} {\bibfnamefont {Y.}~\bibnamefont
  {{Luo}}}, \bibinfo {author} {\bibfnamefont {M.}~\bibnamefont {{Lyutikov}}},
  \bibinfo {author} {\bibfnamefont {T.}~\bibnamefont {{Temim}}},\ and\ \bibinfo
  {author} {\bibfnamefont {L.}~\bibnamefont {{Comisso}}},\ }\bibfield  {title}
  {\bibinfo {title} {{Turbulent model of Crab nebula radiation}},\ }\href
  {https://doi.org/10.3847/1538-4357/ab93c0} {\bibfield  {journal} {\bibinfo
  {journal} {Astrophys. J.}\ }\textbf {\bibinfo {volume} {896}},\ \bibinfo
  {eid} {147} (\bibinfo {year} {2020})}\BibitemShut {NoStop}%
\bibitem [{\citenamefont {{Lu}}\ \emph {et~al.}(2023)\citenamefont {{Lu}},
  \citenamefont {{Zhu}}, \citenamefont {{Hu}},\ and\ \citenamefont
  {{Zhang}}}]{2023ApJ...953..116L}%
  \BibitemOpen
  \bibfield  {author} {\bibinfo {author} {\bibfnamefont {F.-W.}\ \bibnamefont
  {{Lu}}}, \bibinfo {author} {\bibfnamefont {B.-T.}\ \bibnamefont {{Zhu}}},
  \bibinfo {author} {\bibfnamefont {W.}~\bibnamefont {{Hu}}},\ and\ \bibinfo
  {author} {\bibfnamefont {L.}~\bibnamefont {{Zhang}}},\ }\bibfield  {title}
  {\bibinfo {title} {{Turbulent diffusion of the particles within pulsar wind
  nebulae}},\ }\href {https://doi.org/10.3847/1538-4357/ace0c2} {\bibfield
  {journal} {\bibinfo  {journal} {Astrophys. J.}\ }\textbf {\bibinfo {volume}
  {953}},\ \bibinfo {eid} {116} (\bibinfo {year} {2023})}\BibitemShut {NoStop}%
\bibitem [{\citenamefont {{Tanaka}}\ and\ \citenamefont
  {{Ishizaki}}(2024)}]{2024PTEP.2024e3E03T}%
  \BibitemOpen
  \bibfield  {author} {\bibinfo {author} {\bibfnamefont {S.~J.}\ \bibnamefont
  {{Tanaka}}}\ and\ \bibinfo {author} {\bibfnamefont {W.}~\bibnamefont
  {{Ishizaki}}},\ }\bibfield  {title} {\bibinfo {title} {{A Self-regulated
  stochastic acceleration model of pulsar wind nebulae}},\ }\href
  {https://doi.org/10.1093/ptep/ptae069} {\bibfield  {journal} {\bibinfo
  {journal} {Prog. Th. Exp. Phys.}\ }\textbf {\bibinfo {volume} {2024}},\
  \bibinfo {eid} {053E03} (\bibinfo {year} {2024})}\BibitemShut {NoStop}%
\bibitem [{\citenamefont {{Werner}}\ \emph {et~al.}(2016)\citenamefont
  {{Werner}}, \citenamefont {{Uzdensky}}, \citenamefont {{Cerutti}},
  \citenamefont {{Nalewajko}},\ and\ \citenamefont
  {{Begelman}}}]{2016ApJ...816L...8W}%
  \BibitemOpen
  \bibfield  {author} {\bibinfo {author} {\bibfnamefont {G.~R.}\ \bibnamefont
  {{Werner}}}, \bibinfo {author} {\bibfnamefont {D.~A.}\ \bibnamefont
  {{Uzdensky}}}, \bibinfo {author} {\bibfnamefont {B.}~\bibnamefont
  {{Cerutti}}}, \bibinfo {author} {\bibfnamefont {K.}~\bibnamefont
  {{Nalewajko}}},\ and\ \bibinfo {author} {\bibfnamefont {M.~C.}\ \bibnamefont
  {{Begelman}}},\ }\bibfield  {title} {\bibinfo {title} {{The extent of
  power-law energy spectra in collisionless relativistic magnetic reconnection
  in pair plasmas}},\ }\href {https://doi.org/10.3847/2041-8205/816/1/L8}
  {\bibfield  {journal} {\bibinfo  {journal} {Astrophys. J.}\ }\textbf
  {\bibinfo {volume} {816}},\ \bibinfo {eid} {L8} (\bibinfo {year}
  {2016})}\BibitemShut {NoStop}%
\bibitem [{\citenamefont {{Bykov}}\ \emph {et~al.}(2012)\citenamefont
  {{Bykov}}, \citenamefont {{Pavlov}}, \citenamefont {{Artemyev}},\ and\
  \citenamefont {{Uvarov}}}]{2012MNRAS.421L..67B}%
  \BibitemOpen
  \bibfield  {author} {\bibinfo {author} {\bibfnamefont {A.~M.}\ \bibnamefont
  {{Bykov}}}, \bibinfo {author} {\bibfnamefont {G.~G.}\ \bibnamefont
  {{Pavlov}}}, \bibinfo {author} {\bibfnamefont {A.~V.}\ \bibnamefont
  {{Artemyev}}},\ and\ \bibinfo {author} {\bibfnamefont {Y.~A.}\ \bibnamefont
  {{Uvarov}}},\ }\bibfield  {title} {\bibinfo {title} {{Twinkling pulsar wind
  nebulae in the synchrotron cut-off regime and the {\ensuremath{\gamma}}-ray
  flares in the Crab Nebula}},\ }\href
  {https://doi.org/10.1111/j.1745-3933.2011.01208.x} {\bibfield  {journal}
  {\bibinfo  {journal} {Mon.\ Not.\ R.\ Astron.\ Soc.}\ }\textbf {\bibinfo
  {volume} {421}},\ \bibinfo {pages} {L67} (\bibinfo {year}
  {2012})}\BibitemShut {NoStop}%
\bibitem [{\citenamefont {{Camus}}\ \emph {et~al.}(2009)\citenamefont
  {{Camus}}, \citenamefont {{Komissarov}}, \citenamefont {{Bucciantini}},\ and\
  \citenamefont {{Hughes}}}]{2009MNRAS.400.1241C}%
  \BibitemOpen
  \bibfield  {author} {\bibinfo {author} {\bibfnamefont {N.~F.}\ \bibnamefont
  {{Camus}}}, \bibinfo {author} {\bibfnamefont {S.~S.}\ \bibnamefont
  {{Komissarov}}}, \bibinfo {author} {\bibfnamefont {N.}~\bibnamefont
  {{Bucciantini}}},\ and\ \bibinfo {author} {\bibfnamefont {P.~A.}\
  \bibnamefont {{Hughes}}},\ }\bibfield  {title} {\bibinfo {title}
  {{Observations of `wisps' in magnetohydrodynamic simulations of the Crab
  Nebula}},\ }\href {https://doi.org/10.1111/j.1365-2966.2009.15550.x}
  {\bibfield  {journal} {\bibinfo  {journal} {Mon.\ Not.\ R.\ Astron.\ Soc.}\
  }\textbf {\bibinfo {volume} {400}},\ \bibinfo {pages} {1241} (\bibinfo {year}
  {2009})}\BibitemShut {NoStop}%
\bibitem [{\citenamefont {{Porth}}\ \emph {et~al.}(2014)\citenamefont
  {{Porth}}, \citenamefont {{Komissarov}},\ and\ \citenamefont
  {{Keppens}}}]{2014MNRAS.438..278P}%
  \BibitemOpen
  \bibfield  {author} {\bibinfo {author} {\bibfnamefont {O.}~\bibnamefont
  {{Porth}}}, \bibinfo {author} {\bibfnamefont {S.~S.}\ \bibnamefont
  {{Komissarov}}},\ and\ \bibinfo {author} {\bibfnamefont {R.}~\bibnamefont
  {{Keppens}}},\ }\bibfield  {title} {\bibinfo {title} {{Three-dimensional
  magnetohydrodynamic simulations of the Crab nebula}},\ }\href
  {https://doi.org/10.1093/mnras/stt2176} {\bibfield  {journal} {\bibinfo
  {journal} {Mon.\ Not.\ R.\ Astron.\ Soc.}\ }\textbf {\bibinfo {volume}
  {438}},\ \bibinfo {pages} {278} (\bibinfo {year} {2014})}\BibitemShut
  {NoStop}%
\bibitem [{\citenamefont {{Olmi}}\ \emph {et~al.}(2016)\citenamefont {{Olmi}},
  \citenamefont {{Del Zanna}}, \citenamefont {{Amato}}, \citenamefont
  {{Bucciantini}},\ and\ \citenamefont {{Mignone}}}]{2016JPlPh..82f6301O}%
  \BibitemOpen
  \bibfield  {author} {\bibinfo {author} {\bibfnamefont {B.}~\bibnamefont
  {{Olmi}}}, \bibinfo {author} {\bibfnamefont {L.}~\bibnamefont {{Del Zanna}}},
  \bibinfo {author} {\bibfnamefont {E.}~\bibnamefont {{Amato}}}, \bibinfo
  {author} {\bibfnamefont {N.}~\bibnamefont {{Bucciantini}}},\ and\ \bibinfo
  {author} {\bibfnamefont {A.}~\bibnamefont {{Mignone}}},\ }\bibfield  {title}
  {\bibinfo {title} {{Multi-D magnetohydrodynamic modelling of pulsar wind
  nebulae: recent progress and open questions}},\ }\href
  {https://doi.org/10.1017/S0022377816000957} {\bibfield  {journal} {\bibinfo
  {journal} {J. Plasma Phys.}\ }\textbf {\bibinfo {volume} {82}},\ \bibinfo
  {eid} {635820601} (\bibinfo {year} {2016})}\BibitemShut {NoStop}%
\bibitem [{\citenamefont {{Zrake}}\ and\ \citenamefont
  {{Arons}}(2017)}]{2017ApJ...847...57Z}%
  \BibitemOpen
  \bibfield  {author} {\bibinfo {author} {\bibfnamefont {J.}~\bibnamefont
  {{Zrake}}}\ and\ \bibinfo {author} {\bibfnamefont {J.}~\bibnamefont
  {{Arons}}},\ }\bibfield  {title} {\bibinfo {title} {{Turbulent magnetic
  relaxation in Pulsar wind nebulae}},\ }\href
  {https://doi.org/10.3847/1538-4357/aa826d} {\bibfield  {journal} {\bibinfo
  {journal} {Astrophys. J.}\ }\textbf {\bibinfo {volume} {847}},\ \bibinfo
  {eid} {57} (\bibinfo {year} {2017})}\BibitemShut {NoStop}%
\bibitem [{\citenamefont {{Tanaka}}\ \emph {et~al.}(2018)\citenamefont
  {{Tanaka}}, \citenamefont {{Toma}},\ and\ \citenamefont
  {{Tominaga}}}]{2018MNRAS.478.4622T}%
  \BibitemOpen
  \bibfield  {author} {\bibinfo {author} {\bibfnamefont {S.~J.}\ \bibnamefont
  {{Tanaka}}}, \bibinfo {author} {\bibfnamefont {K.}~\bibnamefont {{Toma}}},\
  and\ \bibinfo {author} {\bibfnamefont {N.}~\bibnamefont {{Tominaga}}},\
  }\bibfield  {title} {\bibinfo {title} {{Confinement of the Crab nebula with
  tangled magnetic field by its supernova remnant}},\ }\href
  {https://doi.org/10.1093/mnras/sty1356} {\bibfield  {journal} {\bibinfo
  {journal} {Mon.\ Not.\ R.\ Astron.\ Soc.}\ }\textbf {\bibinfo {volume}
  {478}},\ \bibinfo {pages} {4622} (\bibinfo {year} {2018})}\BibitemShut
  {NoStop}%
\bibitem [{\citenamefont {{Bucciantini}}\ \emph {et~al.}(2023)\citenamefont
  {{Bucciantini}}, \citenamefont {{Ferrazzoli}}, \citenamefont {{Bachetti}},
  \citenamefont {{Rankin}}, \citenamefont {{Di Lalla}}, \citenamefont
  {{Sgr{\`o}}}, \citenamefont {{Omodei}}, \citenamefont {{Kitaguchi}},
  \citenamefont {{Mizuno}}, \citenamefont {{Gunji}}, \citenamefont
  {{Watanabe}}, \citenamefont {{Baldini}}, \citenamefont {{Slane}},
  \citenamefont {{Weisskopf}}, \citenamefont {{Romani}}, \citenamefont
  {{Possenti}}, \citenamefont {{Marshall}}, \citenamefont {{Silvestri}},
  \citenamefont {{Pacciani}}, \citenamefont {{Negro}}, \citenamefont
  {{Muleri}}, \citenamefont {{de O{\~n}a Wilhelmi}}, \citenamefont {{Xie}},
  \citenamefont {{Heyl}}, \citenamefont {{Pesce-Rollins}}, \citenamefont
  {{Wong}}, \citenamefont {{Pilia}}, \citenamefont {{Agudo}}, \citenamefont
  {{Antonelli}}, \citenamefont {{Baumgartner}}, \citenamefont {{Bellazzini}},
  \citenamefont {{Bianchi}}, \citenamefont {{Bongiorno}}, \citenamefont
  {{Bonino}}, \citenamefont {{Brez}}, \citenamefont {{Capitanio}},
  \citenamefont {{Castellano}}, \citenamefont {{Cavazzuti}}, \citenamefont
  {{Chen}}, \citenamefont {{Ciprini}}, \citenamefont {{Costa}}, \citenamefont
  {{De Rosa}}, \citenamefont {{Del Monte}}, \citenamefont {{Di Gesu}},
  \citenamefont {{Di Marco}}, \citenamefont {{Donnarumma}}, \citenamefont
  {{Doroshenko}}, \citenamefont {{Dov{\v{c}}iak}}, \citenamefont {{Ehlert}},
  \citenamefont {{Enoto}}, \citenamefont {{Evangelista}}, \citenamefont
  {{Fabiani}}, \citenamefont {{Garcia}}, \citenamefont {{Hayashida}},
  \citenamefont {{Iwakiri}}, \citenamefont {{Jorstad}}, \citenamefont
  {{Kaaret}}, \citenamefont {{Karas}}, \citenamefont {{Kislat}}, \citenamefont
  {{Kolodziejczak}}, \citenamefont {{Krawczynski}}, \citenamefont {{La
  Monaca}}, \citenamefont {{Latronico}}, \citenamefont {{Liodakis}},
  \citenamefont {{Maldera}}, \citenamefont {{Manfreda}}, \citenamefont
  {{Marin}}, \citenamefont {{Marinucci}}, \citenamefont {{Marscher}},
  \citenamefont {{Massaro}}, \citenamefont {{Matt}}, \citenamefont
  {{Mitsuishi}}, \citenamefont {{Ng}}, \citenamefont {{O'Dell}}, \citenamefont
  {{Oppedisano}}, \citenamefont {{Papitto}}, \citenamefont {{Pavlov}},
  \citenamefont {{Peirson}}, \citenamefont {{Perri}}, \citenamefont
  {{Petrucci}}, \citenamefont {{Poutanen}}, \citenamefont {{Puccetti}},
  \citenamefont {{Ramsey}}, \citenamefont {{Ratheesh}}, \citenamefont
  {{Roberts}}, \citenamefont {{Soffitta}}, \citenamefont {{Spandre}},
  \citenamefont {{Swartz}}, \citenamefont {{Tamagawa}}, \citenamefont
  {{Tavecchio}}, \citenamefont {{Taverna}}, \citenamefont {{Tawara}},
  \citenamefont {{Tennant}}, \citenamefont {{Thomas}}, \citenamefont
  {{Tombesi}}, \citenamefont {{Trois}}, \citenamefont {{Tsygankov}},
  \citenamefont {{Turolla}}, \citenamefont {{Vink}}, \citenamefont {{Wu}},\
  and\ \citenamefont {{Zane}}}]{2023NatAs...7..602B}%
  \BibitemOpen
  \bibfield  {author} {\bibinfo {author} {\bibfnamefont {N.}~\bibnamefont
  {{Bucciantini}}}, \bibinfo {author} {\bibfnamefont {R.}~\bibnamefont
  {{Ferrazzoli}}}, \bibinfo {author} {\bibfnamefont {M.}~\bibnamefont
  {{Bachetti}}}, \bibinfo {author} {\bibfnamefont {J.}~\bibnamefont
  {{Rankin}}}, \bibinfo {author} {\bibfnamefont {N.}~\bibnamefont {{Di
  Lalla}}}, \bibinfo {author} {\bibfnamefont {C.}~\bibnamefont {{Sgr{\`o}}}},
  \bibinfo {author} {\bibfnamefont {N.}~\bibnamefont {{Omodei}}}, \bibinfo
  {author} {\bibfnamefont {T.}~\bibnamefont {{Kitaguchi}}}, \bibinfo {author}
  {\bibfnamefont {T.}~\bibnamefont {{Mizuno}}}, \bibinfo {author}
  {\bibfnamefont {S.}~\bibnamefont {{Gunji}}}, \bibinfo {author} {\bibfnamefont
  {E.}~\bibnamefont {{Watanabe}}}, \bibinfo {author} {\bibfnamefont
  {L.}~\bibnamefont {{Baldini}}}, \bibinfo {author} {\bibfnamefont
  {P.}~\bibnamefont {{Slane}}}, \bibinfo {author} {\bibfnamefont {M.~C.}\
  \bibnamefont {{Weisskopf}}}, \bibinfo {author} {\bibfnamefont {R.~W.}\
  \bibnamefont {{Romani}}}, \bibinfo {author} {\bibfnamefont {A.}~\bibnamefont
  {{Possenti}}}, \bibinfo {author} {\bibfnamefont {H.~L.}\ \bibnamefont
  {{Marshall}}}, \bibinfo {author} {\bibfnamefont {S.}~\bibnamefont
  {{Silvestri}}}, \bibinfo {author} {\bibfnamefont {L.}~\bibnamefont
  {{Pacciani}}}, \bibinfo {author} {\bibfnamefont {M.}~\bibnamefont {{Negro}}},
  \bibinfo {author} {\bibfnamefont {F.}~\bibnamefont {{Muleri}}}, \bibinfo
  {author} {\bibfnamefont {E.}~\bibnamefont {{de O{\~n}a Wilhelmi}}}, \bibinfo
  {author} {\bibfnamefont {F.}~\bibnamefont {{Xie}}}, \bibinfo {author}
  {\bibfnamefont {J.}~\bibnamefont {{Heyl}}}, \bibinfo {author} {\bibfnamefont
  {M.}~\bibnamefont {{Pesce-Rollins}}}, \bibinfo {author} {\bibfnamefont
  {J.}~\bibnamefont {{Wong}}}, \bibinfo {author} {\bibfnamefont
  {M.}~\bibnamefont {{Pilia}}}, \bibinfo {author} {\bibfnamefont
  {I.}~\bibnamefont {{Agudo}}}, \bibinfo {author} {\bibfnamefont {L.~A.}\
  \bibnamefont {{Antonelli}}}, \bibinfo {author} {\bibfnamefont {W.~H.}\
  \bibnamefont {{Baumgartner}}}, \bibinfo {author} {\bibfnamefont
  {R.}~\bibnamefont {{Bellazzini}}}, \bibinfo {author} {\bibfnamefont
  {S.}~\bibnamefont {{Bianchi}}}, \bibinfo {author} {\bibfnamefont {S.~D.}\
  \bibnamefont {{Bongiorno}}}, \bibinfo {author} {\bibfnamefont
  {R.}~\bibnamefont {{Bonino}}}, \bibinfo {author} {\bibfnamefont
  {A.}~\bibnamefont {{Brez}}}, \bibinfo {author} {\bibfnamefont
  {F.}~\bibnamefont {{Capitanio}}}, \bibinfo {author} {\bibfnamefont
  {S.}~\bibnamefont {{Castellano}}}, \bibinfo {author} {\bibfnamefont
  {E.}~\bibnamefont {{Cavazzuti}}}, \bibinfo {author} {\bibfnamefont {C.-T.}\
  \bibnamefont {{Chen}}}, \bibinfo {author} {\bibfnamefont {S.}~\bibnamefont
  {{Ciprini}}}, \bibinfo {author} {\bibfnamefont {E.}~\bibnamefont {{Costa}}},
  \bibinfo {author} {\bibfnamefont {A.}~\bibnamefont {{De Rosa}}}, \bibinfo
  {author} {\bibfnamefont {E.}~\bibnamefont {{Del Monte}}}, \bibinfo {author}
  {\bibfnamefont {L.}~\bibnamefont {{Di Gesu}}}, \bibinfo {author}
  {\bibfnamefont {A.}~\bibnamefont {{Di Marco}}}, \bibinfo {author}
  {\bibfnamefont {I.}~\bibnamefont {{Donnarumma}}}, \bibinfo {author}
  {\bibfnamefont {V.}~\bibnamefont {{Doroshenko}}}, \bibinfo {author}
  {\bibfnamefont {M.}~\bibnamefont {{Dov{\v{c}}iak}}}, \bibinfo {author}
  {\bibfnamefont {S.~R.}\ \bibnamefont {{Ehlert}}}, \bibinfo {author}
  {\bibfnamefont {T.}~\bibnamefont {{Enoto}}}, \bibinfo {author} {\bibfnamefont
  {Y.}~\bibnamefont {{Evangelista}}}, \bibinfo {author} {\bibfnamefont
  {S.}~\bibnamefont {{Fabiani}}}, \bibinfo {author} {\bibfnamefont {J.~A.}\
  \bibnamefont {{Garcia}}}, \bibinfo {author} {\bibfnamefont {K.}~\bibnamefont
  {{Hayashida}}}, \bibinfo {author} {\bibfnamefont {W.}~\bibnamefont
  {{Iwakiri}}}, \bibinfo {author} {\bibfnamefont {S.~G.}\ \bibnamefont
  {{Jorstad}}}, \bibinfo {author} {\bibfnamefont {P.}~\bibnamefont {{Kaaret}}},
  \bibinfo {author} {\bibfnamefont {V.}~\bibnamefont {{Karas}}}, \bibinfo
  {author} {\bibfnamefont {F.}~\bibnamefont {{Kislat}}}, \bibinfo {author}
  {\bibfnamefont {J.~J.}\ \bibnamefont {{Kolodziejczak}}}, \bibinfo {author}
  {\bibfnamefont {H.}~\bibnamefont {{Krawczynski}}}, \bibinfo {author}
  {\bibfnamefont {F.}~\bibnamefont {{La Monaca}}}, \bibinfo {author}
  {\bibfnamefont {L.}~\bibnamefont {{Latronico}}}, \bibinfo {author}
  {\bibfnamefont {I.}~\bibnamefont {{Liodakis}}}, \bibinfo {author}
  {\bibfnamefont {S.}~\bibnamefont {{Maldera}}}, \bibinfo {author}
  {\bibfnamefont {A.}~\bibnamefont {{Manfreda}}}, \bibinfo {author}
  {\bibfnamefont {F.}~\bibnamefont {{Marin}}}, \bibinfo {author} {\bibfnamefont
  {A.}~\bibnamefont {{Marinucci}}}, \bibinfo {author} {\bibfnamefont {A.~P.}\
  \bibnamefont {{Marscher}}}, \bibinfo {author} {\bibfnamefont
  {F.}~\bibnamefont {{Massaro}}}, \bibinfo {author} {\bibfnamefont
  {G.}~\bibnamefont {{Matt}}}, \bibinfo {author} {\bibfnamefont
  {I.}~\bibnamefont {{Mitsuishi}}}, \bibinfo {author} {\bibfnamefont {C.~Y.}\
  \bibnamefont {{Ng}}}, \bibinfo {author} {\bibfnamefont {S.~L.}\ \bibnamefont
  {{O'Dell}}}, \bibinfo {author} {\bibfnamefont {C.}~\bibnamefont
  {{Oppedisano}}}, \bibinfo {author} {\bibfnamefont {A.}~\bibnamefont
  {{Papitto}}}, \bibinfo {author} {\bibfnamefont {G.~G.}\ \bibnamefont
  {{Pavlov}}}, \bibinfo {author} {\bibfnamefont {A.~L.}\ \bibnamefont
  {{Peirson}}}, \bibinfo {author} {\bibfnamefont {M.}~\bibnamefont {{Perri}}},
  \bibinfo {author} {\bibfnamefont {P.-O.}\ \bibnamefont {{Petrucci}}},
  \bibinfo {author} {\bibfnamefont {J.}~\bibnamefont {{Poutanen}}}, \bibinfo
  {author} {\bibfnamefont {S.}~\bibnamefont {{Puccetti}}}, \bibinfo {author}
  {\bibfnamefont {B.~D.}\ \bibnamefont {{Ramsey}}}, \bibinfo {author}
  {\bibfnamefont {A.}~\bibnamefont {{Ratheesh}}}, \bibinfo {author}
  {\bibfnamefont {O.~J.}\ \bibnamefont {{Roberts}}}, \bibinfo {author}
  {\bibfnamefont {P.}~\bibnamefont {{Soffitta}}}, \bibinfo {author}
  {\bibfnamefont {G.}~\bibnamefont {{Spandre}}}, \bibinfo {author}
  {\bibfnamefont {D.}~\bibnamefont {{Swartz}}}, \bibinfo {author}
  {\bibfnamefont {T.}~\bibnamefont {{Tamagawa}}}, \bibinfo {author}
  {\bibfnamefont {F.}~\bibnamefont {{Tavecchio}}}, \bibinfo {author}
  {\bibfnamefont {R.}~\bibnamefont {{Taverna}}}, \bibinfo {author}
  {\bibfnamefont {Y.}~\bibnamefont {{Tawara}}}, \bibinfo {author}
  {\bibfnamefont {A.~F.}\ \bibnamefont {{Tennant}}}, \bibinfo {author}
  {\bibfnamefont {N.~E.}\ \bibnamefont {{Thomas}}}, \bibinfo {author}
  {\bibfnamefont {F.}~\bibnamefont {{Tombesi}}}, \bibinfo {author}
  {\bibfnamefont {A.}~\bibnamefont {{Trois}}}, \bibinfo {author} {\bibfnamefont
  {S.}~\bibnamefont {{Tsygankov}}}, \bibinfo {author} {\bibfnamefont
  {R.}~\bibnamefont {{Turolla}}}, \bibinfo {author} {\bibfnamefont
  {J.}~\bibnamefont {{Vink}}}, \bibinfo {author} {\bibfnamefont
  {K.}~\bibnamefont {{Wu}}},\ and\ \bibinfo {author} {\bibfnamefont
  {S.}~\bibnamefont {{Zane}}},\ }\bibfield  {title} {\bibinfo {title}
  {{Simultaneous space and phase resolved X-ray polarimetry of the Crab pulsar
  and nebula}},\ }\href {https://doi.org/10.1038/s41550-023-01936-8} {\bibfield
   {journal} {\bibinfo  {journal} {Nat. Astron.}\ }\textbf {\bibinfo {volume}
  {7}},\ \bibinfo {pages} {602} (\bibinfo {year} {2023})}\BibitemShut {NoStop}%
\bibitem [{\citenamefont {{Pshirkov}}\ \emph {et~al.}(2020)\citenamefont
  {{Pshirkov}}, \citenamefont {{Nizamov}}, \citenamefont {{Bykov}},\ and\
  \citenamefont {{Uvarov}}}]{2020MNRAS.496.5227P}%
  \BibitemOpen
  \bibfield  {author} {\bibinfo {author} {\bibfnamefont {M.~S.}\ \bibnamefont
  {{Pshirkov}}}, \bibinfo {author} {\bibfnamefont {B.~A.}\ \bibnamefont
  {{Nizamov}}}, \bibinfo {author} {\bibfnamefont {A.~M.}\ \bibnamefont
  {{Bykov}}},\ and\ \bibinfo {author} {\bibfnamefont {Y.~A.}\ \bibnamefont
  {{Uvarov}}},\ }\bibfield  {title} {\bibinfo {title} {{Gamma-ray flux
  depressions of the Crab Nebula in the high-energy range}},\ }\href
  {https://doi.org/10.1093/mnras/staa1921} {\bibfield  {journal} {\bibinfo
  {journal} {Mon.\ Not.\ R.\ Astron.\ Soc.}\ }\textbf {\bibinfo {volume}
  {496}},\ \bibinfo {pages} {5227} (\bibinfo {year} {2020})}\BibitemShut
  {NoStop}%
\bibitem [{\citenamefont {{Yeung}}\ and\ \citenamefont
  {{Horns}}(2020)}]{2020A&A...638A.147Y}%
  \BibitemOpen
  \bibfield  {author} {\bibinfo {author} {\bibfnamefont {P.~K.~H.}\
  \bibnamefont {{Yeung}}}\ and\ \bibinfo {author} {\bibfnamefont
  {D.}~\bibnamefont {{Horns}}},\ }\bibfield  {title} {\bibinfo {title} {{Fermi
  Large Area Telescope observations of the fast-dimming Crab Nebula in 60-600
  MeV}},\ }\href {https://doi.org/10.1051/0004-6361/201936740} {\bibfield
  {journal} {\bibinfo  {journal} {Astron.\ Astrophys.}\ }\textbf {\bibinfo
  {volume} {638}},\ \bibinfo {eid} {A147} (\bibinfo {year} {2020})}\BibitemShut
  {NoStop}%
\bibitem [{\citenamefont {{Huang}}\ \emph {et~al.}(2021)\citenamefont
  {{Huang}}, \citenamefont {{Yuan}},\ and\ \citenamefont
  {{Fan}}}]{2021ApJ...908...65H}%
  \BibitemOpen
  \bibfield  {author} {\bibinfo {author} {\bibfnamefont {X.}~\bibnamefont
  {{Huang}}}, \bibinfo {author} {\bibfnamefont {Q.}~\bibnamefont {{Yuan}}},\
  and\ \bibinfo {author} {\bibfnamefont {Y.-Z.}\ \bibnamefont {{Fan}}},\
  }\bibfield  {title} {\bibinfo {title} {{A systematic study of gamma-rafy
  flares from the Crab nebula with Fermi-LAT. I. Flare detection}},\ }\href
  {https://doi.org/10.3847/1538-4357/abd2b7} {\bibfield  {journal} {\bibinfo
  {journal} {Astrophys. J.}\ }\textbf {\bibinfo {volume} {908}},\ \bibinfo
  {eid} {65} (\bibinfo {year} {2021})}\BibitemShut {NoStop}%
\end{thebibliography}%

\end{document}